\documentclass[twocolumn,
               superscriptaddress,
               floatfix,
               longbibliography, 
               showkeys,apl]{revtex4-1}

\usepackage{graphicx} 
\usepackage{bm}
\usepackage{color}                     
\usepackage{xcolor}
\usepackage{epsfig}
\usepackage{amsmath} 
\usepackage{amssymb} 
\usepackage{longtable} 
\usepackage{float}
\usepackage{mathtools}
\usepackage{xparse}
\usepackage{hyperref}
\usepackage{times}
\usepackage{bbold}

\usepackage[normalem]{ulem}

\usepackage[ruled]{algorithm}
\usepackage{algpseudocode}

\usepackage{enumitem}
\usepackage{tabularx,lipsum,environ} 
\usepackage{sidecap}
\sidecaptionvpos{figure}{t}

\newcommand{\be}{\begin{equation}}
\newcommand{\ee}{\end{equation}}
\newcommand{\bd}{\begin{displaymath}}
\newcommand{\ed}{\end{displaymath}}
\newcommand{\BE}{\begin{eqnarray}}
\newcommand{\EE}{\end{eqnarray}}

\definecolor{darkgreen}{rgb}{0.0, 0.5, 0.0}

\makeatletter

\newcolumntype{\expand}{}
\long\@namedef{NC@rewrite@\string\expand}{\expandafter\NC@find}

\NewEnviron{problem}[2][]{%
  \def\problem@arg{#1}%
  \def\problem@framed{framed}%
  \def\problem@lined{lined}%
  \def\problem@doublelined{doublelined}%
  \ifx\problem@arg\@empty%
    \def\problem@hline{}%
  \else%
    \ifx\problem@arg\problem@doublelined%
      \def\problem@hline{\hline\hline}%
    \else%
      \def\problem@hline{\hline}%
    \fi%
  \fi%
  \ifx\problem@arg\problem@framed%
    \def\problem@tablelayout{|>{\bfseries}lX|c}%
    \def\problem@title{\multicolumn{2}{|%
      >{\raisebox{-\fboxsep}}%
      p{\dimexpr\textwidth-4\fboxsep-2\arrayrulewidth\relax}%
      |}{%
        \textsc{\Large #2}%
      }}%
  \else
    \def\problem@tablelayout{>{\bfseries}lXc}%
    \def\problem@title{\multicolumn{2}{>%
      {\raisebox{-\fboxsep}}%
      p{\dimexpr\textwidth-4\fboxsep\relax}%
      }{%
        \textsc{\large #2}%
      }}%
  \fi%
  \bigskip\par\noindent%
  \begin{tabularx}{\textwidth}{\expand\problem@tablelayout}%
    \problem@hline%
    \problem@title\\[2\fboxsep]%
    \BODY\\\problem@hline%
  \end{tabularx}%
  \medskip\par%
}
\makeatother

\begin{document}

\setlist[enumerate,1]{label=\arabic*, start=0}

\title{Symmetric Tensor Networks for Generative Modeling and Constrained Combinatorial Optimization} 

\author{Javier Lopez-Piqueres}
\affiliation{Zapata Computing Inc., 100 Federal Street, Boston, MA 02110, USA}
\affiliation{Department of Physics, University of Massachusetts, Amherst, Massachusetts 01003, USA }

\author{Jing Chen}
\email{jing.chen@zapatacomputing.com}
\affiliation{Zapata Computing Inc., 100 Federal Street, Boston, MA 02110, USA}

\author{Alejandro Perdomo-Ortiz}
\email{alejandro@zapatacomputing.com}
\affiliation{Zapata Computing Canada Inc., 25 Adelaide St East, M5C 3A1, Toronto, ON, Canada}

\date{\today} 

\begin{abstract}
Constrained combinatorial optimization problems abound in industry, from portfolio optimization to logistics. One of the major roadblocks in solving these problems is the presence of non-trivial hard constraints which limit the valid search space. In some heuristic solvers, these are typically addressed by introducing certain Lagrange multipliers in the cost function, by relaxing them in some way, or worse yet, by generating many samples and only keeping valid ones, which leads to very expensive and inefficient searches. In this work, we encode arbitrary integer-valued equality constraints of the form $A \vec{x} = \vec{b}$, directly into $U(1)$ symmetric tensor networks and leverage their applicability as quantum-inspired generative models to assist in the search of solutions to combinatorial optimization problems within the \textit{generator-enhanced optimization} (GEO) framework of Ref. \cite{alcazar2021enhancing}. This allows us to exploit the generalization capabilities of TN generative models while constraining them so that they only output valid samples. Our constrained TN generative model efficiently captures the constraints by reducing number of parameters and computational costs. We find that at tasks with constraints given by arbitrary equalities, symmetric Matrix Product States outperform their standard unconstrained counterparts at finding novel and better solutions to combinatorial optimization problems.
\end{abstract}

\maketitle

\section{Introduction}\label{s:intro}
Tensor networks (TNs) have emerged in the past few years as a very powerful tool to address challenging problems in high-dimensional spaces, such as quantum many-body physics \cite{orus2019tensor},  electronic structure of chemical compounds \cite{chen2022using}, supervised learning \cite{stoudenmire2016supervised}, generative modeling \cite{han2018unsupervised}, anomaly detection \cite{wang2020anomaly}, combinatorial optimization \cite{liu2021tropical,liu2022computing,hao2022quantum}, and privacy leaking \cite{pozas2022physics}, to name a few. The reason for their wide applicability lies in part for their well-understood mathematical structure and well-controlled expressibility.

Just as artificial neural networks (ANNs) are universal approximators of generic distributions \cite{hornik1989multilayer}, TNs can in principle model any (discrete) distribution with enough computational resources \cite{glasser2019}. While ANNs are inspired by the functioning of the brain (in particular their expressibility determined by the number of activation functions, mimicking the firing of neurons in the brain), themselves nonlinear functions of their input, TNs are inspired by quantum mechanics, arguably a theory with richer mathematical structure than that of ANNs. For instance, the Hilbert space structure of quantum mechanics allows to exploit global \textit{internal} symmetries in quantum states through the use of representation theory. This can have important consequences when we aim to efficiently express probability distributions with redundancies that appear as a result of such symmetries.

It may appear at first sight that exploiting internal symmetries such as $U(1)$ or $SU(2)$ would bear little importance in problems outside physics. In this work, we show that by exploiting $U(1)$ symmetry in TN states it is possible to encode arbitrary integer-valued equalities of the form $A\vec{x}=\vec{b}$ in a probabilistic model, extending the family of so-called tensor network Born machine (TNBM) models~\cite{han2018unsupervised}. Therefore, we refer to our new family of symmetric quantum-inspired TN generative models as s-TNBMs.

This type of equalities appear in many combinatorial optimization problems with \textit{hard} constraints, such as in portfolio optimization, warehouse location, scheduling, transportation, to name a few~\cite{Markowitz52, conforti2014integer}. Current state-of-the-art methods at dealing with combinatorial optimization problems with hard-constraints, such as \textsc{Mixed-integer Linear Programming} (\textsc{MILP}) solvers \cite{gleixner2021miplib, gurobi2018gurobi}, simulated annealing solvers \cite{hauke2020perspectives}, or more recently, Graph Neural Networks (GNNs) \cite{nair2020solving, cappart2021combinatorial, schuetz2022combinatorial} and fully quantum models such as the Quantum Approximate Optimization Algorithm (QAOA) \cite{Farhi2014}, suffer from at least one of the following two drawbacks: they must explicitly break the integrality constraint to then project back to the nearest integer solution (as done in standard MILP solvers relying on the \textit{branch-and-bound} method \cite{conforti2014integer}), and/or they only work for cost functions expressed as polynomials in the binary variables, with linear or quadratic being the most common ones. Here, we focus on a recent proposal, the generator-enhanced optimization (GEO) framework~\cite{alcazar2021enhancing}, which bypasses both of these limitations.

GEO solvers leverage classical or quantum generative models to assist the solution of combinatorial optimization problems, by learning the correlations from observations (i.e., bitstrings and their corresponding costs) and proposing new bitstring solution candidates which resemble those in the lowest cost sector. This is done by reweighting the seen information, which is set as the dataset to train the generative model. This reweighting helps to give more importance to those samples in the training dataset with the lowest cost (see Sec.~\ref{s:prob_statement} for more details). It was shown in Ref.~\cite{alcazar2021enhancing} the power of quantum-inspired generative models based on TNBMs within GEO, matching the performance of state-of-the-art metaheuristics tailored in the last 30 years for a specific variant of the portfolio optimization problem. These remarkable results were obtained without imposing any bias in the TNBM model to enforce the cardinality constrains (i.e., fixed desired number $k$ of assets in the portfolio out of $N$) of the combinatorial problems.

In this work, we exploit the rather simple structure of TNs, in particular their linearity and local connectivity, to encode arbitrary equality constraints, cardinality being a specific realization of these, for the purpose of both generative modeling and combinatorial optimization. We do so by mapping the problem of finding solutions satisfying the constraints $A\vec{x}=\vec{b}$ to that of finding the set of \textit{quantum numbers} (QNs) appearing in symmetric TNs. In this context, QNs effectively correspond to labels of different irreducible representations, in this case of $U(1)$, defined on each local Hilbert space where each tensor is supported.

For equality constraints with high-degree of symmetry, where the coefficients in matrix $A$ have low variance, we find that symmetric TNs correspond to an efficient encoding of the valid space. For a subset of these, such as cardinality type constraints appearing in problems such as portfolio optimization~\cite{alcazar2021enhancing}, one can write down explicitly the TN ansatz for the entire valid space, which can be used as an initial ansatz for finding optimal solutions. For arbitrary integer-valued equalities finding the solution space is likely \textsc{\#P}-complete (since its associated decision problem, \textsc{(0,1)-Integer Programming}, is already \textsc{NP}-complete \cite{schrijver1998theory}), but we provide a novel message-passing algorithm with QNs acting as \textit{messages}, that is able to encode a large fraction of these solutions within a few QNs, thereby allowing to generate many unseen solutions satisfying $A \vec{x}=\vec{b}$ when given access to some (valid) training data, even when the entries of $A$ and $\vec{b}$ are random integers. At the same time, by exploiting the \textit{dimensionality} of each QN, the resulting TN ansatz is able to capture not only the constraints efficiently, but also biases in the dataset, therefore favoring certain solutions over others.

Moreover, by constraining the TN so as to only sample within the valid subspace, we guarantee computational savings, in both memory usage by exploiting block-sparsity of the tensors as well as computational time by guaranteeing that only the needed tensor components get updated. Beyond the ability to impose arbitary equality constraints, TNs enjoy sophisticated optimization algorithms, such as the Density Matrix Renormalization Group (DMRG) algorithm \cite{white1992density,schollwock2011density}  for Matrix Product States (MPS), as well as perfect sampling \cite{ferris2012perfect} (which avoid long autocorrelation times typical of Monte Carlo based sampling algorithms).

While we envision other uses of constrained TNs as efficient representations of constrained datasets in various contexts, we find that some of its most crucial advantages are captured within the GEO framework of Ref. \cite{alcazar2021enhancing}. This is because GEO exploits the presence of existing training data to generate new and better solutions to combinatorial optimization problems by capturing biases in the training set (corresponding to bitstrings of different costs), which our constrained TN approach is capable of incorporating when given data subject to equality constraints.

The structure of the paper is as follows. In Sec.\ref{s:prob_statement} we state the problem we are interested in, namely combinatorial optimization problems subject to equality constraints, and describe our approach to the problem using a generative model based on tensor networks. In Sec. \ref{s:symmetric_MPS} we give a brief introduction to symmetric TNs placing emphasis on symmetric MPS, a special TN structure which will be used in our numerical experiments. In Sec. \ref{s:generative_algo} we explain the main encoding and generative algorithms. In Sec. \ref{s:results} we present various results showcasing the types of canonical problems symmetric MPS can be applied to and compare its performance against vanilla versions (i.e. without imposing any constraint on them). In Sec. \ref{s:efficient} we give strong arguments in favor of using TNs for describing symmetric states over ANN architectures, implying in particular the efficiency of symmetric TNs at capturing arbitrary equality constraints when restricted to $U(1)$ symmetric states. Sec. \ref{s:conclusions} presents our conclusion, and Sec.\ref{s_outlook} gives an outlook for future work.

\section{Problem Statement}~\label{s:prob_statement}
In combinatorial optimization problems, one is interested in finding the minimum of a cost function over a specific set of instances, called valid (or \textit{feasible}) space. Often the set of valid instances grows exponentially with the problem size (such as in the presence of cardinality constraints), which prevents listing exhaustively all possible candidates. Although not all combinatorial problems have explicit constraints, the class of real-world problems that we focus on here are those which can be written as:
\begin{align}
\begin{split} \label{eq_minimization}
&\text{min } \mathcal{C}(\vec{x}), \\
&\text{subject to } A\vec{x}=\vec{b},
\end{split}
\end{align}
where $\mathcal{C}:\{0,1\}^N \to \mathbb{R}$, $\vec{x}\in \{0,1\}^N$, and $A\in \mathbb{Z}^{m \times N}$, $\vec{b} \in \mathbb{Z}^m$. (Some of these conditions may be relaxed in principle and our construction still applies; for instance, we may convert each integer coefficient to a rational and viceversa via a rescaling of all coefficients in each equality. Similarly, we may work with general integer-valued variables instead of just binary ones by accordingly increasing the dimension of each site vector space in the TN as we will describe in the next section). We will denote by $\mathcal{S}$ the valid or solution space, i.e. the set of all bitstrings satisfying the equality constraints. While not all hard constraints are of the form appearing in Eq.~\ref{eq_minimization} \footnote{in particular, a large class of optimization problems contain inequality type constraints and even nonlinear constraints} our focus is on equality type constraints because these are the ones that best exploit the TN ansatz used in this work.

When the cost function $\mathcal{C}$ is linear, the above statement reduces to the \textit{standard form} of \textsc{(0,1)-Integer Linear Programming} (see Appendix~\ref{s:0-1} \sout{Sec.  \ref{s:0-1} of the Appendix} for an introduction into this class of problems), and when $\mathcal{C}$ is at most quadratic, the corresponding problem can be recasted to a \textsc{Quadratic Unconstrained Binary Optimization (QUBO)} with penalty terms enforcing the equality constraints \cite{glover2018tutorial}. In the GEO framework used here, the form of $\mathcal{C}(\vec{x})$ can be arbitrary, i.e., GEO is a black-box solver. While many powerful methods exist for dealing with the minimization part of the algorithm such as \textit{branch-and-bound} based methods for linear cost functions, and \textit{simulated annealing} for problems with arbitrary cost functions, a major roadblock in many of these solvers is dealing precisely with equality constraints (\textit{hard} constraints) that restrict the valid space, and the standard approaches taken are to relax both the integrality and equality constraints as in branch-and-bound based solvers \cite{schrijver1998theory}, to impose such constraints via introducing Lagrange multipliers directly into the cost function as in most \textsc{QUBO} solvers \cite{glover2018tutorial} (which for many equalities results in an overparameterized cost function), or worse yet, to generate many samples only to keep those that meet the constraints.

\begin{figure}
    \centering
    \includegraphics[width=\linewidth, scale=0.5]{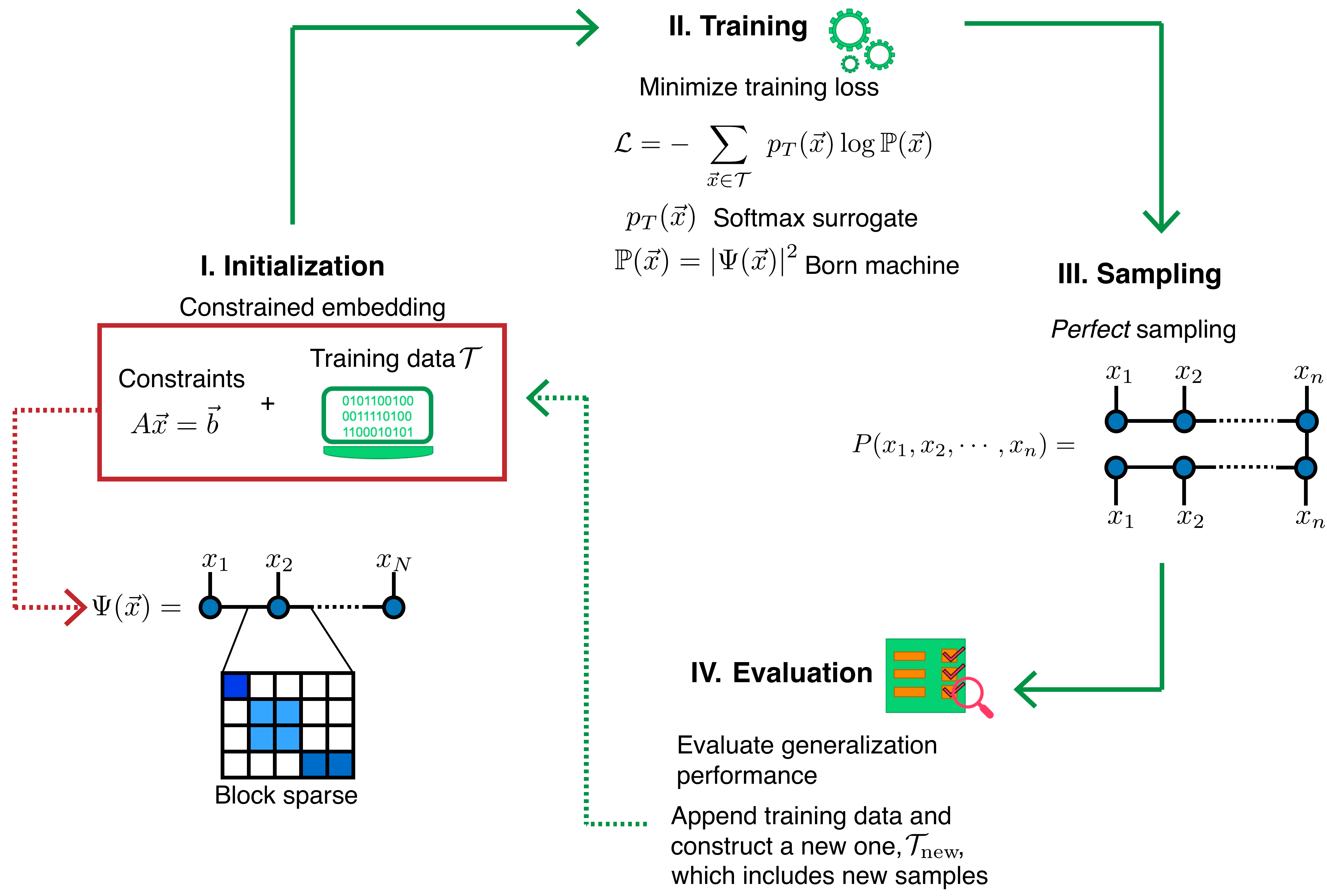}
    \caption{\textbf{The proposed \textit{constrained-GEO} framework.} Our framework consists of four steps. Step I involves encoding bitstrings satisfying $A\vec{x}=\vec{b}$ type constraints into a block sparse tensor network. An efficient embedding of the constraints can be performed in the presence of some training data satisfying the constraints. In Step II, the initialized tensor network is trained by minimizing the cross-entropy or negative log likelihood where the training distribution corresponds to a \textit{softmax} which serves as a \textit{surrogate} to the optimization cost function of interest (see Eq. (\ref{eq_softmax})). The training procedure guarantees block-sparsity of tensors. Step III involves collecting new samples from the model distribution by perfect sampling. In Step IV, new samples can be merged with the training dataset to repeat the process or stop when the user is satisfied with the amount of novel samples or a target desired cost is reached.}
    \label{fig: constrained_GEO_birds_view}
\end{figure}

Inspired by current generative modeling approaches to the task of combinatorial optimization such as the ones from Refs. \cite{alcazar2021enhancing, HibatAllah2022, Bengio2021}, we leverage the generalization capabilities of Tensor Network Born Machines (TNBMs) to address the optimization task in Eq. (\ref{eq_minimization}) while constraining them so that they only output valid samples. Crucially, we do not make any assumption on the nature of the cost function (hence our approach is not limited to linear/quadratic cost functions).  Our approach builds on top of the \textit{generator-enhanced optimization} (GEO) approach of Ref. \cite{alcazar2021enhancing} when subject to arbitrary integer-valued equality constraints. Specifically, denoting \textit{constrained-GEO} the constrained extension of GEO, the problem statement we are interested in can be phrased as follows
\begin{problem}{constrained-GEO}
  Input: & A training dataset $\mathcal{T} \subset \{0,1\}^N$ satisfying $A\vec{x}=\vec{b}$, \\
  & $\forall \vec{x} \in \mathcal{T}$, with $A\in \mathbb{Z}^{m\times N}$, $\vec{b} \in \mathbb{Z}^m$. \\
  Task: & Find new, valid samples, $\vec{x} \in \mathcal{S} \setminus \mathcal{T}$, while\\&  minimizing $\mathcal{L}=-\sum_{\vec{x} \in \mathcal{T}} p_T(\vec{x})\log \mathbb{P}(\vec{x})$.
\end{problem}
Here, $\mathcal{L}$ is the training loss function in the form of the Cross-Entropy or Negative Log Likelihood (NLL), $\mathbb{P}(\vec{x})$ is the \textit{model} distribution represented as a TNBM, i.e. $\mathbb{P}(\vec{x})=|\Psi(\vec{x})|^2/Z$ with $\Psi(\vec{x})$ expressed as a TN and $Z=\sum_{\vec{x}\in \{0,1\}^N}|\Psi(\vec{x})|^2$, and $p_T(\vec{x})$ is the training distribution given by the \textit{softmax}
\begin{equation} \label{eq_softmax}
    p_T(\vec{x})=\frac{1}{M}e^{-\mathcal{C}(\vec{x})/T}, \hspace{0.05in} \vec{x} \in \mathcal{T},
\end{equation}
where $M=\sum_{\vec{x} \in \mathcal{T}}e^{-\mathcal{C}(\vec{x})/T}$ is a normalization factor, and the cost function $\mathcal{C}(\vec{x})$ is in principle arbitrary (not restricted to \textit{polynomial} cost functions) and $T$ is a Lagrange multiplier favoring lower cost samples as $T \to 0$. The motivation for the use of symmetric MPS (and in general, symmetric TNs) is that, as we shall show next, they permit to incorporate equality constraints \textit{directly} into a generative model given in terms of a symmetric MPS, where tensors acquire a block sparse form. In Fig. \ref{fig: constrained_GEO_birds_view} we schematically describe our proposed constrained-GEO framework for solving arbitrary combinatorial optimization problems subject to equality constraints which we will describe in more depth in Sec. \ref{s:generative_algo}.

\section{Symmetric Tensor Networks}~\label{s:symmetric_MPS}
In this section we give a brief introduction to $U(1)$ symmetric MPS (for more details we refer to the Appendix \ref{s:symmetric_TNs_app}). Our motivation here is to set up the notation and terminology that will be used in the following sections as well as to showcase how symmetric MPS allows to encode arbitrary integer-valued equality constraints.
\subsection{Vanilla and $U(1)$ Symmetric Matrix Product States}~\label{s:vn_vs_sym_MPS}
Tensor networks (TNs) are a special ansatz to represent high dimensional data, such as quantum states, based on the contraction of tensors defined on networks. The tensors, whose entries are tunable parameters, are located at the sites of the network, while the links between tensors capture the correlations among different sites and  represent the amount of entanglement shared among them. It has been shown that many states of physical relevance most prominently those obeying the so-called area law of entanglement \cite{hastings2006solving, wolf2008area}, can be efficiently represented by such ansatz. Among the different classes of TNs, MPS corresponding to one-dimensional TNs, has found the widest application, in great part due to its low connectivity as well as the presence of a \textit{canonical form}, which permits very optimized updating and contracting algorithms \cite{schollwock2011density, orus2014practical}. A perspective that is useful when introducing \textit{symmetries} in a tensor network is to view each tensor as a linear map from some \textit{input} vector space $\mathbb{V}^{\rm in}$ to \textit{output} vector space $\mathbb{V}^{\rm out}$. For instance, a 3-rank tensor in an MPS may be of the form $|\alpha\rangle = \sum_{a,\beta} T_{\alpha \beta}^a |a,\beta\rangle$ where $|\alpha\rangle \in \mathbb{V}^{\rm in}$ and $|a,\beta\rangle \in \mathbb{V}^{\rm out}$. Here we make a distinction between upper indices appearing in each tensor component, referring to them as site (or \textit{physical}) indices (indices that are not contracted in the TN), and lower indices corresponding to link (or \textit{virtual}) indices (indices that are contracted and that control the expressibility of the TN ansatz). The dimension of each link's vector space is determined by the bond dimension $\chi$, controlling the amount of \textit{entanglement} shared across nearest neighbors. Its size determines in turn the expressibility of the resulting MPS ansatz. One of the benefits of TNs is that one can reason with them solely based on their diagrammatic representation. For the tensor above, this is represented as

$\begin{gathered}
\includegraphics[scale=0.25]{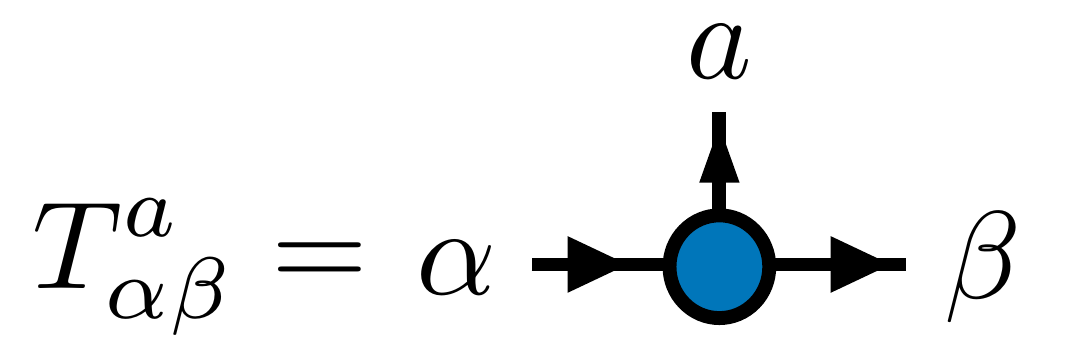}
\end{gathered}$,

where the direction of the arrows reflect whether they refer to incoming/outgoing states.

TNs not only permit to construct states based on their entanglement structure, but also allow to exploit global internal symmetries such as $U(1)$ and $SU(2)$, widely present in many problems of physical relevance. This essentially constrains the form of each tensor so as to fulfill this global requirement. The simplest kind of symmetric TN is the $U(1)$ symmetric MPS. An arbitrary vector $|v\rangle \in \mathbb{V}$ can be expanded as a linear combination of $|n,t_n\rangle \in \mathbb{V}_n$ vectors of fixed QN or \textit{charge} $n$, and degeneracy label $t_n=1,\cdots, d_n$, where $d_n$ is the degeneracy of charge $n$. By going to this basis of well-defined charge, a symmetric MPS can be written in terms of \textit{constrained} tensors of the form \cite{singh2010tensor, singh2011tensor}
\begin{equation} \label{eq_invariant_tensors_main}
    T^{a}_{\alpha,\beta}=(T^{n_a}_{n_\alpha,n_\beta})^{t_{n_a}}_{t_{n_\alpha},t_{n_\beta}}\cdot \delta_{n+N_{\rm in} , N_{\rm out}},
\end{equation}
where
\begin{equation}
N_{\rm in}=\sum_{i\in \mathcal{I}} n_i, \hspace{0.1in} N_{\rm out}=\sum_{i \in \mathcal{O}} n_i,
\end{equation}
where $\mathcal{I}$ ($\mathcal{O}$) denotes the set of incoming (outgoing) indices, and $n$ denotes the \textit{flux} of the tensor, a label carrying the charge of the tensor. When using the canonical form (as will be the case throughout in this work), a symmetric MPS is usually chosen so that the flux of the entire MPS is only carried by one of the tensors (instead of spread across various tensors), the canonical center (also known as \textit{orthogonality center}), while the remaining ones have flux zero ($n=0$ in Eq. \ref{eq_invariant_tensors_main}). Thus, by exploiting global $U(1)$ symmetry we see that each tensor in the MPS factorizes into the product of two tensors. The first one, $(T^{n_a}_{n_\alpha,n_\beta})^{t_a}_{t_\alpha,t_\beta}$, gives the components of tensor $T$ w.r.t. the degeneracy vectors $\{|t_n\rangle\}$, for that reason this tensor goes by the name of \textit{degeneracy tensor} \cite{singh2010tensor}. The second tensor, $\delta_{N_{\rm in},N_{\rm out}}$ depends solely on the conservation of $U(1)$ QNs, and it is referred to as \textit{structural tensor} (for an arbitrary symmetry group $\mathcal{G}$ the structural tensor can take a rather nontrivial form) \cite{singh2010tensor}. For fixed physical QN (fixed $n_a$ appearing in Eq. (\ref{eq_invariant_tensors_main})), the resulting matrix $T^a_{\alpha,\beta}$ will be block diagonal, with blocks of order $d_{n_\alpha}\times d_{n_\beta}$. These blocks are sometimes referred to as \textit{quantum number} (QN) blocks. In other words, charge conservation constrains the form of the tensors to become block sparse, leading to a more efficient representation of the MPS.
\begin{figure}    \centering
    \includegraphics[width=\linewidth, scale=0.5]{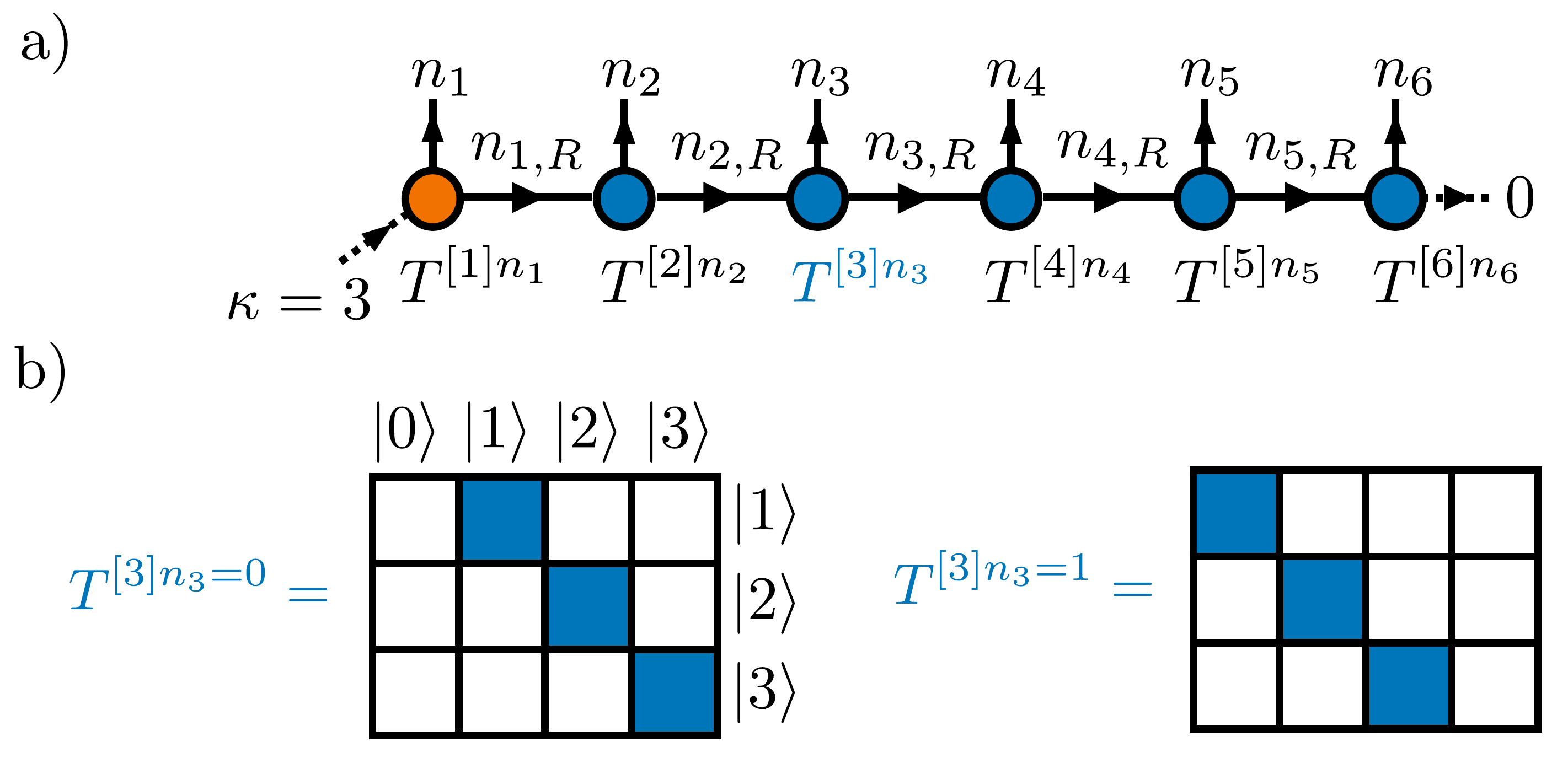}
    \caption{\textbf{Cardinality constrained MPS.} a) The canonical center is chosen at the first site, corresponding to the site where we insert the flux $\kappa$, which for concreteness we choose to be of value $\kappa=3$. Tensors connected with each other  share the same tensor indices.  b) Charge conservation imposes a block-sparse structure in the tensors appearing in the MPS. Example shown for tensor $T^{[3]}$.}
    \label{fig_cardinality_MPS}
\end{figure}
\subsection{Constructing $U(1)$ symmetric MPS} \label{subs_constr_u1}
Having discussed the formalism of $U(1)$ symmetric MPS, we now explain how to explicitly construct a symmetric MPS that only generates bitstrings fulfilling different equality constraints of the form $A\vec{x}=\vec{b}$, with $\vec{x} \in \{0,1\}^N$, $A\in \mathbb{Z}^{m\times N}$, and $\vec{b} \in \mathbb{Z}^m$. Our construction relies on identifying the site (\textit{physical}) charges as the entries of the $A$ matrix together with the empty charge $\varnothing$ (i.e. the charge $0$ in the case of integer valued scalar charges), and the total charge of the MPS (i.e. the flux) as the entries of $\vec{b}$. The remaining task consists in determining all link charges consistent with global charge conservation. This will automatically give us the components of each structural tensor (the degeneracy tensor can always be trivially constructed by augmenting the dimension of each QN block). We remark, however, that finding the set of all possible configurations of link charges that add up to flux $\vec{b}$ is likely \textsc{\#P}-hard, since its associated decision problem, \textsc{(Multiple) Subset Sum}, is at least $\textsc{NP}$-hard (the variant with $A$ and $\vec{b}$ being one dimensional and having non-negative integer entries is for instance \textsc{NP}-complete \cite{kleinberg2006algorithm}). Therefore, in general we will not be interested in finding all such charges for a given $A$, $\vec{b}$, but rather in finding a way to incorporate them into the MPS once these are found (finding all charges for small-sized problems may be done via exhaustive search, random search or dynamic programming).\\

\textit{I. Cardinality constraint -- an exactly solvable case.}

The original motivation behind the construction of symmetric MPS arose from capturing the conservation law of total spin $\hat{S}_z$ along the $z$ direction in spin chains (or alternatively, the conservation of particle number in systems of spinless bosons and fermions). These two types of conservation laws can be viewed as cardinality constraints of the form $\sum_i x_i = \kappa$, where $x_i \in \{0,1\}$, and $\kappa \in \mathbb{N}_0$ s.t. $0 \leq \kappa \leq N$. We shall focus on the case $\kappa \leq N/2$ (assuming $N$ even) since once we find a way to generate bitstrings fulfilling $\sum_i x_i = \kappa \leq N/2$ we can simply flip all bits so as to generate $\sum_i x_i = N-\kappa \geq N/2$.

 Given the simplicity of the constraint, we can construct the set of all charges explicitly. For concreteness, we select the canonical center of the MPS to be located at the first site, where we insert the flux of charge $\kappa$; see Fig. \ref{fig_cardinality_MPS}. The construction of the MPS is accomplished once we determine the charges at the link indices. To distinguish between site charges and link charges we will denote the former via $n_i$ with $i=1,\cdots,N$, and the link charges as $n_{i,R}$, $i=1,\cdots,N-1$ for the link charge immediately to the right of charge $n_i$. As mentioned at the beginning of this subsection, the site charges will be given by the coefficients of the $A$ matrix, together with the $\varnothing$ charge. For cardinality, $n_i \in \{\varnothing, 1\}$.

The construction of the link charges requires solving the set of equations $n_{i+1,R}=n_{i,R}-n_{i+1}$ with the boundary terms $n_{1,R}=\kappa-n_1$ and $n_{N-1,R}=n_N$. The solution is given by
\begin{equation}
    n_{i,R} \in \begin{cases}
    \{\kappa,\kappa-1,\cdots, \kappa-i\}, \hspace{0.05in} &1 \leq i< \kappa, \\
    \{\kappa,\kappa-1,\cdots,\varnothing\}, \hspace{0.05in} &\kappa \leq i \leq N-\kappa, \\
    \{N-i,N-i-1,\cdots,\varnothing\}, \hspace{0.05in} &N-\kappa < i \leq N-1.
    \end{cases}
\end{equation}

Crucially, while the solution space is given by the binomial coefficient $|S|=\binom{N}{\kappa}$ (which in particular for $\kappa \approx N/2$ and $N \gg 1$ grows as $|S| \sim \mathcal{O}(e^N/\sqrt{N})$), the number of parameters in the symmetric MPS ansatz grows as $\mathcal{O}(N (\kappa+1))$. Thus, the symmetric MPS ansatz is an \textit{efficient encoding} of the solution space. \\

\textit{II. Arbitrary single equality.}

Next we consider an equality of the type $\sum_i a_i x_i =b$ with $a_i,b \in \mathbb{Z}$, which appears for instance in \textsc{Subset Sum}-type problems. In this setting, we can still impose this constraint into an MPS, which is accomplished by first setting the local site charges as $n_i \in \{\varnothing, a_i\}$. As opposed to the cardinality constraint above, we cannot a priori determine the set of link charges for a general equality. A standard approach for this problem would be to use backtracking: construct all possible link charges solving $n_{i,R}=n_{i+1}+n_{i+1,R}$ in a forward sweep, and discard those that are not consistent with charge conservation in a backward sweep. In practice, one can do better than this and apply a \textit{meet-in-the-middle} technique \cite{horowitz1974computing} whereby we locate the canonical center near the middle of the MPS and enumerate exhaustively all configurations of charges to the left and to the right, as well as their corresponding sums. Using a binary search tree on top to match configurations of charges to the left and to the right adding up to the total charge of interest brings down the time complexity from $\mathcal{O}(2^N)$ to $\mathcal{O}(N2^{N/2})$. Alternatively, when the coefficients $a_i$, $b$ are non-negative one may use a standard dynamic programming approach with pseudopolynomial runtime $\mathcal{O}(Nb)$ \cite{bellman1957dynamic}.\\

\textit{III. Arbitrary set of equalities.}

The most general case of arbitrary number of equalities of the form $A\vec{x}=\vec{b}$, with $A \in \mathbb{Z}^{m\times N}$, $\vec{b} \in \mathbb{Z}^m$, $\vec{x} \in \{0,1\}^N$ is just a small extension of the previous case. Now, each site charge is instead given by a vector of length $m$ as $n_i \in \{\varnothing, \vec{A_i}\}$ where $\vec{A}_i \equiv (A_{1,i},A_{2,i},\cdots,A_{m,i})$, while the flux is given by the vector $\vec{b}$. In Appendix \ref{sec:assignment} we consider \textsc{Assignment}-type equalities which appear also in many combinatorial optimization problems and that can be encoded in an MPS of bond dimension scaling at worst exponentially with the number of equalities (i.e. these do not scale with system size $N$).

\section{Constrained-GEO}~\label{s:generative_algo}
Our framework is divided into three stages (see Fig. \ref{fig: constrained_GEO_birds_view}): 1) Initialization, 2) Training, and 3) Sampling, followed by Evaluation of the generalization performance. First, we construct the structural tensors based on the training data, or, if an efficient MPS representation of the valid space is available (such as in cardinality type constraints), we may also generate training data uniformly from it. Second, we minimize the training loss function via the DMRG-inspired algorithm put forward in Ref. \cite{han2018unsupervised}. After a few sweeps we sample from the resulting MPS and compute the costs of the samples, and retrain the model with those samples together with the initial training dataset, as we detail next.

\subsection{Initialization: Constrained Embedding} \label{sec_initialization}
\begin{figure}
    \centering
    \includegraphics[width=\linewidth, scale=0.5]{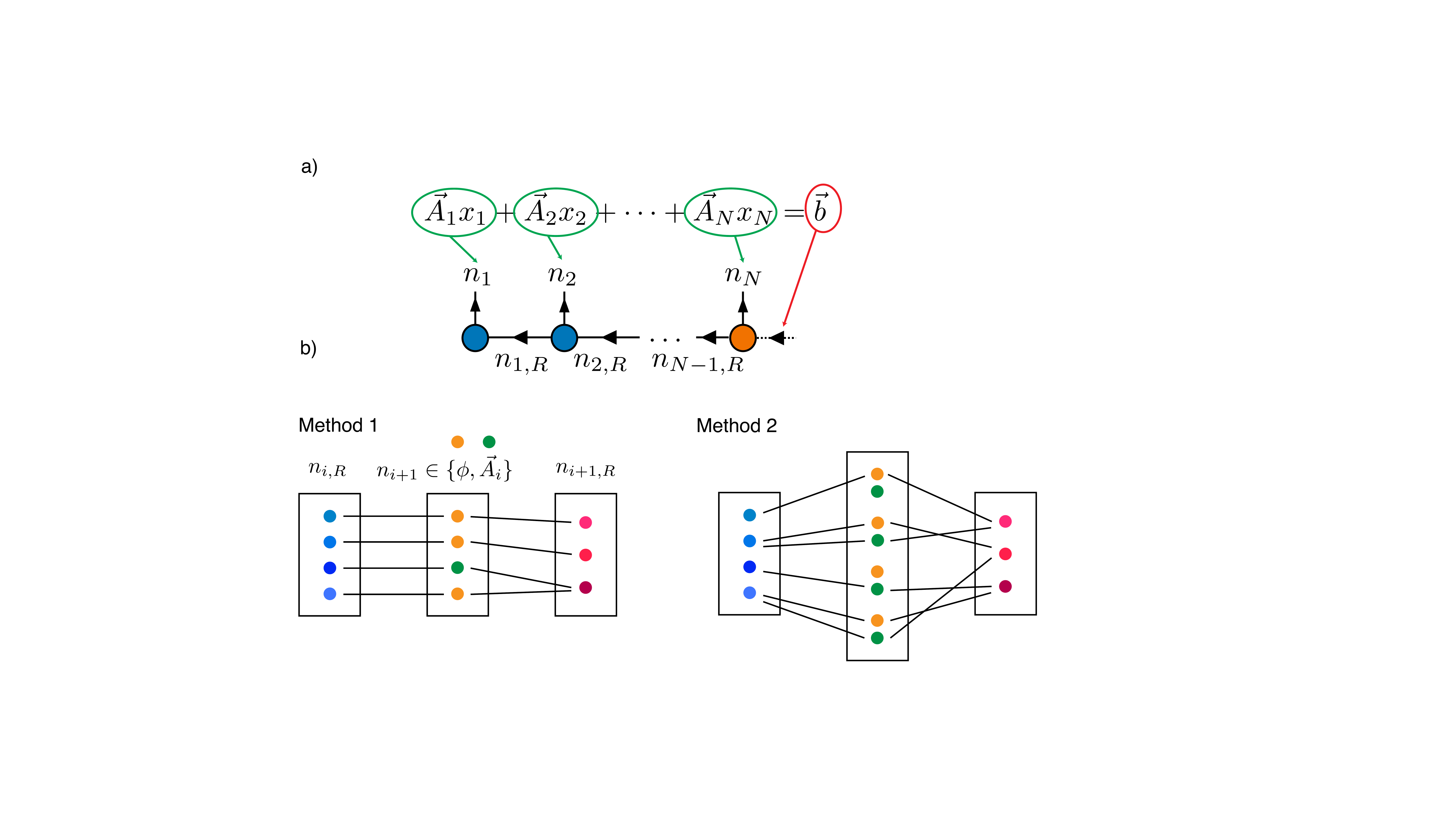}
    \caption{\textbf{Initialization in the generative algorithm.} a) Given a valid bitstring $\vec{x}$ satisfying $A\vec{x}=\vec{b}$, we assign to each site charge the vector $x_i \vec{A}_i$, with $\vec{A}_i$ corresponding to the $i$'th column of matrix $A$. With fixed site charges and the flux, corresponding to the vector $\vec{b}$, this uniquely determines the set of link charges $n_{i,R}$ in $(N-1)m$ steps. Carrying this procedure for all bitstrings allows us to construct the structural tensors $\delta_{N_{\rm in}, N_{\rm out}}$ appearing in our symmetric MPS ansatz. b) Method 1: we map each configuration of charges $(n_{i,R}, n_{i+1}, n_{i+1,R})$ to a nonzero entry in the structural tensor. Different dots correspond to different charges. Black lines indicate compatible configurations. Method 2: same as in Method 1 but we now determine all possible site charges $n_i$ consistent with the set of link charges $\{(n_{i,R}, n_{i+1,R})\}$. Method 2 contains more compatible configurations of charges, hence more bitstrings are generated using Method 2.}
    \label{fig:MPS_layout}
\end{figure}

The first step consists in selecting all unique bitstrings from the dataset. Next we construct the set of link charges by first fixing the position of the canonical center as well as that of the flux. For the purposes of illustration we set both to be at the last site as shown in Fig. \ref{fig:MPS_layout} a). The determination of all link charges for each bitstring is performed in $\mathcal{O}(Nm)$ steps by solving recursively $n_{i+1,R}=x_{i+1}\vec{A}_{i+1}+n_{i,R}$ starting from either end of the MPS. We store in memory each local configuration of charges $i \to (n_{i,R},n_{i+1},n_{i+1,R})$), and move on to the next bitstring. We repeat the same procedure for the next bitstring in the set, and obtain a new configuration of charges. Once we loop over all distinct bitstrings from the training set we are ready to set up the structural tensors by reading off the configuration of charges. At this step, there are two alternative ways of proceeding forward, as illustrated in Fig. \ref{fig:MPS_layout} b). The first one is to directly map each configuration of charges $(n_{i,R},n_{i+1},n_{i+1,R})$ to a nonzero component of the structural tensor. Together with the finding of charges, this preprocessing step takes $\mathcal{O}(N m |\mathcal{T}|)$ steps when using a hashtable for storing the configuration of charges. The second approach is to only store the configuration of all link charges consistent with the training data plus constraints, $i \to (n_{i,R}, n_{i+1,R})$, and subsequently determine all configurations of charges $(n_{i,R}, n_{i+1}, n_{i+1,R})$ consistent with the flux of each tensor. In contrast with the previous method, this preprocessing step scales as $\mathcal{O}(dNm|\mathcal{T}|)$ when using a hashtable, where $d$ is the dimension of each site, which for our purposes will be $2$ given that we are dealing with bitstrings. Thus we see that this latter method takes in general twice as much time as the first one. The benefit nevertheless from using this latter method is that this permits us to find more configuration of charges, which in turn will lead to better generalization, i.e. more (unique) bitstrings can be generated from the second method than with the first one.

\begin{figure}
    \centering
    \includegraphics[width=\linewidth, scale=0.5]{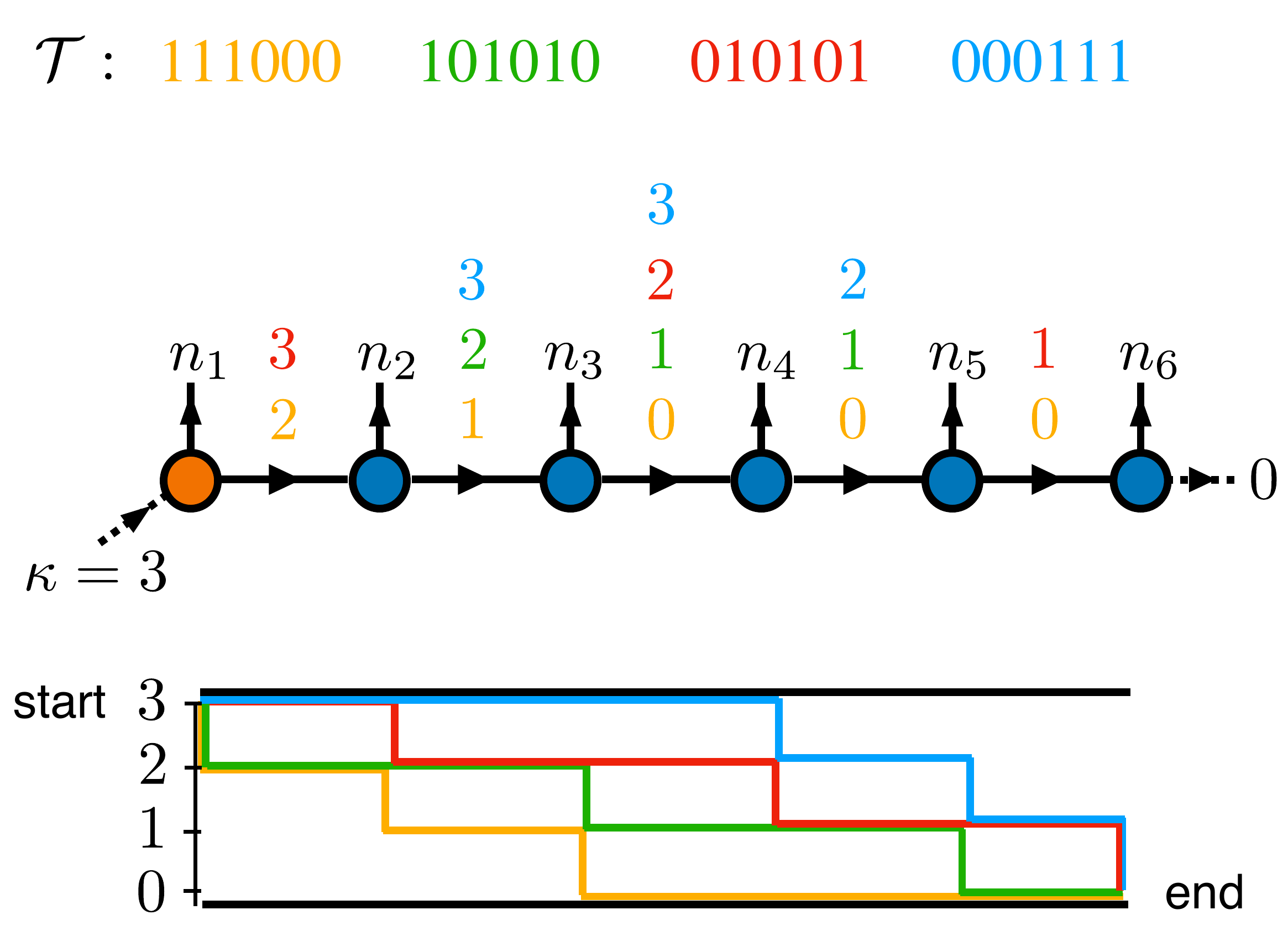}
    \caption{\textbf{MPS in the cardinality dataset.} The training dataset is given by the four bitstrings $\mathcal{T}=\{111000,101010,010101,000111\}$ fulfilling $\sum_i x_i=\kappa$, with $N=6$, $\kappa=3$. Numbers on top of each link in the MPS correspond to new link charges $n_{i,R}$ extracted from each bitstring going from the left one (in yellow) to the right one (in blue). The colored paths in the bottom diagram correspond to the link quantum numbers extracted from each bitstring as we move from left to right in each bitstring (every jump corresponds to a bit taking value 1).}
    \label{fig:cardinality_bitstrings}
\end{figure}

To illustrate these two methods we consider the cardinality dataset formed by bistrings fulfilling $\sum_i x_i=\kappa$. Focusing on bitstrings of length $N=6$ and cardinality $\kappa=3$ we choose the training data $\mathcal{T}=\{111000,101010,010101,000111\}$. The resulting MPS using the first method will generate 6 new bitstrings (10 in total), while the second method will generate the entire solution space of $\binom{6}{3}=20$ bitstrings. The extraction of link charges from data is illustrated in Fig. \ref{fig:cardinality_bitstrings}. There we show all possible quantum numbers for each link index, $n_{i,R}$, consistent with the cardinality constraint. In total, there are $\binom{6}{3}=20$ bitstrings. Out of them, only $4$ are needed to extract all link charges. For arbitrary cardinality constraint $\kappa$ and bitstring length $N$, it can be shown that only $\min(\kappa,N-\kappa)+1$ bitstrings, are needed to exactly construct the fixed-cardinality MPS using the second method. In our work we shall use the second method when initializing the MPS.

We remark that the structural tensor crucially depends on the location where we insert the flux, which corresponds to the location of the canonical center. This is important to bear in mind because the training procedure will move the canonical center when optimizing each local tensor as we shall describe next (however the most costly step is setting up the structural tensors at the beginning, as updates during training in TNs are local, which will only change the link charges around each site, and these charge updates can be determined efficiently).

Lastly, we note that encoding arbitrary number of equality constraints (without training data) into a TN may lead to an exponential scaling of the bond dimension (see Sec. \ref{sec:assignment} of the Appendix for the particular case of \textsc{Assignment}-type equalities). The present approach, guarantees that, in the worst case scenario, the bond dimension be lower bounded by the number of bitstrings in the training set, which allows for better scalability of the algorithm.
\subsection{Training} \label{sec_training}
\begin{figure}
    \centering
    \includegraphics[width=\linewidth, scale=0.5]{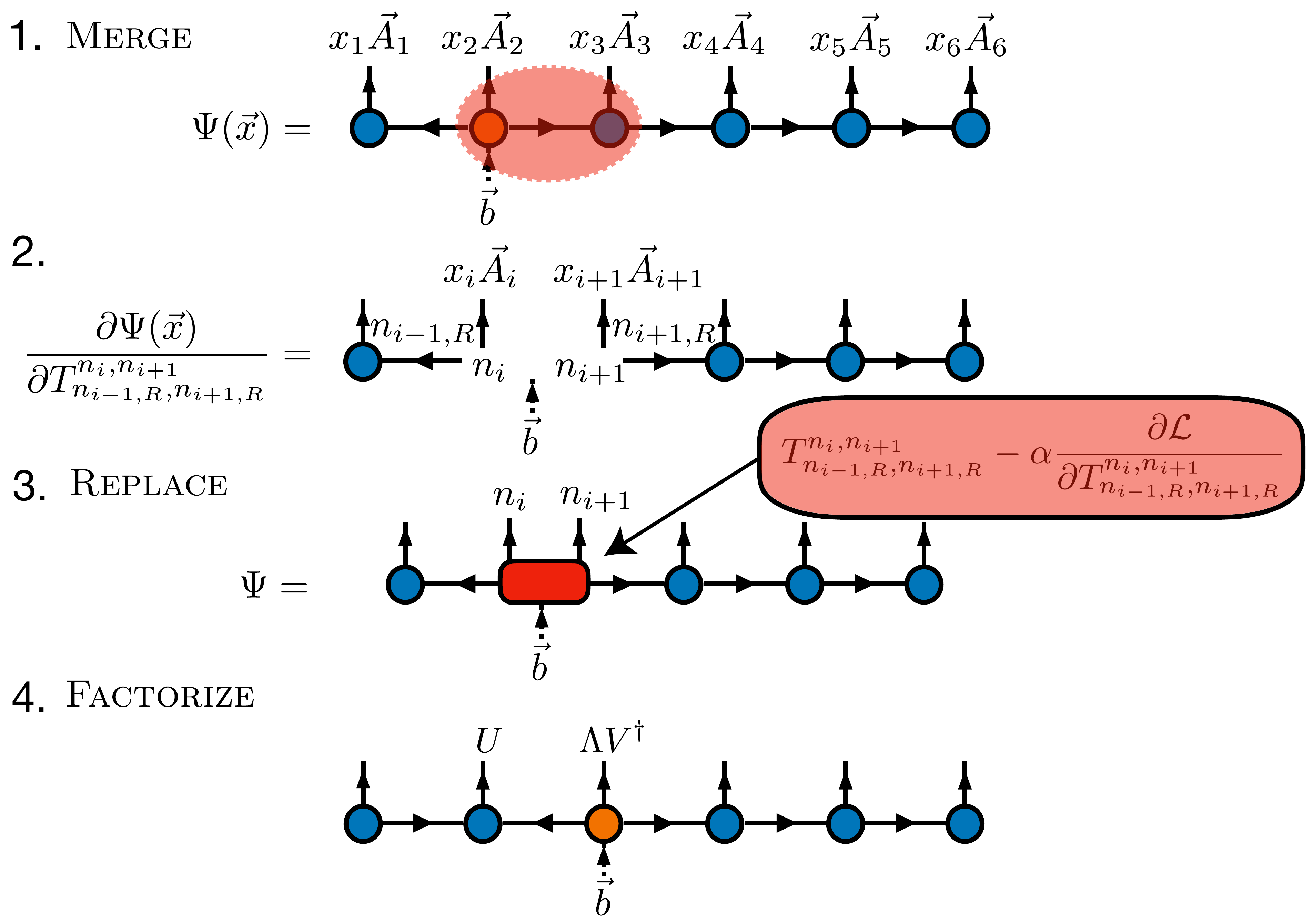}
    \caption{\textbf{Training steps in the generative algorithm}. $1.$ Two neighboring tensors are merged in the s-TNBM. $2.$ Computing the gradient of the s-TNBM w.r.t. the merged tensor. $3.$ The two neighboring tensors get updated by the gradient term. $4.$ The new tensor is factorized back into two neighboring tensors. The canonical center (orange tensor) has been shifted one site in the process. We repeat this four steps across all sites in the MPS from left to right and back corresponding to one full sweep. }
    \label{fig:DMRG_steps}
\end{figure}

Once we have set up the structural tensors deterministically, either via the initialization step described in Sec. \ref{sec_initialization} or by exact construction of the MPS, we can proceed with the \textit{training} step. The process is an extension of the Density Matrix Renormalization Group (DMRG) inspired algorithm described in Ref. \cite{han2018unsupervised} in the presence of equality constraints. The training process aims to minimize the loss function
\begin{equation}
    \mathcal{L}=-\sum_{\vec{x} \in \mathcal{T}}p_T(\vec{x})\log \mathbb{P}(\vec{x}),
\end{equation}
where $\mathbb{P}(\vec{x})=|\Psi(\vec{x})|^2/Z$, $Z=\sum_{\vec{x}\in \{0,1\}^N}|\Psi(\vec{x})|^2$, and $p_T$ is a softmax surrogate to the cost function, and is given as in Eq. (\ref{eq_softmax}). Since the Born machine $|\Psi(\vec{x})|^2$ is constructed out of a symmetric MPS $\Psi$, we shall henceforth refer to this as symmetric-TNBM (s-TNBM). The various steps involved in the training are summarized in Fig. \ref{fig:DMRG_steps} and in more detail consist of the following:
\begin{enumerate}
\setcounter{enumi}{0}
    \item For fixed location of the canonical center, construct the corresponding s-TNBM for a given bitstring $\vec{x} \in \mathcal{T}$ by fixing the site charges to $x_i\vec{A}_i$. Merge the two neighboring tensors at the canonical center. In TN notation: $\sum_{n_{i,R}}T^{[i]x_i\vec{A}_i}_{n_{i-1,R},n_{i,R}}T^{[i+1]x_{i+1}\vec{A}_{i+1}}_{n_{i,R},n_{i+1,R}}=\textsc{Merge}\left(\begin{gathered} \includegraphics[scale=0.25]{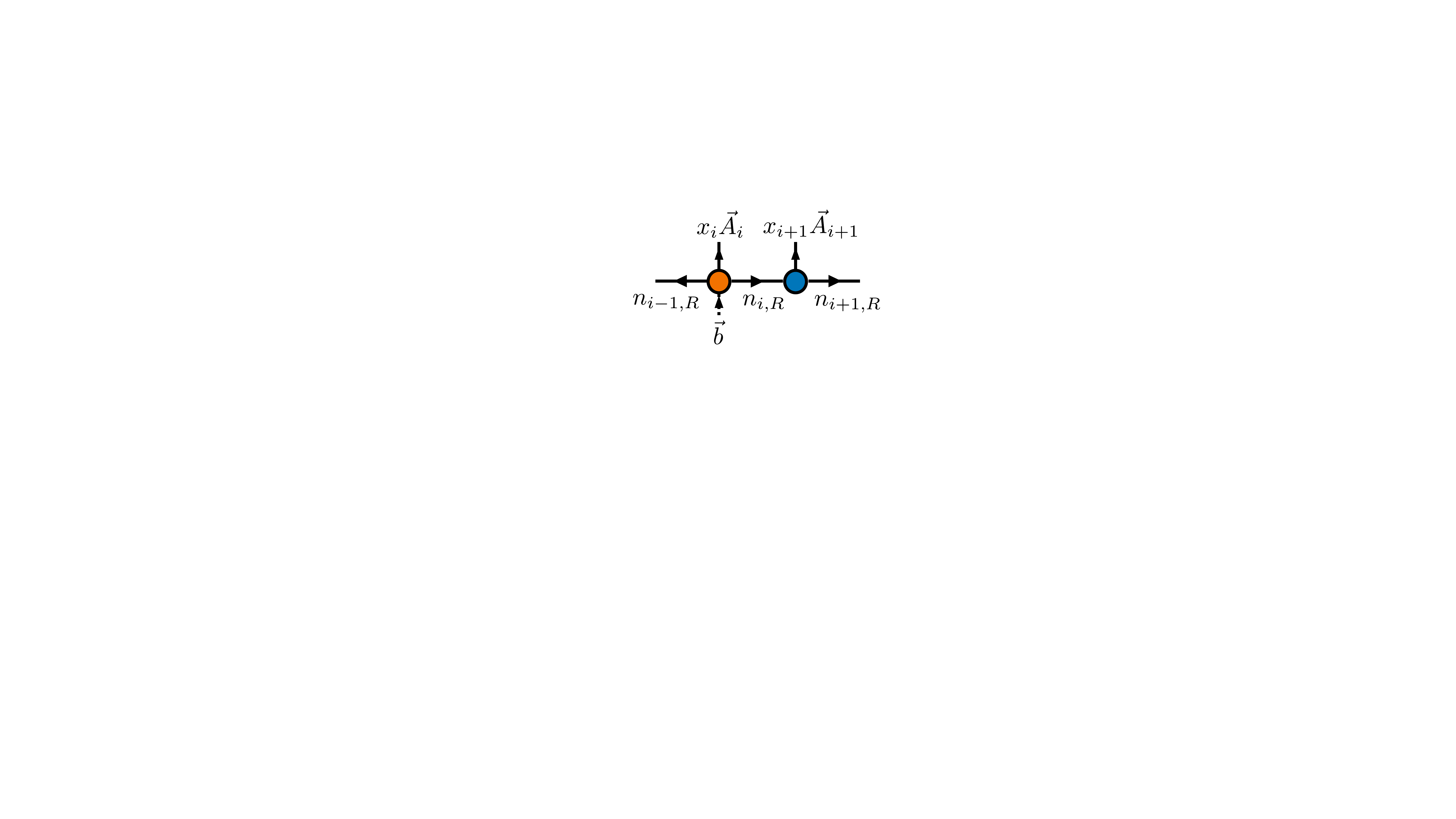} \end{gathered}\right)=\begin{gathered} \includegraphics[scale=0.25]{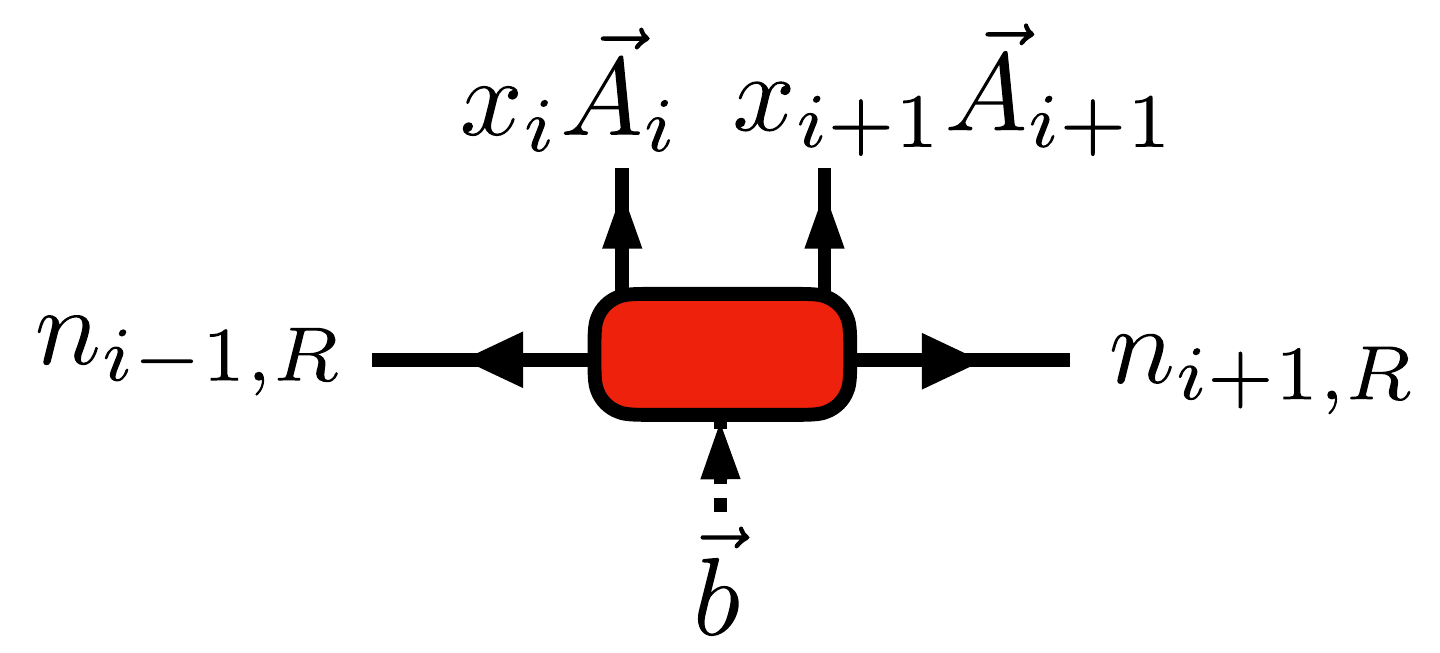} \end{gathered}\equiv T^{x_i\vec{A}_i,x_{i+1}\vec{A}_{i+1}}_{n_{i-1,R},n_{i+1,R}}$, where the link charges are constrained by charge conservation as $n_{i-1,R}+n_{i+1,R}=\vec{b}-x_i\vec{A}_i-x_{i+1}\vec{A}_{i+1}$. Because of charge conservation, the merged tensor will preserve block-sparsity.
    \item Compute gradient of NLL as
    \begin{equation}
        \frac{\partial \mathcal{L}}{\partial T_{n_{i-1,R},n_{i+1,R}}^{n_i,n_{i+1}}}=\frac{Z'}{Z}-2\sum_{\vec{x}\in \mathcal{T}}p_T(\vec{x})\frac{\Psi'(\vec{x})}{\Psi(\vec{x})},
    \end{equation}
    where $\Psi'(\vec{x})=\frac{\partial \Psi(\vec{x})}{\partial T_{n_{i-1,R},n_{i+1,R}}^{n_i,n_{i+1}}}$. The term $\Psi'(\vec{x})$ forces the site charges $n_i$, $n_{i+1}$ to take the values $n_i=x_i\vec{A}_i$, $n_{i+1}=x_{i+1}\vec{A}_{i+1}$. The $Z'$ term is exactly given as $Z'=T^{n_i,n_{i+1}}_{n_{i-1,R},n_{i+1,R}}$, while $Z$ can be evaluated efficiently via TN contraction.
    \item Replace the merged tensor $T^{n_i,n_{i+1}}_{n_{i-1,R},n_{i+1,R}}$ from the MPS by $T^{n_i,n_{i+1}}_{n_{i-1,R},n_{i+1,R}} - \alpha       \frac{\partial \mathcal{L}}{\partial T_{n_{i-1,R},n_{i+1,R}}^{n_i,n_{i+1}}}$, where $\alpha$ is the learning rate.
    \item Factorize the newly updated merged tensor. This involves two steps. $\textsc{Reshape}\left(\begin{gathered} \includegraphics[scale=0.25]{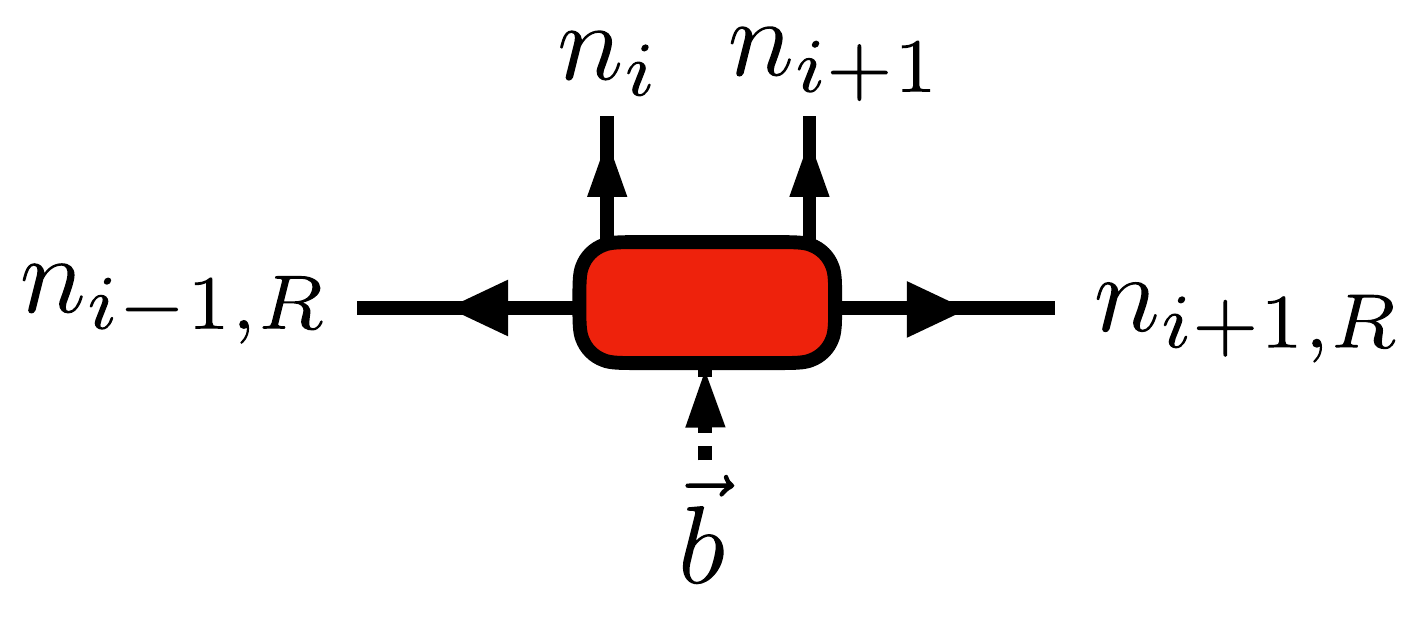}\end{gathered}\right)=\begin{gathered}\includegraphics[scale=0.25]{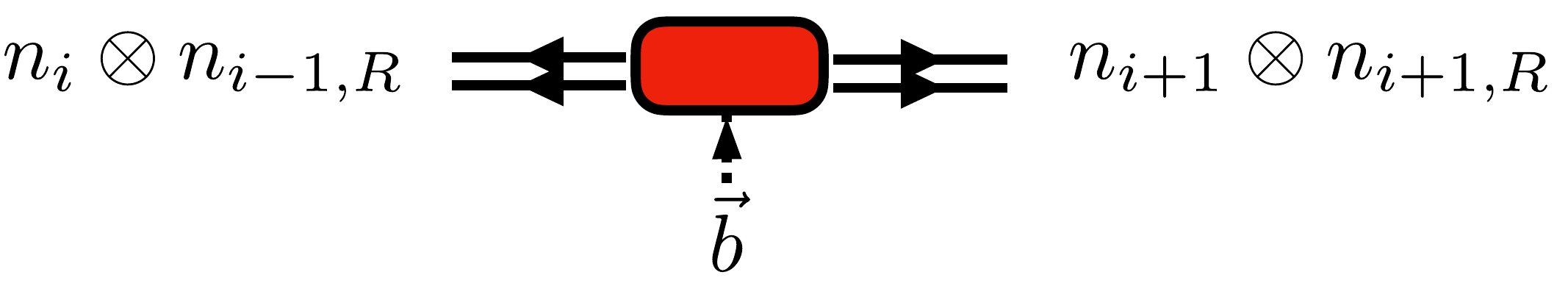}\end{gathered}$, i.e. the merged order-4 tensor is reshaped onto a matrix. This is followed by a singular value decomposition $\textsc{SVD}\left(\begin{gathered}\includegraphics[scale=0.25]{reshape_tensor.pdf}\end{gathered}\right)=\begin{gathered} \includegraphics[scale=0.25]{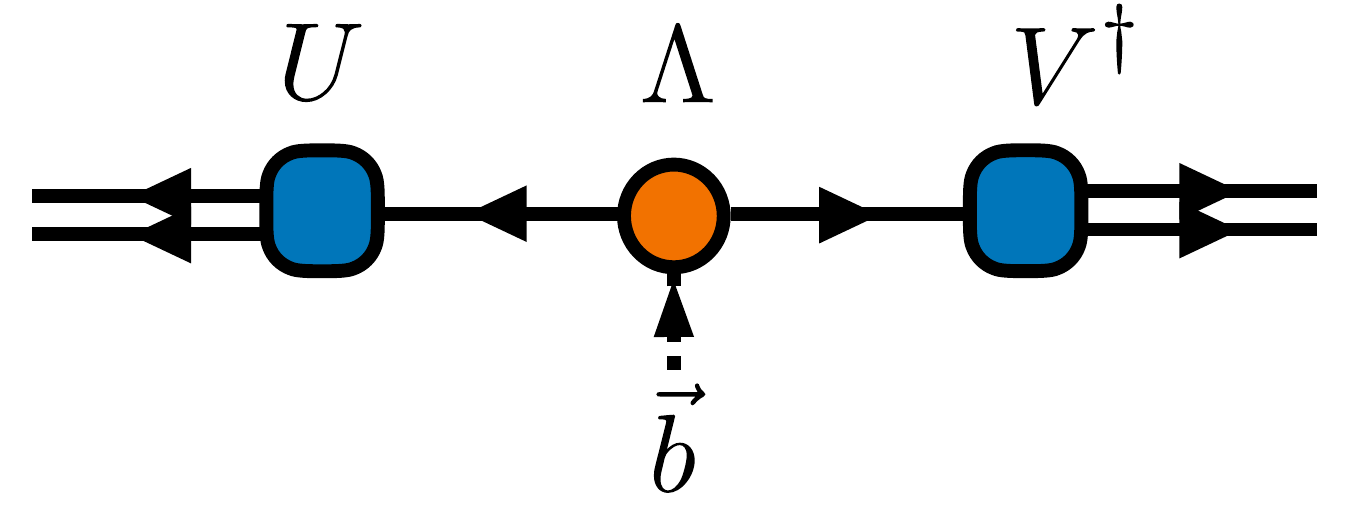}\end{gathered}$. The resulting spectrum $\Lambda$ is truncated, keeping the $\chi$ greatest singular values. This results in a compressed MPS. We reshape the tensors back to their original shapes by applying the inverse of the reshape step: $\textsc{Reshape}^{-1}\left(\begin{gathered} \includegraphics[scale=0.25]{svd.pdf}\end{gathered}\right)= \begin{gathered} \includegraphics[scale=0.25]{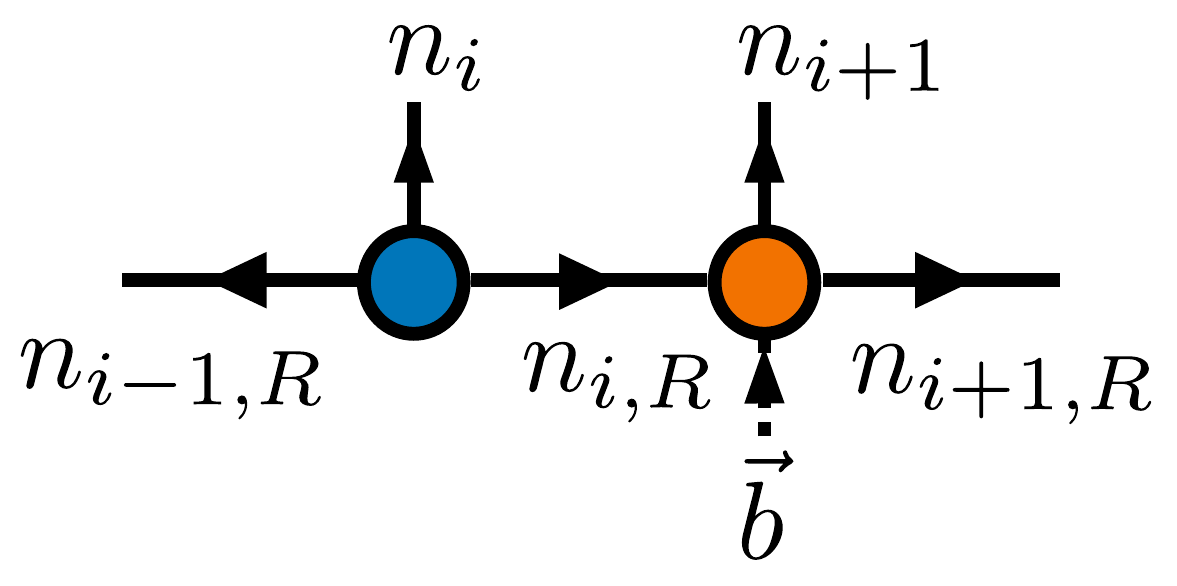}\end{gathered}$, where we have identified $T^{[i]n_i}_{n_{i-1,R},n_{i,R}}\simeq U$, $T^{[i+1]n_{i+1}}_{n_{i,R},n_{i+1,R}}\simeq \Lambda V^{\dagger}$. Note the canonical center has shifted to the right at this point.
\end{enumerate}
We repeat these four steps at every site of the MPS when going left-to-right, and an analogous series of steps when going right-to-left except that now the canonical center moves one site to the left after every update.

The main advantage of using symmetric MPS for tasks constrained by equalities is that most of the linear algebra operations such as merging tensors and factorizing just described  can be carried out block-wise, with each block corresponding to a QN block.This leads to a more efficient optimization procedure than when using dense (unconstrained) tensors.

\begin{figure}
    \centering
    \includegraphics[width=\linewidth, scale=0.5]{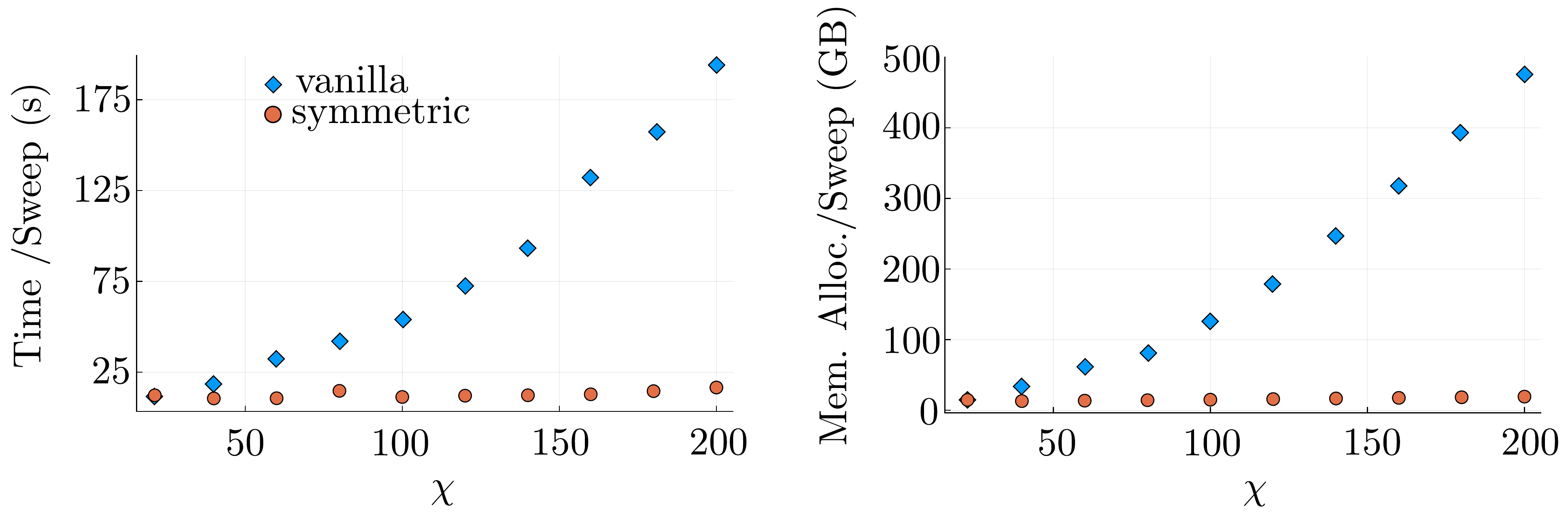}
    \caption{\textbf{Profiling computational resources for computing the average training step using vanilla \textit{vs.} symmetric MPS in the cardinality dataset.} The system size chosen is $N=50$ and the cardinality $\kappa=25$. We train both vanilla and symmetric MPS (initialized in the uniform superposition of fixed cardinality) using $|\mathcal{T}|=1000$ bitstrings. The results are averaged over 3 sweeps. Left (right) panel: Wall-time elapsed (memory allocations) as a function of maximum bond dimension in the MPS.}
    \label{fig:resources_v_vs_s}
\end{figure}

To showcase the benefits of using symmetric MPS in practice we consider a cardinality dataset with $N=50$, $\kappa=25$ and $|\mathcal{T}|=1000$ training samples. Our benchmark experiments were carried out on a Macbook Pro with a 1.4 GHz Quad-Core Intel Core i5, using  4 threads and on \texttt{Julia} and the \texttt{ITensors.jl} package for smart index tensor network contractions \cite{ITensor_julia}. The results showcased in Fig. \ref{fig:resources_v_vs_s} indicate that a considerable advantage is obtained in using symmetric MPS over vanilla MPS in terms of computational resources. We report memory allocations as well as the wall-time employed during the entire  training algorithm (including computing the gradients, replacing the tensors by the gradients, factorizing the resulting tensors, as well as computing the NLL after each sweep) averaged over 3 sweeps. In both cases we see the substantial savings on computing time and space memory: while for vanilla MPS these resources scale polynomially as we increase the bond dimension $\chi$, for symmetric MPS the increase in resources is barely present, as a result of a total of 21 QN blocks at the middle link.
\subsection{Sampling} \label{sec_sampling}
Drawing samples from ANNs can be challenging, especially when using Markov Chain Monte Carlo (MCMC) sampling based algorithms due to very long autocorrelation times that limit their efficiency.

In contrast, TNs enjoy the property of \textit{perfect sampling} \cite{ferris2012perfect}. The bit-wise sampling of MPS is done without rejections and sequentially as one moves from one end of the MPS to the other, consisting on computing a series of single-site conditional density matrices and conditional single-site probabilities. The resulting sampling algorithm avoids long autocorrelation times typical of MCMC sampling-based algorithms that are needed to extract reliable samples. Since there is no correlation between samples, it is straight-forward to parallelize the sampling process by, for instance, making use of multi-threading.

Once samples are drawn, we evaluate the generalization performance of the algorithm by computing the costs of the bitstrings. The performance of the generative model will be measured according to 1) the \textit{quantity} of new, valid samples (i.e. bitstrings not contained in the dataset, but that still fulfill the constraints), and 2) the \textit{quality} of new samples as measured by the cost (higher-quality samples corresponding to lower-cost samples). These samples may be merged with the training dataset to form a new training set which we can be used for initialization and training in a new iteration of the algorithm. Various strategies can be devised for selecting this new training set as detailed in Ref. \cite{alcazar2021enhancing}.

\section{Results and Discussion}~\label{s:results}
To test the performance of s-TNBMs we consider the following tasks. In the first part, Sec. \ref{sec_validity}, we will analyze how well symmetric MPS are able to generate new valid solutions to $A\vec{x}=\vec{b}$ from a given set of known solutions. This will test the ability of the structural tensors at capturing QNs that are shared among different bitstrings, which in turn will lead to generalization. In the second part, we will endow QN blocks with a degeneracy factor, taking full advantage of both QNs and their dimensions, so as to model arbitrary probability distributions (not just uniform on the space of solutions) subject to arbitrary equality constraints. We will test this on two tasks: first, in Sec. \ref{sec_quality_I}, we shall consider the training dataset to be sampled from the ground state of a well-known Hamiltonian in condensed matter physics: the one dimensional Transverse Field Ising (TFI) model \cite{sachdev1999quantum}, subject to the constraint that only bitstrings of fixed cardinality are valid. Next, in Sec. \ref{sec_quality_II}, we will test the ability of s-TNBMs at finding new, \textit{high-quality} solutions to combinatorial optimization problems subject to equality constraints by mapping this problem to a generative modeling task, as briefly discussed in Sec. \ref{s:prob_statement}, and detailed in Sec. \ref{s:generative_algo}.
\subsection{Validity-based generalization: Entanglement assisted generation of solutions to arbitrary equality constraints} \label{sec_validity}
\begin{figure}
    \centering
    \includegraphics[width=\linewidth, scale=0.5]{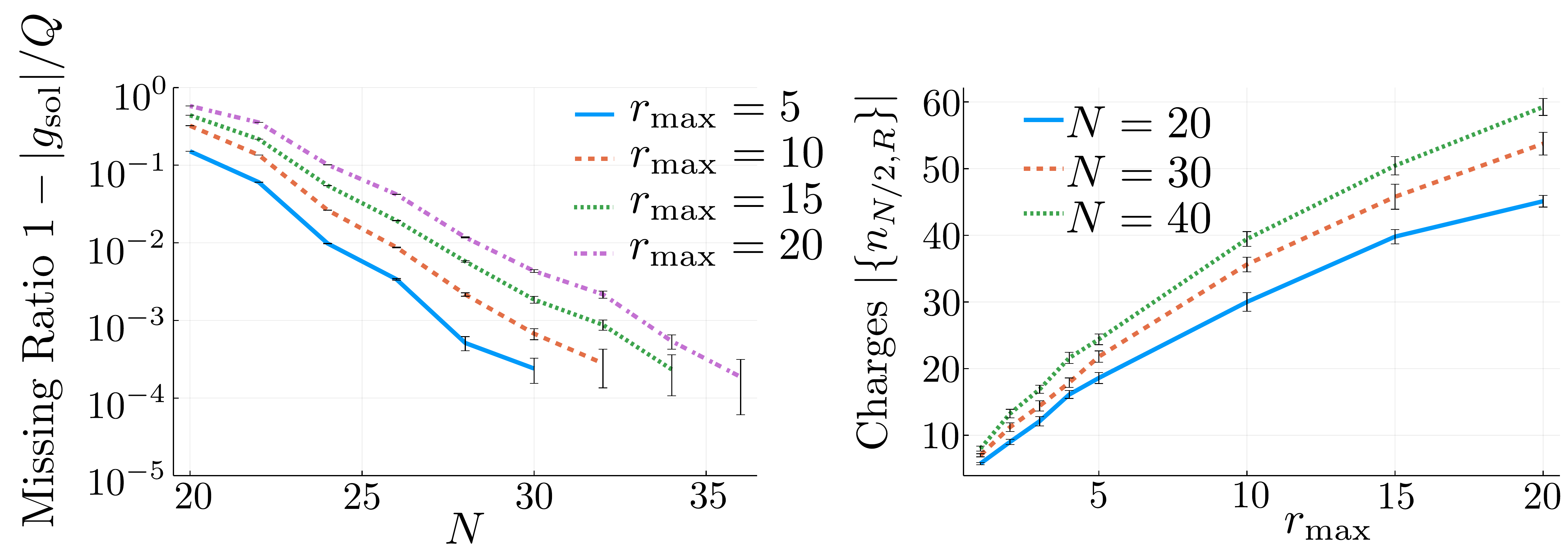}
    \caption{\textbf{Generalization performance using the s-TNBM: one equation $\sum_i \alpha_i x_i = b$}, where $b$, $\alpha_i$ are random integer coefficients uniformly distributed in the range $[-r_{\rm max}, r_{\rm max}]$. The input is 100 bitstrings found by random search, and the goal is to generate as many new, unique solutions after $Q=10^4$ queries on the s-TNBM. Left panel: Missing ratio of new, unique solutions $|g_{\rm sol}|$ found after $Q$ queries, as a function of system size $N$ for various choices of $r_{\rm max}$. Data averaged over 10 samples and error bars corresponding to standard errors. Right: growth in the number of distinct charges at the middle link for various choices of $N$.}
    \label{fig:one-eq-random}
\end{figure}
Evaluating the generalization performance of unsupervised generative models is a challenging task. For the following experiments, a candidate generalization metric should favor valid samples that are not only new, but unique as well. Our focus will be thus to measure variants of the set of new, and unique solutions, denoted as $|g_{\rm sol}|$ (see Ref. \cite{gili2022evaluating} where this and other various metrics were introduced).

In the next experiments we shall be interested in generating new solutions to a system of linear constraints of the form $A\vec{x} = \vec{b}$, where $\vec{x} \in \{0,1\}^N$, $A\in \mathbb{Z}^{m\times N}$, $\vec{b}\in \mathbb{Z}^m$ from a set of known solutions. In our first experiment we consider a single equation of the form $\sum_i \alpha_i x_i = b$, where $\alpha_i$ and $b$ are random integers uniformly distributed in the range $[-r_{\rm max}, r_{\rm max}]$. We start with 100 unique known solutions to this equation, which we shall also call \textit{seeds} (found e.g. by random search, or \textsc{Subset Sum} solvers when possible) and the goal is to generate as many solutions as possible. Since there is no bias in the training dataset (i.e. all solution bitstrings carry the same probability), we do not need to carry any explicit training, and instead we only make use of the initialization step explained in Sec. \ref{sec_initialization}, followed by the sampling step on the resulting s-TNBM. Since the solution space may grow with system size (this growth could be in principle exponential), we bound the number of solutions found by the number of total queries to the s-TNBM generator, which we set to be $Q=10^4$. The results of this experiment are shown in Fig. \ref{fig:one-eq-random}. Remarkably we find that, despite the absence of obvious global $U(1)$ symmetry given the nature of the equality constraint, we do find very good generalization performance. It is interesting to see that the performance increases as we scale up system size $N$. The number of generated solutions grows at a pace which would be consistent with exponential (we do not rule out power law) as a function of system size $N$ for fixed $r_{\rm max}$, even for large values of $r_{\rm max}$. Importantly, the set of QNs at the middle link, $|\{n_{N/2,R}\}|$ (which corresponds to the largest structural tensor link dimension in the s-TNBM), is of size at worst linear as a function of $r_{\rm max}$ for fixed $N$, at least for the system sizes and choices of $r_{\rm max}$ simulated here. Note this set size is a measure of the complexity of datasets constrained by equalities, and it provides a lower bound on the minimum bond dimension needed for an MPS to generate a given number of new, distinct solutions. Clearly, real-world equality constraints will have a greater degree of symmetry so the results shown here for a single equality with random coefficients $A$ are already encouraging. This would in turn translate into a smaller bond dimension for a given fixed system size, and therefore better scalability as we go to larger system sizes.

To have a better intuition on these results and on the role played by the bond dimension, it is helpful to consider the most \textit{symmetric} equality constraint, that of cardinality, where $\alpha_i=1$ $\forall i$. As argued in Sec. \ref{subs_constr_u1}, an exact MPS representation of the entire valid space requires a bond dimension of at least $\chi = \mathcal{O}(\kappa)$ (in fact $\chi=\mathrm{min}(\kappa,N-\kappa)+1$), while the solution space grows exponentially in $N$ (for $\kappa \approx N/2$). How does this relate to the generalization capabilities of s-TNBMs? In Fig. \ref{fig:cardinality_bitstrings} we gave an intuition of this mechanism, focusing on $N=6$, $\kappa=3$. For concreteness we choose the canonical center to be located at the first site. We label all link indices by their QNs. Next, we apply the initialization step of the algorithm explained in Sec. \ref{s:generative_algo}, with the set of four colored bitstrings in Fig. \ref{fig:cardinality_bitstrings}. These four samples   already covers all possible QNs for each link with bond dimension $2,3,4,3,2$ respectively. The corresponding MPS is able to generate all valid $\binom{6}{3}=20$ bitstrings. The choice of the four samples is not unique as long as they \textit{patch} the entire set of link QNs from start, at $n_{0,R}=\kappa$ (viewing the extra dashed index on the left of the MPS in Fig. \ref{fig:cardinality_bitstrings} as having charge $\kappa$) to finish, at $n_{6,R}=0$ (viewing the extra dashed leg on the right of the MPS in Fig. \ref{fig:cardinality_bitstrings} as having charge $0$). Of course this experiment carries over to any cardinality and system size. The upshot is that the minimum number of bitstrings needed to generate all solutions to a given equation is lower bounded by the number of QNs at the middle link. In other words, how well our s-TNBM is able to generalize depends crucially on the number of distinct paths that can traverse this middle link.

\begin{figure}
    \centering
    \includegraphics[width=\linewidth, scale=0.5]{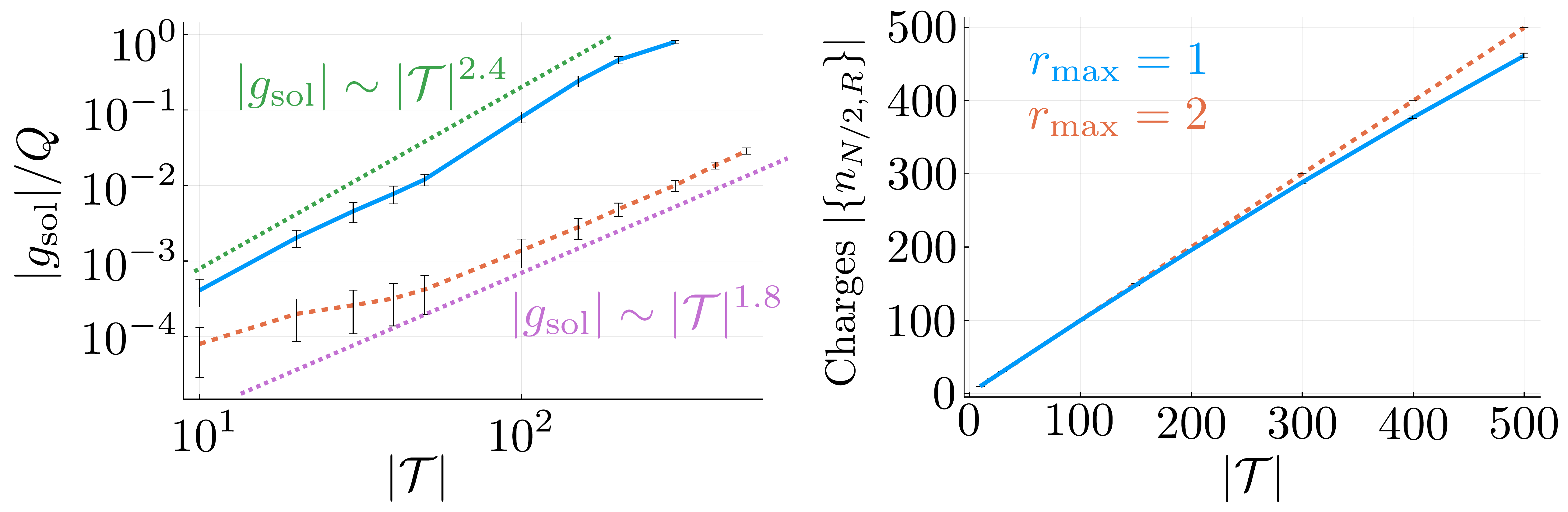}
    \caption{\textbf{Generalization using the s-TNBM: five equations of the form $A\vec{x}=\vec{b}$.} Left: Number of new and unique solutions generated when sampling $Q=10^4$ times on our s-TNBM as a function of training set size $|\mathcal{T}|$ of unique training bitstrings for a system size $N=50$ and $r_{\rm max}\in \{1, 2\}$ (solid and dashed lines, respectively). Dotted lines correspond to best power-law fit for each case. Right: Corresponding QN set size at the middle link as a function of $|\mathcal{T}|$. }
    \label{fig:n_50_random}
\end{figure}

We scale up the previous experiment from one to five equations, where now $A$, $\vec{b}$ entries are chosen uniformly random in $[-r_{\rm max}, r_{\rm max}]$. The results of the experiment are shown in Fig. \ref{fig:n_50_random}. We are interested here in checking how efficiently does our s-TNBM generalize as a function of training set size, i.e. how many new, unique solutions, as a function of the training set size $|\mathcal{T}|$ are found by the s-TNBM. The training set is provided by random search and keeping valid samples. We find that, even for random instances as the ones considered in here, where $r_{\rm max} \in \{1, 2\}$, our s-TNBM is able to generate new solutions with a power-law exponent close to $2$, as shown by the best-fit power-law exponent $1.8$ and $2.4$ in Fig. \ref{fig:n_50_random}. The corresponding QN set size at the middle link grows linearly, and in fact scales roughly as $|\mathcal{T}|$, $|\{n_{N/2,R}\}| \approx |\mathcal{T}|$.

\begin{figure}[h!]
\includegraphics[width=\linewidth, scale=0.5]{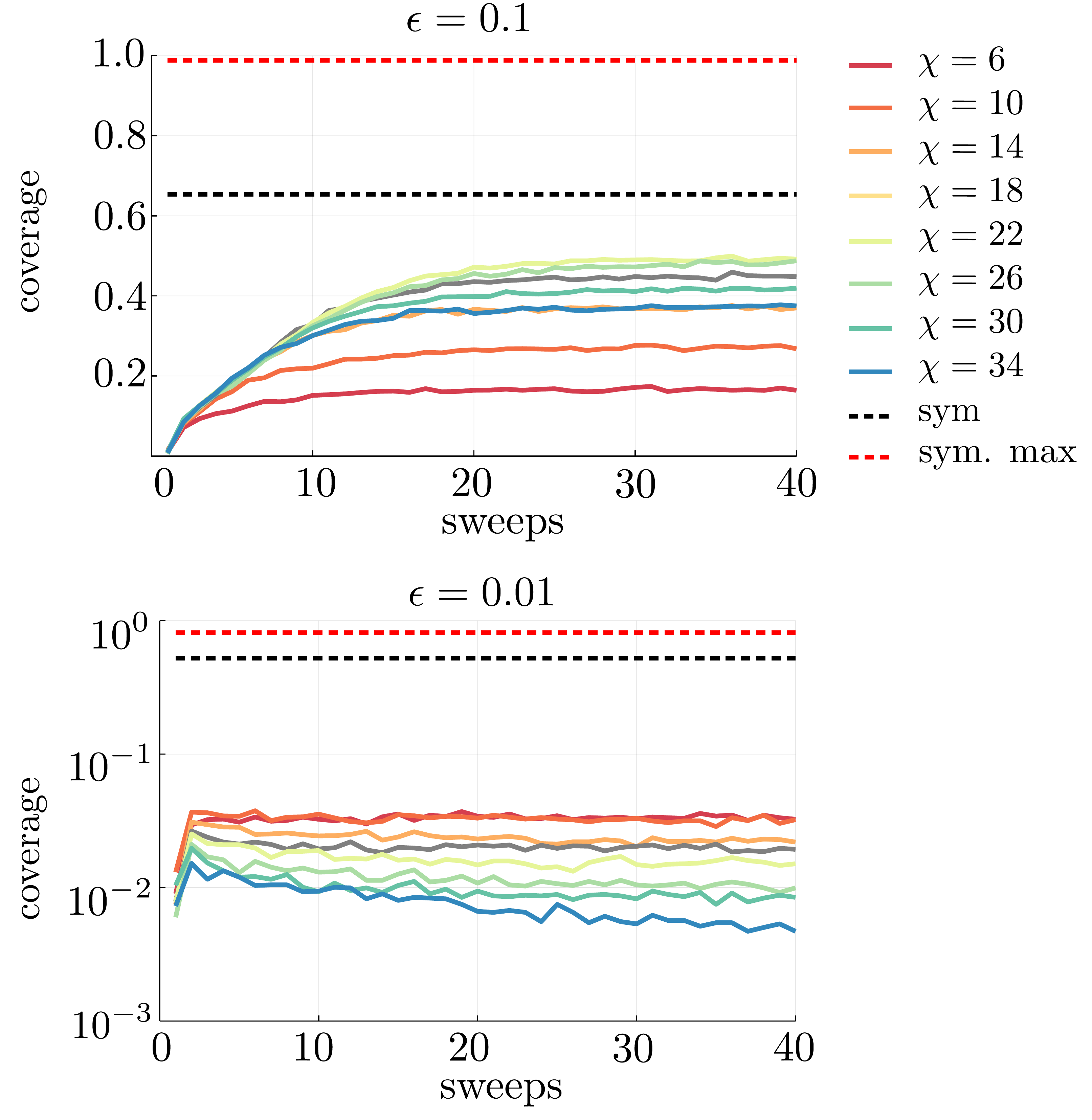}
\caption{\textbf{Comparison between vanilla and symmetric TNBM at generating new solutions to $A\vec{x} = \vec{b}$ as measured by the coverage}. We consider two equations with $N=20$ binary variables, where the entries in $A$ and $\vec{b}$ are randomly selected integers in the set $\{-2,-1,0,1,2\}$. The goal is to generate new solutions from a fixed $|\mathcal{T}|=\epsilon |\mathcal{S}|$ number of samples, where $|\mathcal{S}|$ is the total number of solutions fulfilling the two equations, which is $|\mathcal{S}|=9452$. Results shown after $Q=10^4$ queries. Different bond dimensions correspond to best results out of 10 trials using vanilla MPS (color lines). The symmetric MPS results do not require training, only the initialization step explained in Sec. \ref{sec_initialization}. Black dashed lines correspond to the coverage after sampling from the resulting symmetric MPS, whereas red dashed lines correspond to the maximum coverage using symmetric MPS, obtained after exact tensor network contraction. Top: coverage for a training dataset with $\epsilon=0.1$, and bond dimension $\chi=37$ for the symmetric MPS. Bottom: same but with $\epsilon=0.01$, and bond dimension $\chi=26$ for the symmetric MPS.
}
\label{fig:s_vs_v_novelty}
\end{figure}

So far our focus has been solely on the generalization capabilities of s-TNBMs at generating new solutions to $A \vec{x}=\vec{b}$, without any comparison to the performance of vanilla TNBMs at the same task. In Fig. \ref{fig:s_vs_v_novelty} we consider that comparison. A vanilla TNBM does not include any bias from the constraints, so one must train them using the DMRG-like algorithm from Ref. \cite{han2018unsupervised} (and extended to the presence of equality constraints in Sec. \ref{sec_training}). The experiment now considers a set of two equations with integer coefficients randomly selected in the range $[-2,2]$. We are interested in the number of new, unique solutions, $|g_{\rm sol}|$, found after a given number of queries $Q$ to the generator. A useful metric in this sense is the coverage, as introduced in Ref. \cite{gili2022evaluating}, and defined as \begin{equation}
    C=\frac{|g_{\rm sol}|}{|\mathcal{S}|-|\mathcal{T}|},
\end{equation}
where $|\mathcal{S}|$ is the number of solutions to the system of equations. This metric measures effectively the number of new, unique solutions found by the generator within the remaining available solution space (not contained in the training set). Note that this quantity is implicitly a function of the number of queries to the generator. Given the system size chosen, $N=20$, we can compute explicitly all solutions to random system of equations, which for our experiment turns out to be $|\mathcal{S}|=9452$. We feed in both symmetric and vanilla MPS the same amount of training data, given as $\mathcal{T}=\epsilon |\mathcal{S}|$, with $\epsilon <1$, which consist of unique, valid samples. When the training set contains $10\%$ of the solutions ($\epsilon=0.1$), both vanilla and symmetric MPS are able to generate new solutions with comparative performance. Two crucial differences between the vanilla and symmetric MPS results is that a priori we do not know which bond dimension gives the best performance using vanilla MPS and so one is left with trying many of them only to pick up the one that gives the best results, in this case $\chi=22$. This is not the case when using symmetric MPS, where the bond dimension is fixed entirely by the training set and the constraints, i.e. it is given by the QN set size at the middle link, $\chi=|\{n_{N/2,R}\}|$. Perhaps more importantly, there is no training required when using symmetric MPS because the ansatz already contains the inductive bias coming from the constraints, and we only need the initialization step from Sec. \ref{sec_initialization}. The benefits of using symmetric MPS at learning datasets with equality constraints become more evident in the presence of scarcity of training samples: when we lower this amount down to $1 \%$ ($\epsilon=0.01$), vanilla MPS produces about $3 \%$ of the remaining  unique solutions after $Q$ queries. This is true even for various choices of learning rates used in here, with the results of $\alpha=0.02$ displayed in Fig. \ref{fig:s_vs_v_novelty}, whereas the symmetric version does so at around $50 \%$, down from around $55 \%$ when $\epsilon=0.1$.

\subsection{Quality-based generalization: Ground state reconstruction of local symmetric Hamiltonians} \label{sec_quality_I}
\begin{figure*}
    \centering
    \includegraphics[width=\linewidth, scale=0.5]{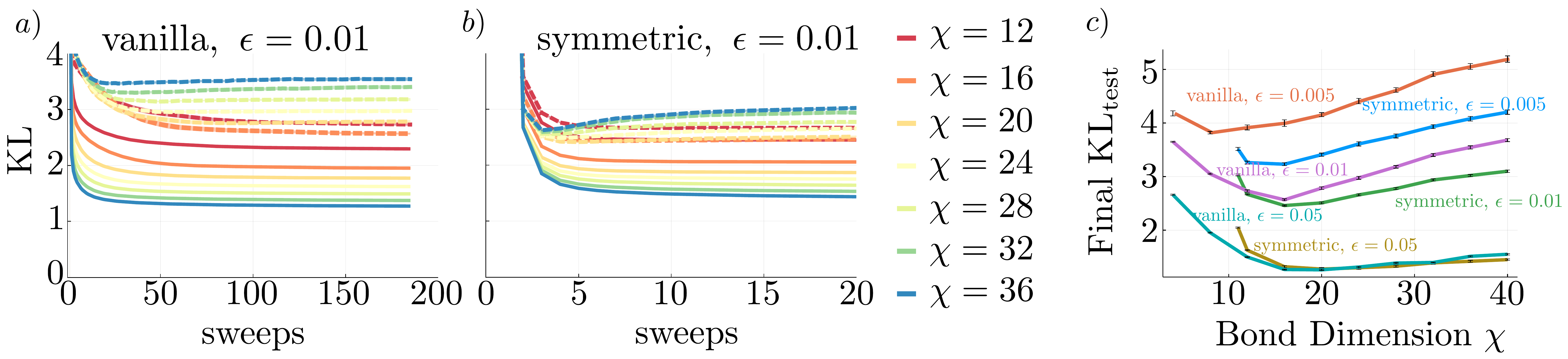}
    \caption{\textbf{Loss curves for symmetric \textit{vs.} vanilla MPS in the TFI Hamiltonian subject to cardinality constraint in the presence of scarcity of data.} a) KL divergence using vanilla MPS for different bond dimensions. Solid lines correspond to training KL, dashed lines correspond to testing KL. $\chi=16$ corresponds to the minimum testing KL. b) same but for symmetric MPS. c) \textit{U-curves} corresponding to final testing losses for $\epsilon \in \{0.005, 0.01, 0.05\}$, as a function of bond dimension $\chi$. Error bars correspond to standard error after 10 independent training iterations.}
    \label{fig:u-curve}
\end{figure*}
In this part we consider a generative model capturing equality constraints as well as a distribution bias. Our goal here is to learn the distribution corresponding to the ground state of the Transverse-Field Ising (TFI) model subject to an equality constraint which for simplicity we take it to be of cardinality type. The TFI Hamiltonian reads
\begin{equation}
    \mathcal{H}=-s\sum_{i=1}^{N-1} Z_i Z_{i+1} -(1-s) \sum_{i=1}^N X_i,
\end{equation}
where $s \in [0,1]$, and $\{X_i, Z_i\}$ are Pauli matrices. Here we shall take $s=1/2$, corresponding to the critical point of the TFI model, and the system size to be $N=20$. This Hamiltonian by itself is not $U(1)$ symmetric (however it does have an inherent global $\mathbb{Z}_2$ symmetry, since the Hamiltonian is invariant under the parity operator $P=\prod_{i=1}^N X_i$), but its ground state is nevertheless efficiently captured by an MPS of low bond dimension,  $\chi\approx 8$ (note however that for general $N$, the MPS ansatz is suboptimal at the critical point since entanglement grows logarithmically $S \sim \log N$, and instead the Multi-scale Entanglement Renormalization Ansatz (MERA) is a better choice, see Ref. \cite{vidal2007entanglement}). The target distribution corresponds to that constructed from taking the Born machine associated with this ground state followed by a \textit{fixed cardinality filter}, that is, we project the resulting MPS onto the manifold of states constrained to only sample bitstrings fulfilling $\sum_i x_i=\kappa$, where for concreteness we choose the cardinality $\kappa=N/2=10$. Thus the entire solution space (composed of all valid bitstrings fulfilling the cardinality constraint) is $|\mathcal{S}|=\binom{20}{10} \approx 185\times 10^3$.

In Fig. \ref{fig:u-curve} we show the training and testing loss curves in terms of their respective Kullback-Leibler (KL) divergences using vanilla \textit{vs.} symmetric MPS for a training dataset size of $|\mathcal{T}|=\epsilon |\mathcal{S}|$ (note that the set may include repeated samples, in which case their probabilities are weighted according to their frequency). We see that while for $\epsilon=0.05$ both vanilla and symmetric MPS are able to learn the underlying distribution just as well, when data is scarce, $\epsilon \lesssim 0.01$, the symmetric counterpart performs better. These results are obtained without major refinements, in particular the learning rate is fixed throughout and of value $\alpha=0.02$ (in principle it would be advisable to adjust this as a function of the size of the tensors \cite{han2018unsupervised}), no DMRG noise is used, and no minibatching or multiple \textit{inner steps} (i.e. multiple consecutive gradient updates at the same site) has been used. These tricks can bring down the value of the loss function by escaping local minima and may be crucial steps for larger scale models.

It is a well-known fact in ML models, that the testing loss may develop a \textit{U-shaped} curve as a function of model capacity \cite{hastie2009elements}. This is a consequence of the \textit{bias-variance tradeoff}. In Fig. \ref{fig:u-curve} (right panel) we show that both vanilla and symmetric MPS display such a U-curve, as a function of bond dimension (see also \cite{strashko2022generalization}). This type of overfitting behavior is \textit{static} in that it only depends on the complexity of the MPS as parameterized by its bond dimension $\chi$. On the left and central panel in the same figure we also report a different kind of overfitting behavior which we dub \textit{dynamic} in that it develops as a function of sweeps, and is more pronounced in the case of symmetric MPS, where at large enough $\chi$, the best testing loss is localized at around a minimum after a few sweeps -- such dynamic U-curve is also present in vanilla MPS, but its convexity is less striking.

We remark that we have also tried more suitable Hamiltonians with an inherent cardinality constraint, such as the Heisenberg model. This considerably improves the testing loss (since the Hamiltonian already satisfies the constraint) but we found the minimum in the U-curve to be located at the minimum bond dimension needed to learn the constraint, $\chi=11$ for $N=20$, which of course is not the general case for arbitrary cost functions (not just the ones stemming from local Hamiltonians). 

\subsection{Quality-based generalization: Finding optimal solutions to constrained combinatorial optimization problems} \label{sec_quality_II}
\begin{figure*}
    \centering
    \includegraphics[width=\linewidth, scale=0.5]{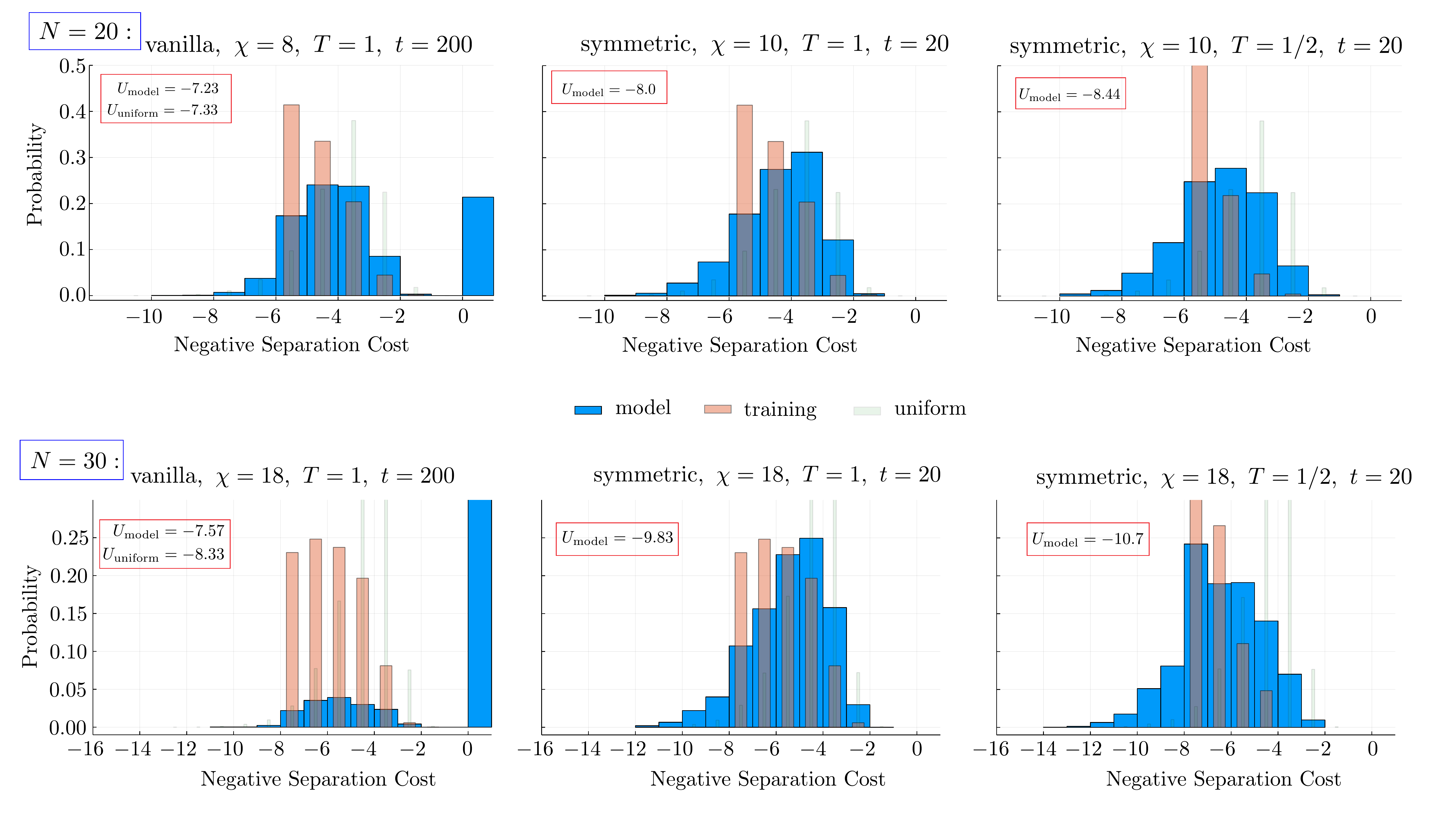}
    \caption{\textbf{Constrained combinatorial optimization with cardinality constraint and negative separation cost function.} Final bitstring distributions using the optimal choices of bond dimensions for vanilla MPS \textit{vs.}  symmetric MPS, for a given softmax temperature choice $T$. The cardinality constraint is \sout{cardinality} $\sum_i x_i =N/2$, with $N=20$ (top panels) and $N=30$ (bottom panels), being the size of the bitstrings. We are interested in the final utility metric after training convergence. Blue histograms correspond to the model distributions, extracted after sampling $Q=10000$ times from the TNBM; orange histograms correspond to the distribution of $|\mathcal{T}|=1848$ training bitstrings, which is formed of uniformly distributed, unique bitstrings satisfying the cardinality constraint but restricted to the subspace of bitstrings of cost greater or equal to $-6$ for $N=20$, and $-8$ for $N=30$; and green histograms correspond to the uniform distribution, extracted from the uniform distribution of \textit{all} bitstrings satisfying the cardinality constraint. Left panels: results for vanilla MPS at $T=1$. Middle panels: results for symmetric MPS at the same temperature $T=1$. Right panels: results for symmetric MPS at $T=1/2$.}
    \label{fig:comb_opt_cardinality}
\end{figure*}
The previous example was motivated by physics, in particular by exploiting the locality of the cost function in the form of the TFI Hamitonian. Here we illustrate how s-TNBMs can also be used to find novel and optimal solutions to combinatorial optimization problems subject to equality constraints, outperforming vanilla TNBMs. As an example, let's consider the cost function to be the \textit{negative separation cost} of bitstrings, which is the negative separation of the farthest two `$1$' separated only by `$0$' bits. For instance, for a bitstring $01011001$ the negative separation cost would be $-3$. The goal of the task is to generate bitstrings with the largest negative separation cost (in absolute value) as possible, subject to an equality constraint which we take to be the cardinality constraint. This is achieved using the constrained-GEO framework described in Sec. \ref{s:generative_algo}, where the cost function appearing in the softmax corresponds to the negative separation cost subject to a cardinality constraint. Because of this constraint, we assign a cost zero to invalid samples (bitstrings not fulfilling this constraint). The starting point is a set of valid solutions that have a negative separation cost at or above $\mathcal{C}_0$. We remark that as opposed to the previous task, with a physics motivation, the current exercise serves as a proxy for typical cardinality constrained optimization problems appearing in industry problems, such as portfolio optimization \cite{Markowitz52} and traveling salesman \cite{garey79}, both \textsc{NP}-hard problems \cite{kellerer2000selecting, garey79}.

The results of this experiment are shown in Fig. \ref{fig:comb_opt_cardinality}, where the learning rate is fixed to $\alpha=0.02$ throughout, and we choose two different system sizes. For $N=20$ and $\kappa=10$, the training set contains only bitstrings of cost greater than or equal to $\mathcal{C}_0=-6$, while for $N=30$ and $\kappa=15$, the training set contains only bitstrings of cost greater than or equal to $\mathcal{C}_0=-8$. To test the performance of both models at capturing low cost samples outside the training set we make use of the utility metric, as introduced in Ref. \cite{gili2022evaluating}. It amounts to computing the average cost of the $5 \%$ lowest cost samples obtained from querying the trained models. The motivation for this metric is that, often times, one is not interested in a unique lowest cost sample (which could have been found by mere luck). By taking instead a small percentage of the lowest cost samples, we make a more robust assessment that the model is indeed obtaining consistently high quality (low cost) samples. Our results indicate that, while the vanilla TNBM is not able to extract any better samples than a \textit{constrained} uniform sampler, corresponding to the model that samples all bitstrings of the given cardinality uniformly, the symmetric version not only does sample already within the valid target space of fixed cardinality (thus avoiding samples with cost zero), but their samples are of higher quality as evidenced by the utility metric $U$. The value of this metric is even further optimized when decreasing the temperature in the softmax function, giving further evidence that the s-TNBM is able to learn the bias and that biasing the training set helps the s-TNBM to propose even better candidates. Crucially, we see that the performance gap between symmetric and vanilla TNBMs increase as we increase the system size from $N=20$ to $N=30$: in the latter case, the vanilla TNBM is not able to capture the cardinality constraint to the extent that less than $20 \%$ of the samples are valid, and more importantly, the quality of the valid ones, as measured by the utility, is still far from that obtained using s-TNBM.

\begin{figure}
    \centering
    \includegraphics[width=\linewidth, scale=0.4]{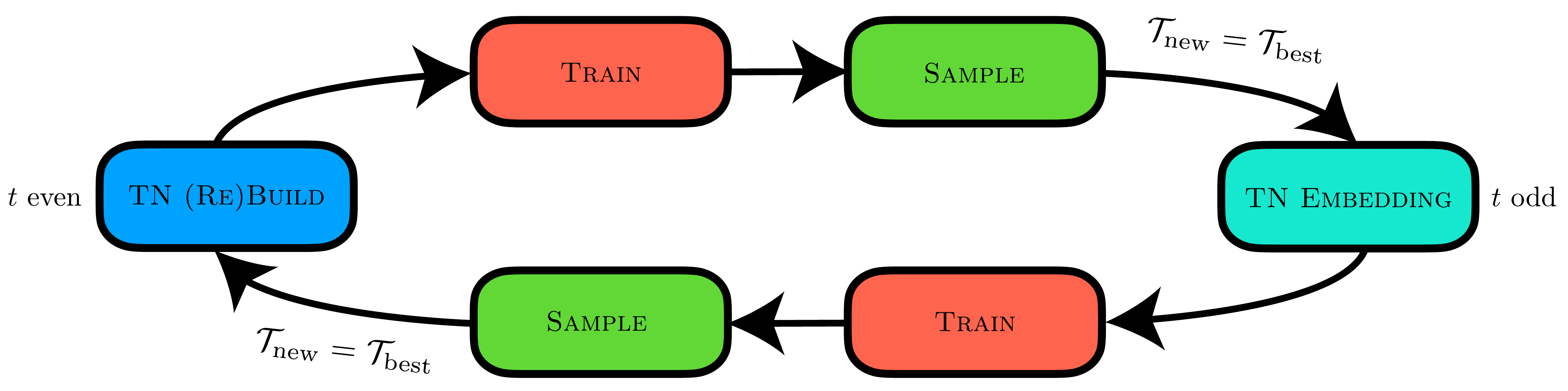}
    \caption{\textbf{Flowchart of all the steps entering in the tensor network generative algorithm for combinatorial optimization problems with hard-constraints.} See also \hyperref[alg:alg1]{Algorithm 1} for a detailed description of the various steps.}
    \label{fig:flowchart}
\end{figure}

\begin{widetext}
    \begin{center}
    \begin{tabular}{p{0.95\textwidth}}
    \hline
    \textbf{Algorithm 1:} Tensor network generative algorithm for hard constraints\\
    \hline
\begin{algorithmic} \label{alg:alg1}
\State \textbf{Input} \hspace{0.1in} Callback cost function $y=C(\vec{x})$, $\vec{x} \in \{0,1\}^{\otimes N}$, and equality constraints $A\vec{x}=\vec{b}$
\State \textbf{Output} \hspace{0.1in} $\vec{x}^{*} \approx \mathrm{argmin} (C(\vec{x}))$
\State $\psi(\vec{x}) \gets \text{Constraints } A\vec{x}=\vec{b}$ \Comment{Build$^*$  uniform TN from constraints (all QN blocks set to same value)}
\State $t \gets 0 $
\State $\mathcal{T} \gets \textsc{Sample}(|\psi|^2)$ \Comment{Extract training set via \hyperref[sec_sampling]{perfect sampling} from s-TNBM}
\State $U|_{t} \gets \mathbf{mean}(5\% $ lowest $\{C(\vec{x})\text{: } \vec{x} \in \mathcal{T}\})$ \Comment{Compute utility from mean of lowest cost 5\% samples}
    \While{$(U|_{t+1}-U|_{t}) > \epsilon $}
        \State $t \gets t+1 $ \;
        \If{$t$ odd}
            \State $\mathcal{T} \gets \mathcal{T}|_{5 \%}$ \Comment{Create new training set from $5\%$ lowest cost samples}
            \State $p_T(\{\vec{x}\in \mathcal{T}\}) \gets \{C(\{\vec{x} \in \mathcal{T}\}), T\}$ \Comment{Construct training distribution $p_T(\vec{x})=e^{-C(\vec{x})/T}/\sum_{\vec{x} \in \mathcal{T}} e^{-C(\vec{x})/T}$}
            \State $\psi(\vec{x}) \gets \{\text{Constraints } A\vec{x}=\vec{b}, \text{ Samples } \mathcal{T}\}$ \Comment{\hyperref[sec_initialization]{Constrained TN embedding} from constraints and samples}

        \Else
            \State $p_T(\{\vec{x}\in \mathcal{T}\}) \gets \{\vec{x} \in \mathcal{T}\}$ \Comment{Construct uniform probability distribution over training set}
            \State $\psi(\vec{x}) \gets \text{Constraints } A\vec{x}=\vec{b}$ \Comment{Rebuild$^*$ uniform TN from constraints (all QN blocks set to same value)}
        \EndIf
     \State $\psi \gets \textsc{Train}(\psi)$ \Comment{\hyperref[sec_training]{Train MPS} with distribution $p_T$ for a few sweeps}
     \State $\mathcal{T} \gets \textsc{Sample}(|\psi|^2)$ \Comment{Extract training set via \hyperref[sec_sampling]{perfect sampling} from s-TNBM}
     \State $U|_{t+1} \gets \mathbf{mean}(5\% $ lowest $\{C(\vec{x})\text{: } \forall \vec{x} \in \mathcal{T}\})$ \Comment{Compute utility from mean of lowest cost 5\% samples}

    \EndWhile
    \State \textbf{Return} \hspace{0.1in} $\vec{x} = \text{argmin}(C(\vec{x})\text{: } \vec{x} \in \mathcal{T}))$\\
    \State $^*$: When an efficient TN representation of the solution space does not exist, an alternative construction is to search randomly samples satisfying the constraints and applying the \hyperref[sec_initialization]{Constrained TN embedding} step on these. This will only construct an approximate TN to the entire solution space (note that in general the TN ansatz will contain more solutions than are fed in from a set of seeds, as argued in Sec. \ref{sec_initialization}).
\end{algorithmic}\\
    \hline
    \end{tabular}
    \end{center}
\end{widetext}

\begin{figure}
\centering
\includegraphics[width=\linewidth, scale=0.5]{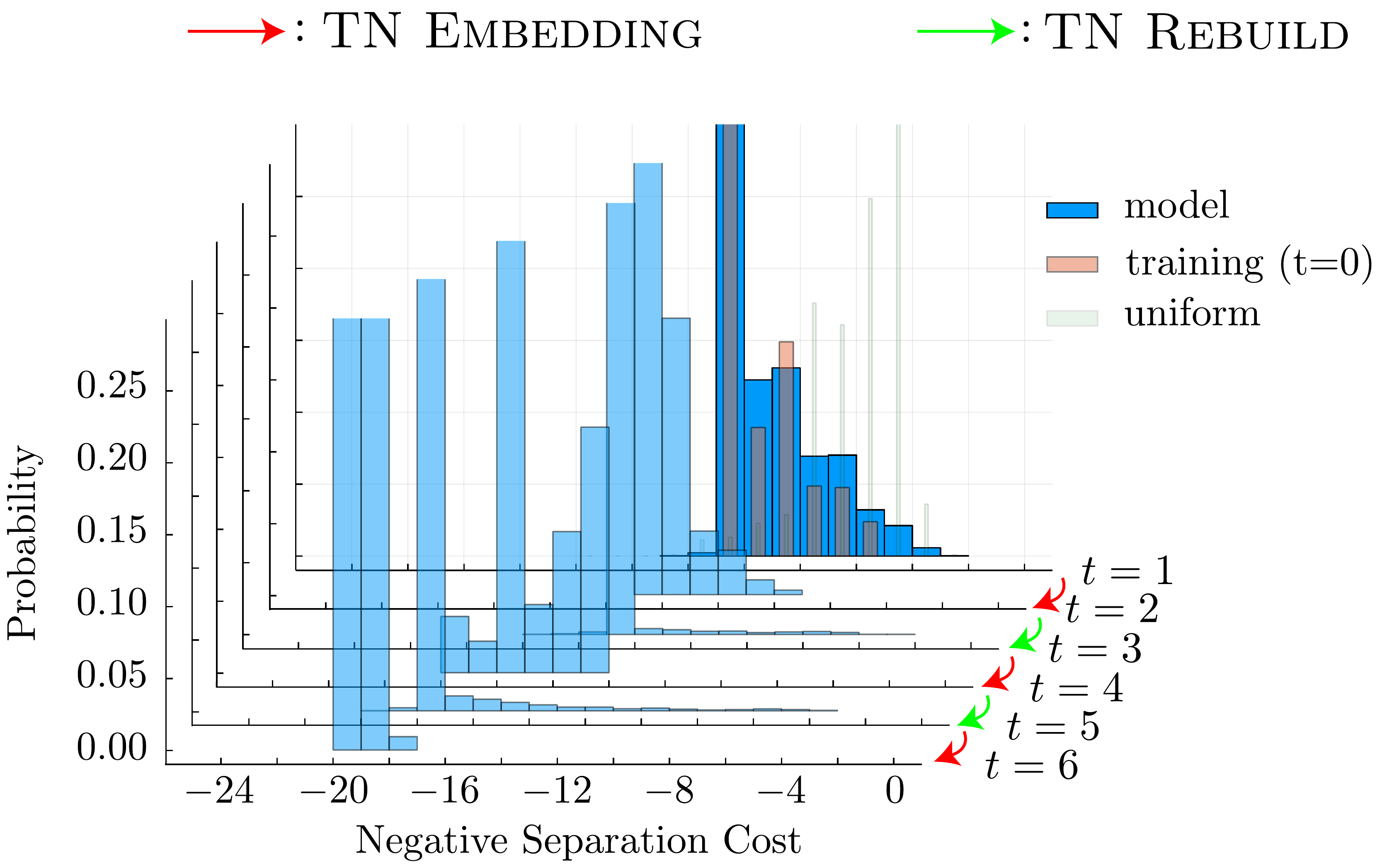}
\caption{\textbf{Finding optimal solutions when efficient MPS representation is available, using \hyperref[alg:alg1]{Algorithm 1}.} Data shown for $N=50$, $\kappa=25$, and negative separation cost as cost function. The utility extracted from the model distribution after the first step is $U_{\rm model}|_{t=1}=-11.05$, while that at the final step is $U_{\rm model}|_{t=6}=-20.0$. The uniform sampler has utility $U_{\rm uniform}=-9.64$. Bond dimension and learning rate are fixed throughout all steps and given by $\chi=30$, $\alpha=0.02$, respectively.}
\label{fig:utility_evol}
\end{figure}

In the next experiment we take full advantage of the knowledge that an exact and efficient representation of the valid space exists in terms of a symmetric MPS of fixed cardinality. We consider here a much larger system with $N=50$ and cardinality $\kappa=25$, which results in a solution space of size $|\mathcal{S}|=\binom{50}{25} \approx 1.26 \times 10^{14}$. We use the following protocol, which is also summarized in Fig. \ref{fig:flowchart} for general equality constraints and detailed in \hyperref[alg:alg1]{Algorithm 1}. The first step is to sample uniformly from the fixed cardinality MPS. To guarantee that this sampling is uniform over the space of fixed cardinality bitstrings, we set all allowed QN blocks to be of size $1\times 1$ and filled in by the the same value (the value will be set by the normalization condition of the TN). This forms our initial training set. We train as usual for one sweep with this initial choice of MPS and sample at least as many times as samples there are in the training set. For our experiments the size of the training set is always fixed to be $|\mathcal{T}|=100$, while the number of samples or queries on the generator is $Q=10^4$ (we emphasize again that sampling on an MPS can be done very efficiently). With the samples at hand we keep the $100$ lowest cost samples as the new training data set. At the next step we apply the initialization step described in Sec. \ref{sec_initialization} so as to extract the most relevant QNs from where we construct our next initial MPS. Next we train for one sweep with an effective temperature which we choose to be $T=1/2\text{std}(C(\{\vec{x} \in \mathcal{T}\}))$, with std the standard deviation. We next sample $Q$ times and keep the lowest cost $100$ samples for our next step training dataset. We repeat this procedure for as long as the results improve. The results of this experiment are shown in Fig. \ref{fig:utility_evol}. After only 6 steps (and only a couple minutes of computational time on a regular laptop), the symmetric MPS is able to reach a minimum cost of $\mathcal{C}=-20$, with the absolute minimum corresponding to $\mathcal{C}=-26$. To put this result into context, we remark that the space of bitstrings with cost $\mathcal{C}\leq 20$, satisfies $|\mathcal{S}|_{\mathcal{C} \leq 20}/|\mathcal{S}| \approx 10^{-7}$ (see App. \ref{s:degeneracy} for an exact analytical expression), so that a constrained uniform sampler over the solution space would require at least of the order of $\mathcal{O}(10^3)$ times more samples than the ones carried out here, indicating that our MPS indeed is able to find its way to a subspace near the global minimum, even when this effectively corresponds to a tiny fraction of the entire valid space, and in turn the cost function is highly nonlocal.

This optimization scheme bears some resemblance with the QAOA and other approaches following an \textit{exploration/exploitation} paradigm: the \textsc{TN Rebuild} step serves as a mixer near the optimum solution (which learns over the best data from the previous iteration), while the TN Embedding step, moves directly along the direction of minimum by learning only the most relevant QNs (\textit{latent features}) from the lowest cost samples from the previous iteration. Our approach, being data driven, is agnostic to the form of the cost function, as illustrated in our experiments.

\section{ Comparison to neural network architectures} \label{s:efficient}

\begin{figure}
\centering
\includegraphics[width=\linewidth]{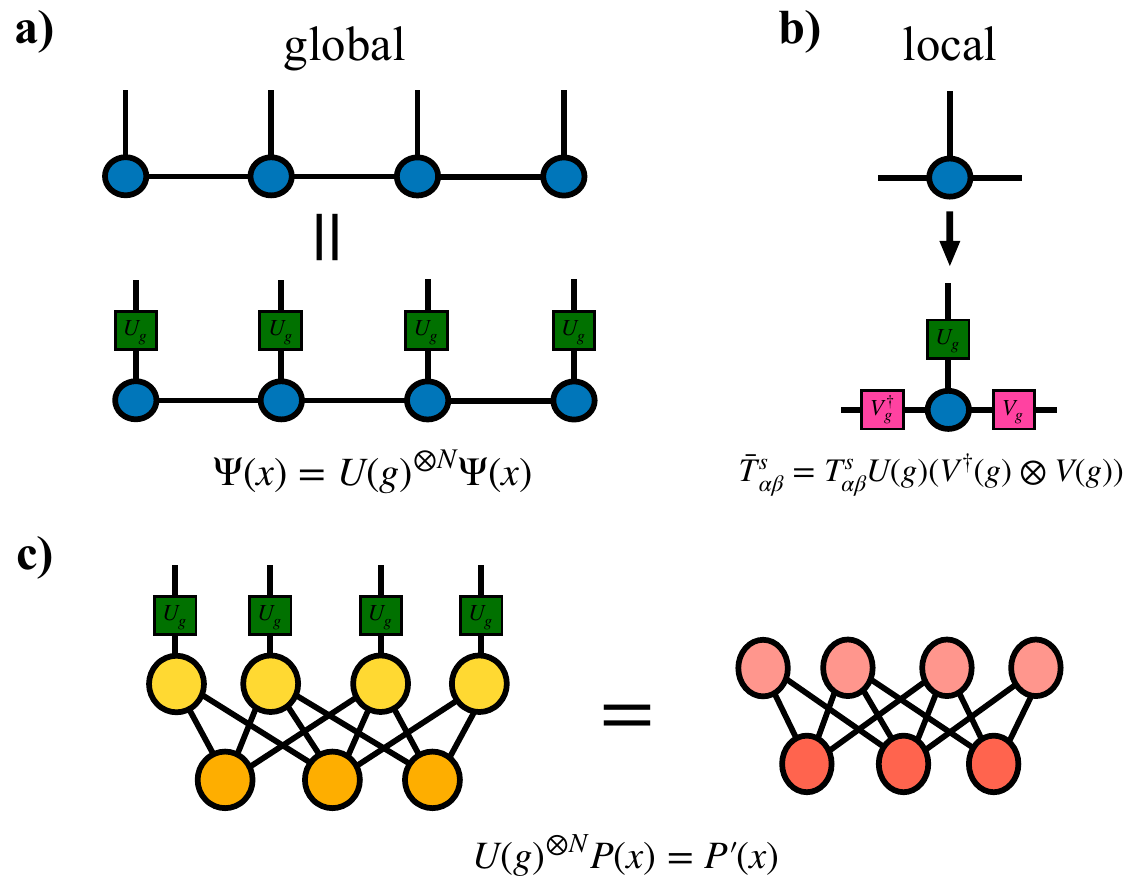}
\caption{\textbf{Illustration of how a TN and RBM are transformed under global symmetry.} (a) $\Psi(x)$ is an MPS with $\mathcal{G}$ symmetry. The MPS remains constant under a global rotation for all sites. (b) The local tensors can be associated with the original ones by contracting local tensor $T$ with gate $U$. (c) Global symmetry transformation on a RBM, which results in a new RBM with updated weights. Due to the sparsity structure of the RBM network, it becomes very challenging to get the updated RBM parameters after the symmetry transformation. }
\label{fig:mpsrbm}
\end{figure}

\begin{figure}
\centering
\includegraphics[width=\linewidth]{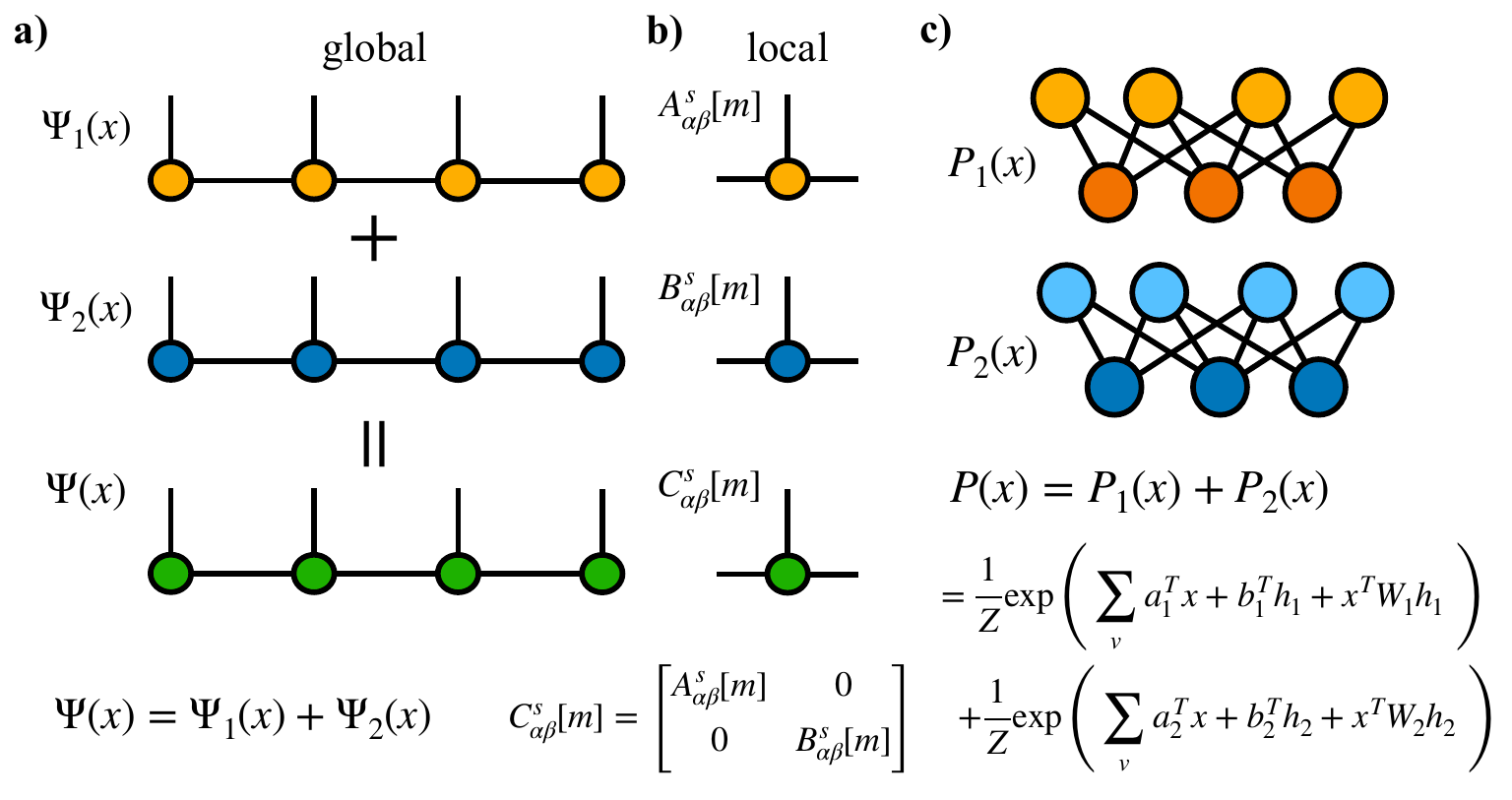}
\caption{\textbf{Linearity of TN vs nonlinearity of RBM.} (a) Linear hybridization of two different MPS $\Psi_1(x)$ and $\Psi_2(x)$ can be accomplished by constructing a new MPS with each tensor taking the direct sum as shown in (b). However, for nonlinear models such as RBM or ANNs, it's very challenging to construct the sum of $P_1(x)$ and $P_2(x)$ directly without any training. }
\label{fig:mpsrbm2}
\end{figure}

TNs are an efficient representation of states with limited amount of entanglement. In recent years there has been a lot of excitement about the possibility of using other representations that may be more efficient than TNs at capturing long-range entanglement. RBMs, the building block of many deep neural network architectures, have been suggested to be a more efficient representation than TNs for certain classes of states displaying long-range entanglement \cite{Deng_2017, huang2021neural}. Here we argue that $U(1)$-symmetric states may be more challenging to construct using RBMs, and likely the same true for any $\mathcal{G}$-symmetric state, with $\mathcal{G}$ any global internal symmetry, when compared to TNs. The intuition behind this argument relies on the fact that TNs are multi-linear models. In particular, the contraction of an MPS with one-site operators can be performed locally. Under a global group transformation represented by a unitary matrix $U(g)^{\otimes N}$, with $g \in \mathcal{G}$, the new MPS can be constructed straight-forwardly, see Fig. \ref{fig:mpsrbm}(a-b). In contrast, for a RBM, after applying such global group transformation, it is very challenging to get the weights of the new RBM. Unlike for a TN, we have to train the new RBM. So it becomes natural to impose such symmetries in TNs, in contrast to RBMs or ANNs in general.

For a TN, group operations on sites will introduce group representations on each link consistent with charge conservation. So a global symmetry transformation will generate group representations on each link of the MPS. While for RBM or ANN, it is very challenging to have this property.

Another argument in favor of TNs for describing symmetric states is precisely due to the linearity of the TN. As shown in Fig. \ref{fig:mpsrbm2}(a,b), it is very easy to construct the linear hybridization of two MPS, the new tensors are a direct sum of these MPS tensors. The bond dimension would be the sum of bond dimensions of $\Psi_1(x)$ and $\Psi_2(x)$, and the construction is exact.  In contrast, for a RBM, we are not able to construct the new RBM directly from the original RBM parameters. The only option left is to train a new RBM, which mimics the summation, which is only an approximation.

\section{Conclusions}\label{s:conclusions}
In this work we have shown how one can use symmetric TNs for the purpose of generative modeling and constrained combinatorial optimization. We have tested their performance on various datasets serving as proxies for the kinds of problems our generative algorithm, dubbed here constrained-GEO, is applicable to. Most prominently, we include combinatorial optimization problems subject to arbitrary integer-valued equality constraints, which appear in many problems of industrial interest. We have found that by mapping the problem of equalities to a problem involving $U(1)$ global charges, a symmetric MPS generative model is able to generalize (even in the presence of random instances of $A$ and $\vec{b}$). This class of models, referred here as s-TNBMs, has a series of benefits which we recount:
\begin{itemize}
    \item By exploiting $U(1)$ global charge conservation, we are able to reduce the space complexity by working with block-sparse tensors, as opposed to dense tensors, as shown on the right panel of Fig. \ref{fig:resources_v_vs_s}.
    \item In turn, given the block-sparse structure of tensors allows for faster learning of the data, often requiring an order of magnitude less number of sweeps to converge when training for the problem sizes studied in here.
    \begin{itemize}
    \item This point translates into less computational time per operation (contraction, merging) as shown on the left panel of Fig. \ref{fig:resources_v_vs_s};
    \item less number of sweeps as shown on Fig. \ref{fig:u-curve}(a-b); and
    \item no time is spent on sampling invalid samples.
    \end{itemize}
    \item It permits to have a better intuition on the complexity of the data, as viewed when counting the number of QN blocks appearing in the MPS (i.e. the minimum bond dimension needed to exactly capture all charges from the data). \item It allows to generalize in the presence of scarcity of data.
    \item It allows to find better solutions in the context of combinatorial optimization problems subject to constraints.
\end{itemize}
Of all these points, the last two ones are perhaps the most relevant ones. Classical ML models require many samples for them to generalize. By exploiting internal global symmetries like $U(1)$, which in turn are hard to impose in ANNs architectures as argued in Sec. \ref{s:efficient} (see also Refs. \cite{choo2018symmetries, dawid2022modern}), we are able to reduce the amount of training data even further.
The last point is also worth stressing: symmetric TNs were originally conceived to efficiently encode quantum states subject to global symmetries, but in the context of finding ground-states of Hamiltonians, as well as in the context of Hamiltonian dynamics. What we have found in this work is that constraining TNs may be a key step at not only decreasing the complexity of the learning algorithm, but crucially, at finding novel and better solutions to constrained combinatorial optimization problems, which would be otherwise out-of-reach by other classical ML and state-of-the-art optimization heuristics.
\section{Outlook} \label{s_outlook}
A clear future extension of the current work is to analyze the role of other symmetries. Trivial extensions include other abelian symmetries, like $\mathbb{Z}_2$, which appear in physical models such as the TFI model discussed in this work, but also in some toy datasets like the \textit{parity} dataset, for which TNs have already shown superior performance over ANNs such as RBMs at both learning and generative modeling performance  \cite{shalev2017failures, romero2019weighted,montufar2011expressive, stokes2019probabilistic, bradley2020modeling}. Perhaps more intriguing would be to study the role of nonabelian symmetries in the context of generative modeling. Of special relevance is $SU(2)$, perhaps the other most relevant \textit{quantum}-inspired symmetry apart from $U(1)$, and for which a way to construct TNs that are symmetric under $SU(2)$ already exists in the literature \cite{mcculloch2007density,singh2012tensor,singh2010simulation}. A more challenging and practically relevant question is to address the presence of inequality constraints (i.e. of the type $A\vec{x} \leq \vec{b}$), which also abound in many combinatorial optimization problems. This is an ongoing work and will be presented in a subsequent publication.

We now put our work into the broader context of current progress in the domains of constrained combinatorial optimization and ML models exploring the role of symmetries and constraints. In the domain of combinatorial optimization, GNNs have become a very promising framework (see e.g., \cite{gasse2019exact, nair2020solving}). In this case, one can encode the coefficients of matrix $A$ and vector $\vec{b}$ as features in a bipartite graph. This approach leverages generative modeling for \textit{node selection} and \textit{variable selection}, steps that appear in branch-and-bound for solving \textsc{MILP} problems. In the context of quantum models addressing \textsc{QUBO} problems with hard-constraints, such as XY-QAOA \cite{wang2018quantum,niroula2022constrained}, current methods suffer from the fact that only a very limited set of equality constraints can be addressed, namely those for which one can implement \textit{local} moves around the valid space (and which can be implemented by at most two-site, local gates), such as cardinality-type constraints. Likewise, a recent tensor network approach was proposed to tackle the open-pit problem subject to equality constraints \cite{hao2022quantum}. While bearing some resemblance to our work, this approach however is only valid for equality constraints with a locality structure, in that only few local variables can appear at any time in any equality, and where in turn the cost function must be mapped to a local Hamiltonian.  Our approach is not limited by specific equality constraints nor by specific choices of cost functions.

On the side of implementing symmetries in the context of ML, there has been very recent interest in equivariant ANNs, both classical (see e.g. Refs. \cite{kondor2018generalization, bronstein2021geometric}), and quantum (see e.g. Refs. \cite{meyer2022exploiting,Larocca_2022,nguyen2022theory,ragone2022representation}). Equivariant ANNs exploit symmetry in data so that the output samples also preserve this symmetry, such as is the case of $E(3)$ equivariant ANNs (which transform data under translations and rotations, with applications including image processing). In contrast, our work exploits symmetry at the level of the generator, in this case a s-TNBM, so that it only outputs samples satisfying arbitrary equality constraints, where different equality constraints correspond to different irreps of $U(1)$ (i.e. different fluxes in TN terminology). Lastly, there has been recent interest in using autoregressive neural networks for modeling physical systems with local gauge symmetries \cite{luo2021gauge, luo2022gauge}, which are fundamentally different from the global internal symmetries we are interested in here. The latter have been studied in works \cite{Hibat2020, morawetz2021u} for the simplest $U(1)$ symmetry corresponding to an equality of cardinality type, using Recurrent Neural Networks (RNNs). However at present it is not clear how to extend their construction to arbitrary equality constraints, and if so, whether the resulting RNN corresponds to a more efficient representation than that of symmetric MPS.

An exciting milestone ahead is to demonstrate the power of our s-TNBM models in the context of large-scale industrial applications. Achieving such quantum-inspired advantage would be a promising step towards demonstrating practical quantum advantage with near-term devices. This could be possible by exploiting recent efficient proposals~\cite{rudolph2022MPSdecomposition, rudolph2022synergistic} to map TNs to quantum circuit generative models such as quantum-circuit Born machines~\cite{Benedetti2019}.

\begin{acknowledgments} 

The authors would like to acknowledge Mohamed Hibat-Allah, Marta Mauri, and Artem Strashko for helpful discussions and insights. We also acknowledge Dax Enshan Koh and Brian Dellabetta for reviewing our manuscript and providing feedback.

\end{acknowledgments}


\bibliography{main.bbl}

\begin{thebibliography}{75}%
\makeatletter
\providecommand \@ifxundefined [1]{%
 \@ifx{#1\undefined}
}%
\providecommand \@ifnum [1]{%
 \ifnum #1\expandafter \@firstoftwo
 \else \expandafter \@secondoftwo
 \fi
}%
\providecommand \@ifx [1]{%
 \ifx #1\expandafter \@firstoftwo
 \else \expandafter \@secondoftwo
 \fi
}%
\providecommand \natexlab [1]{#1}%
\providecommand \enquote  [1]{``#1''}%
\providecommand \bibnamefont  [1]{#1}%
\providecommand \bibfnamefont [1]{#1}%
\providecommand \citenamefont [1]{#1}%
\providecommand \href@noop [0]{\@secondoftwo}%
\providecommand \href [0]{\begingroup \@sanitize@url \@href}%
\providecommand \@href[1]{\@@startlink{#1}\@@href}%
\providecommand \@@href[1]{\endgroup#1\@@endlink}%
\providecommand \@sanitize@url [0]{\catcode `\\12\catcode `\$12\catcode
  `\&12\catcode `\#12\catcode `\^12\catcode `\_12\catcode `\%12\relax}%
\providecommand \@@startlink[1]{}%
\providecommand \@@endlink[0]{}%
\providecommand \url  [0]{\begingroup\@sanitize@url \@url }%
\providecommand \@url [1]{\endgroup\@href {#1}{\urlprefix }}%
\providecommand \urlprefix  [0]{URL }%
\providecommand \Eprint [0]{\href }%
\providecommand \doibase [0]{http://dx.doi.org/}%
\providecommand \selectlanguage [0]{\@gobble}%
\providecommand \bibinfo  [0]{\@secondoftwo}%
\providecommand \bibfield  [0]{\@secondoftwo}%
\providecommand \translation [1]{[#1]}%
\providecommand \BibitemOpen [0]{}%
\providecommand \bibitemStop [0]{}%
\providecommand \bibitemNoStop [0]{.\EOS\space}%
\providecommand \EOS [0]{\spacefactor3000\relax}%
\providecommand \BibitemShut  [1]{\csname bibitem#1\endcsname}%
\let\auto@bib@innerbib\@empty
\bibitem [{\citenamefont {Alcazar}\ \emph {et~al.}(2021)\citenamefont
  {Alcazar}, \citenamefont {Vakili}, \citenamefont {Kalayci},\ and\
  \citenamefont {Perdomo-Ortiz}}]{alcazar2021enhancing}%
  \BibitemOpen
  \bibfield  {author} {\bibinfo {author} {\bibfnamefont {Javier}\ \bibnamefont
  {Alcazar}}, \bibinfo {author} {\bibfnamefont {Mohammad~Ghazi}\ \bibnamefont
  {Vakili}}, \bibinfo {author} {\bibfnamefont {Can~B.}\ \bibnamefont
  {Kalayci}}, \ and\ \bibinfo {author} {\bibfnamefont {Alejandro}\ \bibnamefont
  {Perdomo-Ortiz}},\ }\bibfield  {title} {\enquote {\bibinfo {title} {Geo:
  Enhancing combinatorial optimization with classical and quantum generative
  models},}\ }\href {https://arxiv.org/abs/2101.06250} {\bibfield  {journal}
  {\bibinfo  {journal} {arXiv:2101.06250}\ } (\bibinfo {year}
  {2021})}\BibitemShut {NoStop}%
\bibitem [{\citenamefont {Or{\'u}s}(2019)}]{orus2019tensor}%
  \BibitemOpen
  \bibfield  {author} {\bibinfo {author} {\bibfnamefont {Rom{\'a}n}\
  \bibnamefont {Or{\'u}s}},\ }\bibfield  {title} {\enquote {\bibinfo {title}
  {Tensor networks for complex quantum systems},}\ }\href@noop {} {\bibfield
  {journal} {\bibinfo  {journal} {Nature Reviews Physics}\ }\textbf {\bibinfo
  {volume} {1}},\ \bibinfo {pages} {538--550} (\bibinfo {year}
  {2019})}\BibitemShut {NoStop}%
\bibitem [{\citenamefont {Chen}\ \emph {et~al.}(2022)\citenamefont {Chen},
  \citenamefont {Helms}, \citenamefont {Hale}, \citenamefont {Lee},
  \citenamefont {Li}, \citenamefont {Gray}, \citenamefont {Christou},
  \citenamefont {Zapf}, \citenamefont {Chan},\ and\ \citenamefont
  {Cheng}}]{chen2022using}%
  \BibitemOpen
  \bibfield  {author} {\bibinfo {author} {\bibfnamefont {Dian-Teng}\
  \bibnamefont {Chen}}, \bibinfo {author} {\bibfnamefont {Phillip}\
  \bibnamefont {Helms}}, \bibinfo {author} {\bibfnamefont {Ashlyn~R}\
  \bibnamefont {Hale}}, \bibinfo {author} {\bibfnamefont {Minseong}\
  \bibnamefont {Lee}}, \bibinfo {author} {\bibfnamefont {Chenghan}\
  \bibnamefont {Li}}, \bibinfo {author} {\bibfnamefont {Johnnie}\ \bibnamefont
  {Gray}}, \bibinfo {author} {\bibfnamefont {George}\ \bibnamefont {Christou}},
  \bibinfo {author} {\bibfnamefont {Vivien~S}\ \bibnamefont {Zapf}}, \bibinfo
  {author} {\bibfnamefont {Garnet Kin-Lic}\ \bibnamefont {Chan}}, \ and\
  \bibinfo {author} {\bibfnamefont {Hai-Ping}\ \bibnamefont {Cheng}},\
  }\bibfield  {title} {\enquote {\bibinfo {title} {Using hyperoptimized tensor
  networks and first-principles electronic structure to simulate the
  experimental properties of the giant $\{$Mn84$\}$ torus},}\ }\href@noop {}
  {\bibfield  {journal} {\bibinfo  {journal} {The Journal of Physical Chemistry
  Letters}\ }\textbf {\bibinfo {volume} {13}},\ \bibinfo {pages} {2365--2370}
  (\bibinfo {year} {2022})}\BibitemShut {NoStop}%
\bibitem [{\citenamefont {Stoudenmire}\ and\ \citenamefont
  {Schwab}(2016)}]{stoudenmire2016supervised}%
  \BibitemOpen
  \bibfield  {author} {\bibinfo {author} {\bibfnamefont {Edwin}\ \bibnamefont
  {Stoudenmire}}\ and\ \bibinfo {author} {\bibfnamefont {David~J}\ \bibnamefont
  {Schwab}},\ }\bibfield  {title} {\enquote {\bibinfo {title} {Supervised
  learning with tensor networks},}\ }\href@noop {} {\bibfield  {journal}
  {\bibinfo  {journal} {Advances in Neural Information Processing Systems}\
  }\textbf {\bibinfo {volume} {29}} (\bibinfo {year} {2016})}\BibitemShut
  {NoStop}%
\bibitem [{\citenamefont {Han}\ \emph {et~al.}(2018)\citenamefont {Han},
  \citenamefont {Wang}, \citenamefont {Fan}, \citenamefont {Wang},\ and\
  \citenamefont {Zhang}}]{han2018unsupervised}%
  \BibitemOpen
  \bibfield  {author} {\bibinfo {author} {\bibfnamefont {Zhao-Yu}\ \bibnamefont
  {Han}}, \bibinfo {author} {\bibfnamefont {Jun}\ \bibnamefont {Wang}},
  \bibinfo {author} {\bibfnamefont {Heng}\ \bibnamefont {Fan}}, \bibinfo
  {author} {\bibfnamefont {Lei}\ \bibnamefont {Wang}}, \ and\ \bibinfo {author}
  {\bibfnamefont {Pan}\ \bibnamefont {Zhang}},\ }\bibfield  {title} {\enquote
  {\bibinfo {title} {Unsupervised generative modeling using matrix product
  states},}\ }\href {\doibase 10.1103/physrevx.8.031012} {\bibfield  {journal}
  {\bibinfo  {journal} {PRX}\ }\textbf {\bibinfo {volume} {8}},\ \bibinfo
  {pages} {031012} (\bibinfo {year} {2018})}\BibitemShut {NoStop}%
\bibitem [{\citenamefont {Wang}\ \emph {et~al.}(2020)\citenamefont {Wang},
  \citenamefont {Roberts}, \citenamefont {Vidal},\ and\ \citenamefont
  {Leichenauer}}]{wang2020anomaly}%
  \BibitemOpen
  \bibfield  {author} {\bibinfo {author} {\bibfnamefont {Jinhui}\ \bibnamefont
  {Wang}}, \bibinfo {author} {\bibfnamefont {Chase}\ \bibnamefont {Roberts}},
  \bibinfo {author} {\bibfnamefont {Guifre}\ \bibnamefont {Vidal}}, \ and\
  \bibinfo {author} {\bibfnamefont {Stefan}\ \bibnamefont {Leichenauer}},\
  }\bibfield  {title} {\enquote {\bibinfo {title} {Anomaly detection with
  tensor networks},}\ }\href@noop {} {\bibfield  {journal} {\bibinfo  {journal}
  {arXiv preprint arXiv:2006.02516}\ } (\bibinfo {year} {2020})}\BibitemShut
  {NoStop}%
\bibitem [{\citenamefont {Liu}\ \emph {et~al.}(2021)\citenamefont {Liu},
  \citenamefont {Wang},\ and\ \citenamefont {Zhang}}]{liu2021tropical}%
  \BibitemOpen
  \bibfield  {author} {\bibinfo {author} {\bibfnamefont {Jin-Guo}\ \bibnamefont
  {Liu}}, \bibinfo {author} {\bibfnamefont {Lei}\ \bibnamefont {Wang}}, \ and\
  \bibinfo {author} {\bibfnamefont {Pan}\ \bibnamefont {Zhang}},\ }\bibfield
  {title} {\enquote {\bibinfo {title} {Tropical tensor network for ground
  states of spin glasses},}\ }\href@noop {} {\bibfield  {journal} {\bibinfo
  {journal} {Physical Review Letters}\ }\textbf {\bibinfo {volume} {126}},\
  \bibinfo {pages} {090506} (\bibinfo {year} {2021})}\BibitemShut {NoStop}%
\bibitem [{\citenamefont {Liu}\ \emph {et~al.}(2022)\citenamefont {Liu},
  \citenamefont {Gao}, \citenamefont {Cain}, \citenamefont {Lukin},\ and\
  \citenamefont {Wang}}]{liu2022computing}%
  \BibitemOpen
  \bibfield  {author} {\bibinfo {author} {\bibfnamefont {Jin-Guo}\ \bibnamefont
  {Liu}}, \bibinfo {author} {\bibfnamefont {Xun}\ \bibnamefont {Gao}}, \bibinfo
  {author} {\bibfnamefont {Madelyn}\ \bibnamefont {Cain}}, \bibinfo {author}
  {\bibfnamefont {Mikhail~D}\ \bibnamefont {Lukin}}, \ and\ \bibinfo {author}
  {\bibfnamefont {Sheng-Tao}\ \bibnamefont {Wang}},\ }\bibfield  {title}
  {\enquote {\bibinfo {title} {Computing solution space properties of
  combinatorial optimization problems via generic tensor networks},}\
  }\href@noop {} {\bibfield  {journal} {\bibinfo  {journal} {arXiv preprint
  arXiv:2205.03718}\ } (\bibinfo {year} {2022})}\BibitemShut {NoStop}%
\bibitem [{\citenamefont {Hao}\ \emph {et~al.}(2022)\citenamefont {Hao},
  \citenamefont {Huang}, \citenamefont {Jia},\ and\ \citenamefont
  {Peng}}]{hao2022quantum}%
  \BibitemOpen
  \bibfield  {author} {\bibinfo {author} {\bibfnamefont {Tianyi}\ \bibnamefont
  {Hao}}, \bibinfo {author} {\bibfnamefont {Xuxin}\ \bibnamefont {Huang}},
  \bibinfo {author} {\bibfnamefont {Chunjing}\ \bibnamefont {Jia}}, \ and\
  \bibinfo {author} {\bibfnamefont {Cheng}\ \bibnamefont {Peng}},\ }\bibfield
  {title} {\enquote {\bibinfo {title} {A quantum-inspired tensor network method
  for constrained combinatorial optimization problems},}\ }\href@noop {}
  {\bibfield  {journal} {\bibinfo  {journal} {arXiv preprint arXiv:2203.15246}\
  } (\bibinfo {year} {2022})}\BibitemShut {NoStop}%
\bibitem [{\citenamefont {Pozas-Kerstjens}\ \emph {et~al.}(2022)\citenamefont
  {Pozas-Kerstjens}, \citenamefont {Hern{\'a}ndez-Santana}, \citenamefont
  {Monturiol}, \citenamefont {L{\'o}pez}, \citenamefont {Scarpa}, \citenamefont
  {Gonz{\'a}lez-Guill{\'e}n},\ and\ \citenamefont
  {P{\'e}rez-Garc{\'\i}a}}]{pozas2022physics}%
  \BibitemOpen
  \bibfield  {author} {\bibinfo {author} {\bibfnamefont {Alejandro}\
  \bibnamefont {Pozas-Kerstjens}}, \bibinfo {author} {\bibfnamefont {Senaida}\
  \bibnamefont {Hern{\'a}ndez-Santana}}, \bibinfo {author} {\bibfnamefont
  {Jos{\'e} Ram{\'o}n~Pareja}\ \bibnamefont {Monturiol}}, \bibinfo {author}
  {\bibfnamefont {Marco~Castrill{\'o}n}\ \bibnamefont {L{\'o}pez}}, \bibinfo
  {author} {\bibfnamefont {Giannicola}\ \bibnamefont {Scarpa}}, \bibinfo
  {author} {\bibfnamefont {Carlos~E}\ \bibnamefont {Gonz{\'a}lez-Guill{\'e}n}},
  \ and\ \bibinfo {author} {\bibfnamefont {David}\ \bibnamefont
  {P{\'e}rez-Garc{\'\i}a}},\ }\bibfield  {title} {\enquote {\bibinfo {title}
  {Physics solutions for machine learning privacy leaks},}\ }\href@noop {}
  {\bibfield  {journal} {\bibinfo  {journal} {arXiv preprint arXiv:2202.12319}\
  } (\bibinfo {year} {2022})}\BibitemShut {NoStop}%
\bibitem [{\citenamefont {Hornik}\ \emph {et~al.}(1989)\citenamefont {Hornik},
  \citenamefont {Stinchcombe},\ and\ \citenamefont
  {White}}]{hornik1989multilayer}%
  \BibitemOpen
  \bibfield  {author} {\bibinfo {author} {\bibfnamefont {Kurt}\ \bibnamefont
  {Hornik}}, \bibinfo {author} {\bibfnamefont {Maxwell}\ \bibnamefont
  {Stinchcombe}}, \ and\ \bibinfo {author} {\bibfnamefont {Halbert}\
  \bibnamefont {White}},\ }\bibfield  {title} {\enquote {\bibinfo {title}
  {Multilayer feedforward networks are universal approximators},}\ }\href@noop
  {} {\bibfield  {journal} {\bibinfo  {journal} {Neural networks}\ }\textbf
  {\bibinfo {volume} {2}},\ \bibinfo {pages} {359--366} (\bibinfo {year}
  {1989})}\BibitemShut {NoStop}%
\bibitem [{\citenamefont {Glasser}\ \emph {et~al.}(2019)\citenamefont
  {Glasser}, \citenamefont {Sweke}, \citenamefont {Pancotti}, \citenamefont
  {Eisert},\ and\ \citenamefont {Cirac}}]{glasser2019}%
  \BibitemOpen
  \bibfield  {author} {\bibinfo {author} {\bibfnamefont {Ivan}\ \bibnamefont
  {Glasser}}, \bibinfo {author} {\bibfnamefont {Ryan}\ \bibnamefont {Sweke}},
  \bibinfo {author} {\bibfnamefont {Nicola}\ \bibnamefont {Pancotti}}, \bibinfo
  {author} {\bibfnamefont {Jens}\ \bibnamefont {Eisert}}, \ and\ \bibinfo
  {author} {\bibfnamefont {Ignacio}\ \bibnamefont {Cirac}},\ }\bibfield
  {title} {\enquote {\bibinfo {title} {Expressive power of tensor-network
  factorizations for probabilistic modeling},}\ }\href@noop {} {\bibfield
  {journal} {\bibinfo  {journal} {Advances in neural information processing
  systems}\ }\textbf {\bibinfo {volume} {32}} (\bibinfo {year}
  {2019})}\BibitemShut {NoStop}%
\bibitem [{\citenamefont {Markowitz}(1952)}]{Markowitz52}%
  \BibitemOpen
  \bibfield  {author} {\bibinfo {author} {\bibfnamefont {Harry}\ \bibnamefont
  {Markowitz}},\ }\bibfield  {title} {\enquote {\bibinfo {title} {Portfolio
  selection},}\ }\href {https://www.jstor.org/stable/2975974} {\bibfield
  {journal} {\bibinfo  {journal} {The Journal of Finance}\ }\textbf {\bibinfo
  {volume} {7}},\ \bibinfo {pages} {77--91} (\bibinfo {year}
  {1952})}\BibitemShut {NoStop}%
\bibitem [{\citenamefont {Conforti}\ \emph {et~al.}(2014)\citenamefont
  {Conforti}, \citenamefont {Cornu{\'e}jols}, \citenamefont {Zambelli} \emph
  {et~al.}}]{conforti2014integer}%
  \BibitemOpen
  \bibfield  {author} {\bibinfo {author} {\bibfnamefont {Michele}\ \bibnamefont
  {Conforti}}, \bibinfo {author} {\bibfnamefont {G{\'e}rard}\ \bibnamefont
  {Cornu{\'e}jols}}, \bibinfo {author} {\bibfnamefont {Giacomo}\ \bibnamefont
  {Zambelli}},  \emph {et~al.},\ }\href@noop {} {\emph {\bibinfo {title}
  {Integer programming}}},\ Vol.\ \bibinfo {volume} {271}\ (\bibinfo
  {publisher} {Springer},\ \bibinfo {year} {2014})\BibitemShut {NoStop}%
\bibitem [{\citenamefont {Gleixner}\ \emph {et~al.}(2021)\citenamefont
  {Gleixner}, \citenamefont {Hendel}, \citenamefont {Gamrath}, \citenamefont
  {Achterberg}, \citenamefont {Bastubbe}, \citenamefont {Berthold},
  \citenamefont {Christophel}, \citenamefont {Jarck}, \citenamefont {Koch},
  \citenamefont {Linderoth} \emph {et~al.}}]{gleixner2021miplib}%
  \BibitemOpen
  \bibfield  {author} {\bibinfo {author} {\bibfnamefont {Ambros}\ \bibnamefont
  {Gleixner}}, \bibinfo {author} {\bibfnamefont {Gregor}\ \bibnamefont
  {Hendel}}, \bibinfo {author} {\bibfnamefont {Gerald}\ \bibnamefont
  {Gamrath}}, \bibinfo {author} {\bibfnamefont {Tobias}\ \bibnamefont
  {Achterberg}}, \bibinfo {author} {\bibfnamefont {Michael}\ \bibnamefont
  {Bastubbe}}, \bibinfo {author} {\bibfnamefont {Timo}\ \bibnamefont
  {Berthold}}, \bibinfo {author} {\bibfnamefont {Philipp}\ \bibnamefont
  {Christophel}}, \bibinfo {author} {\bibfnamefont {Kati}\ \bibnamefont
  {Jarck}}, \bibinfo {author} {\bibfnamefont {Thorsten}\ \bibnamefont {Koch}},
  \bibinfo {author} {\bibfnamefont {Jeff}\ \bibnamefont {Linderoth}},  \emph
  {et~al.},\ }\bibfield  {title} {\enquote {\bibinfo {title} {Miplib 2017:
  data-driven compilation of the 6th mixed-integer programming library},}\
  }\href@noop {} {\bibfield  {journal} {\bibinfo  {journal} {Mathematical
  Programming Computation}\ }\textbf {\bibinfo {volume} {13}},\ \bibinfo
  {pages} {443--490} (\bibinfo {year} {2021})}\BibitemShut {NoStop}%
\bibitem [{\citenamefont {Gurobi~Optimization}(2018)}]{gurobi2018gurobi}%
  \BibitemOpen
  \bibfield  {author} {\bibinfo {author} {\bibfnamefont {LLC}\ \bibnamefont
  {Gurobi~Optimization}},\ }\href
  {https://www.gurobi.com/documentation/9.5/refman/index.html} {\enquote
  {\bibinfo {title} {Gurobi optimizer reference manual},}\ } (\bibinfo {year}
  {2018})\BibitemShut {NoStop}%
\bibitem [{\citenamefont {Hauke}\ \emph {et~al.}(2020)\citenamefont {Hauke},
  \citenamefont {Katzgraber}, \citenamefont {Lechner}, \citenamefont
  {Nishimori},\ and\ \citenamefont {Oliver}}]{hauke2020perspectives}%
  \BibitemOpen
  \bibfield  {author} {\bibinfo {author} {\bibfnamefont {Philipp}\ \bibnamefont
  {Hauke}}, \bibinfo {author} {\bibfnamefont {Helmut~G}\ \bibnamefont
  {Katzgraber}}, \bibinfo {author} {\bibfnamefont {Wolfgang}\ \bibnamefont
  {Lechner}}, \bibinfo {author} {\bibfnamefont {Hidetoshi}\ \bibnamefont
  {Nishimori}}, \ and\ \bibinfo {author} {\bibfnamefont {William~D}\
  \bibnamefont {Oliver}},\ }\bibfield  {title} {\enquote {\bibinfo {title}
  {Perspectives of quantum annealing: Methods and implementations},}\
  }\href@noop {} {\bibfield  {journal} {\bibinfo  {journal} {Reports on
  Progress in Physics}\ }\textbf {\bibinfo {volume} {83}},\ \bibinfo {pages}
  {054401} (\bibinfo {year} {2020})}\BibitemShut {NoStop}%
\bibitem [{\citenamefont {Nair}\ \emph {et~al.}(2020)\citenamefont {Nair},
  \citenamefont {Bartunov}, \citenamefont {Gimeno}, \citenamefont {von Glehn},
  \citenamefont {Lichocki}, \citenamefont {Lobov}, \citenamefont {O'Donoghue},
  \citenamefont {Sonnerat}, \citenamefont {Tjandraatmadja}, \citenamefont
  {Wang} \emph {et~al.}}]{nair2020solving}%
  \BibitemOpen
  \bibfield  {author} {\bibinfo {author} {\bibfnamefont {Vinod}\ \bibnamefont
  {Nair}}, \bibinfo {author} {\bibfnamefont {Sergey}\ \bibnamefont {Bartunov}},
  \bibinfo {author} {\bibfnamefont {Felix}\ \bibnamefont {Gimeno}}, \bibinfo
  {author} {\bibfnamefont {Ingrid}\ \bibnamefont {von Glehn}}, \bibinfo
  {author} {\bibfnamefont {Pawel}\ \bibnamefont {Lichocki}}, \bibinfo {author}
  {\bibfnamefont {Ivan}\ \bibnamefont {Lobov}}, \bibinfo {author}
  {\bibfnamefont {Brendan}\ \bibnamefont {O'Donoghue}}, \bibinfo {author}
  {\bibfnamefont {Nicolas}\ \bibnamefont {Sonnerat}}, \bibinfo {author}
  {\bibfnamefont {Christian}\ \bibnamefont {Tjandraatmadja}}, \bibinfo {author}
  {\bibfnamefont {Pengming}\ \bibnamefont {Wang}},  \emph {et~al.},\ }\bibfield
   {title} {\enquote {\bibinfo {title} {Solving mixed integer programs using
  neural networks},}\ }\href@noop {} {\bibfield  {journal} {\bibinfo  {journal}
  {arXiv preprint arXiv:2012.13349}\ } (\bibinfo {year} {2020})}\BibitemShut
  {NoStop}%
\bibitem [{\citenamefont {Cappart}\ \emph {et~al.}(2021)\citenamefont
  {Cappart}, \citenamefont {Ch{\'e}telat}, \citenamefont {Khalil},
  \citenamefont {Lodi}, \citenamefont {Morris},\ and\ \citenamefont
  {Veli{\v{c}}kovi{\'c}}}]{cappart2021combinatorial}%
  \BibitemOpen
  \bibfield  {author} {\bibinfo {author} {\bibfnamefont {Quentin}\ \bibnamefont
  {Cappart}}, \bibinfo {author} {\bibfnamefont {Didier}\ \bibnamefont
  {Ch{\'e}telat}}, \bibinfo {author} {\bibfnamefont {Elias}\ \bibnamefont
  {Khalil}}, \bibinfo {author} {\bibfnamefont {Andrea}\ \bibnamefont {Lodi}},
  \bibinfo {author} {\bibfnamefont {Christopher}\ \bibnamefont {Morris}}, \
  and\ \bibinfo {author} {\bibfnamefont {Petar}\ \bibnamefont
  {Veli{\v{c}}kovi{\'c}}},\ }\bibfield  {title} {\enquote {\bibinfo {title}
  {Combinatorial optimization and reasoning with graph neural networks},}\
  }\href@noop {} {\bibfield  {journal} {\bibinfo  {journal} {arXiv preprint
  arXiv:2102.09544}\ } (\bibinfo {year} {2021})}\BibitemShut {NoStop}%
\bibitem [{\citenamefont {Schuetz}\ \emph {et~al.}(2022)\citenamefont
  {Schuetz}, \citenamefont {Brubaker},\ and\ \citenamefont
  {Katzgraber}}]{schuetz2022combinatorial}%
  \BibitemOpen
  \bibfield  {author} {\bibinfo {author} {\bibfnamefont {Martin~JA}\
  \bibnamefont {Schuetz}}, \bibinfo {author} {\bibfnamefont {J~Kyle}\
  \bibnamefont {Brubaker}}, \ and\ \bibinfo {author} {\bibfnamefont {Helmut~G}\
  \bibnamefont {Katzgraber}},\ }\bibfield  {title} {\enquote {\bibinfo {title}
  {Combinatorial optimization with physics-inspired graph neural networks},}\
  }\href {https://www.nature.com/articles/s42256-022-00468-6} {\bibfield
  {journal} {\bibinfo  {journal} {Nature Machine Intelligence}\ }\textbf
  {\bibinfo {volume} {4}},\ \bibinfo {pages} {367--377} (\bibinfo {year}
  {2022})}\BibitemShut {NoStop}%
\bibitem [{\citenamefont {Edward~Farhi}(2014)}]{Farhi2014}%
  \BibitemOpen
  \bibfield  {author} {\bibinfo {author} {\bibfnamefont {Sam~Gutmann}\
  \bibnamefont {Edward~Farhi}, \bibfnamefont {Jeffrey~Goldstone}},\ }\bibfield
  {title} {\enquote {\bibinfo {title} {A quantum approximate optimization
  algorithm},}\ }\href@noop {} {\bibfield  {journal} {\bibinfo  {journal}
  {arXiv:1411.4028}\ } (\bibinfo {year} {2014})}\BibitemShut {NoStop}%
\bibitem [{\citenamefont {Schrijver}(1998)}]{schrijver1998theory}%
  \BibitemOpen
  \bibfield  {author} {\bibinfo {author} {\bibfnamefont {Alexander}\
  \bibnamefont {Schrijver}},\ }\href@noop {} {\emph {\bibinfo {title} {Theory
  of linear and integer programming}}}\ (\bibinfo  {publisher} {John Wiley \&
  Sons},\ \bibinfo {year} {1998})\BibitemShut {NoStop}%
\bibitem [{\citenamefont {White}(1992)}]{white1992density}%
  \BibitemOpen
  \bibfield  {author} {\bibinfo {author} {\bibfnamefont {Steven~R}\
  \bibnamefont {White}},\ }\bibfield  {title} {\enquote {\bibinfo {title}
  {Density matrix formulation for quantum renormalization groups},}\
  }\href@noop {} {\bibfield  {journal} {\bibinfo  {journal} {Physical review
  letters}\ }\textbf {\bibinfo {volume} {69}},\ \bibinfo {pages} {2863}
  (\bibinfo {year} {1992})}\BibitemShut {NoStop}%
\bibitem [{\citenamefont {Schollw{\"o}ck}(2011)}]{schollwock2011density}%
  \BibitemOpen
  \bibfield  {author} {\bibinfo {author} {\bibfnamefont {Ulrich}\ \bibnamefont
  {Schollw{\"o}ck}},\ }\bibfield  {title} {\enquote {\bibinfo {title} {The
  density-matrix renormalization group in the age of matrix product states},}\
  }\href@noop {} {\bibfield  {journal} {\bibinfo  {journal} {Annals of
  physics}\ }\textbf {\bibinfo {volume} {326}},\ \bibinfo {pages} {96--192}
  (\bibinfo {year} {2011})}\BibitemShut {NoStop}%
\bibitem [{\citenamefont {Ferris}\ and\ \citenamefont
  {Vidal}(2012)}]{ferris2012perfect}%
  \BibitemOpen
  \bibfield  {author} {\bibinfo {author} {\bibfnamefont {Andrew~J}\
  \bibnamefont {Ferris}}\ and\ \bibinfo {author} {\bibfnamefont {Guifre}\
  \bibnamefont {Vidal}},\ }\bibfield  {title} {\enquote {\bibinfo {title}
  {Perfect sampling with unitary tensor networks},}\ }\href@noop {} {\bibfield
  {journal} {\bibinfo  {journal} {Physical Review B}\ }\textbf {\bibinfo
  {volume} {85}},\ \bibinfo {pages} {165146} (\bibinfo {year}
  {2012})}\BibitemShut {NoStop}%
\bibitem [{Note1()}]{Note1}%
  \BibitemOpen
  \bibinfo {note} {In particular, a large class of optimization problems
  contain inequality type constraints and even nonlinear
  constraints}\BibitemShut {NoStop}%
\bibitem [{\citenamefont {Glover}\ \emph {et~al.}(2018)\citenamefont {Glover},
  \citenamefont {Kochenberger},\ and\ \citenamefont {Du}}]{glover2018tutorial}%
  \BibitemOpen
  \bibfield  {author} {\bibinfo {author} {\bibfnamefont {Fred}\ \bibnamefont
  {Glover}}, \bibinfo {author} {\bibfnamefont {Gary}\ \bibnamefont
  {Kochenberger}}, \ and\ \bibinfo {author} {\bibfnamefont {Yu}~\bibnamefont
  {Du}},\ }\bibfield  {title} {\enquote {\bibinfo {title} {A tutorial on
  formulating and using qubo models},}\ }\href@noop {} {\bibfield  {journal}
  {\bibinfo  {journal} {arXiv preprint arXiv:1811.11538}\ } (\bibinfo {year}
  {2018})}\BibitemShut {NoStop}%
\bibitem [{\citenamefont {Hibat-Allah}\ \emph {et~al.}(2021)\citenamefont
  {Hibat-Allah}, \citenamefont {Inack}, \citenamefont {Wiersema}, \citenamefont
  {Melko},\ and\ \citenamefont {Carrasquilla}}]{HibatAllah2022}%
  \BibitemOpen
  \bibfield  {author} {\bibinfo {author} {\bibfnamefont {Mohamed}\ \bibnamefont
  {Hibat-Allah}}, \bibinfo {author} {\bibfnamefont {Estelle~M.}\ \bibnamefont
  {Inack}}, \bibinfo {author} {\bibfnamefont {Roeland}\ \bibnamefont
  {Wiersema}}, \bibinfo {author} {\bibfnamefont {Roger~G.}\ \bibnamefont
  {Melko}}, \ and\ \bibinfo {author} {\bibfnamefont {Juan}\ \bibnamefont
  {Carrasquilla}},\ }\bibfield  {title} {\enquote {\bibinfo {title}
  {Variational neural annealing},}\ }\href {\doibase
  10.1038/s42256-021-00401-3} {\bibfield  {journal} {\bibinfo  {journal}
  {Nature Machine Intelligence}\ }\textbf {\bibinfo {volume} {3}},\ \bibinfo
  {pages} {952--961} (\bibinfo {year} {2021})}\BibitemShut {NoStop}%
\bibitem [{\citenamefont {Bengio}\ \emph {et~al.}(2021)\citenamefont {Bengio},
  \citenamefont {Jain}, \citenamefont {Korablyov}, \citenamefont {Precup},\
  and\ \citenamefont {Bengio}}]{Bengio2021}%
  \BibitemOpen
  \bibfield  {author} {\bibinfo {author} {\bibfnamefont {Emmanuel}\
  \bibnamefont {Bengio}}, \bibinfo {author} {\bibfnamefont {Moksh}\
  \bibnamefont {Jain}}, \bibinfo {author} {\bibfnamefont {Maksym}\ \bibnamefont
  {Korablyov}}, \bibinfo {author} {\bibfnamefont {Doina}\ \bibnamefont
  {Precup}}, \ and\ \bibinfo {author} {\bibfnamefont {Yoshua}\ \bibnamefont
  {Bengio}},\ }\bibfield  {title} {\enquote {\bibinfo {title} {Flow network
  based generative models for non-iterative diverse candidate generation},}\
  }\href {https://arxiv.org/abs/2106.04399} {\bibfield  {journal} {\bibinfo
  {journal} {arXiv:2106.04399}\ } (\bibinfo {year} {2021})}\BibitemShut
  {NoStop}%
\bibitem [{\citenamefont {Hastings}(2006)}]{hastings2006solving}%
  \BibitemOpen
  \bibfield  {author} {\bibinfo {author} {\bibfnamefont {Matthew~B}\
  \bibnamefont {Hastings}},\ }\bibfield  {title} {\enquote {\bibinfo {title}
  {Solving gapped hamiltonians locally},}\ }\href@noop {} {\bibfield  {journal}
  {\bibinfo  {journal} {Physical review b}\ }\textbf {\bibinfo {volume} {73}},\
  \bibinfo {pages} {085115} (\bibinfo {year} {2006})}\BibitemShut {NoStop}%
\bibitem [{\citenamefont {Wolf}\ \emph {et~al.}(2008)\citenamefont {Wolf},
  \citenamefont {Verstraete}, \citenamefont {Hastings},\ and\ \citenamefont
  {Cirac}}]{wolf2008area}%
  \BibitemOpen
  \bibfield  {author} {\bibinfo {author} {\bibfnamefont {Michael~M}\
  \bibnamefont {Wolf}}, \bibinfo {author} {\bibfnamefont {Frank}\ \bibnamefont
  {Verstraete}}, \bibinfo {author} {\bibfnamefont {Matthew~B}\ \bibnamefont
  {Hastings}}, \ and\ \bibinfo {author} {\bibfnamefont {J~Ignacio}\
  \bibnamefont {Cirac}},\ }\bibfield  {title} {\enquote {\bibinfo {title} {Area
  laws in quantum systems: mutual information and correlations},}\ }\href@noop
  {} {\bibfield  {journal} {\bibinfo  {journal} {Physical review letters}\
  }\textbf {\bibinfo {volume} {100}},\ \bibinfo {pages} {070502} (\bibinfo
  {year} {2008})}\BibitemShut {NoStop}%
\bibitem [{\citenamefont {Or{\'u}s}(2014)}]{orus2014practical}%
  \BibitemOpen
  \bibfield  {author} {\bibinfo {author} {\bibfnamefont {Rom{\'a}n}\
  \bibnamefont {Or{\'u}s}},\ }\bibfield  {title} {\enquote {\bibinfo {title} {A
  practical introduction to tensor networks: Matrix product states and
  projected entangled pair states},}\ }\href@noop {} {\bibfield  {journal}
  {\bibinfo  {journal} {Annals of physics}\ }\textbf {\bibinfo {volume}
  {349}},\ \bibinfo {pages} {117--158} (\bibinfo {year} {2014})}\BibitemShut
  {NoStop}%
\bibitem [{\citenamefont {Singh}\ \emph
  {et~al.}(2010{\natexlab{a}})\citenamefont {Singh}, \citenamefont {Pfeifer},\
  and\ \citenamefont {Vidal}}]{singh2010tensor}%
  \BibitemOpen
  \bibfield  {author} {\bibinfo {author} {\bibfnamefont {Sukhwinder}\
  \bibnamefont {Singh}}, \bibinfo {author} {\bibfnamefont {Robert~NC}\
  \bibnamefont {Pfeifer}}, \ and\ \bibinfo {author} {\bibfnamefont
  {Guifr{\'e}}\ \bibnamefont {Vidal}},\ }\bibfield  {title} {\enquote {\bibinfo
  {title} {Tensor network decompositions in the presence of a global
  symmetry},}\ }\href@noop {} {\bibfield  {journal} {\bibinfo  {journal}
  {Physical Review A}\ }\textbf {\bibinfo {volume} {82}},\ \bibinfo {pages}
  {050301} (\bibinfo {year} {2010}{\natexlab{a}})}\BibitemShut {NoStop}%
\bibitem [{\citenamefont {Singh}\ \emph {et~al.}(2011)\citenamefont {Singh},
  \citenamefont {Pfeifer},\ and\ \citenamefont {Vidal}}]{singh2011tensor}%
  \BibitemOpen
  \bibfield  {author} {\bibinfo {author} {\bibfnamefont {Sukhwinder}\
  \bibnamefont {Singh}}, \bibinfo {author} {\bibfnamefont {Robert~NC}\
  \bibnamefont {Pfeifer}}, \ and\ \bibinfo {author} {\bibfnamefont {Guifre}\
  \bibnamefont {Vidal}},\ }\bibfield  {title} {\enquote {\bibinfo {title}
  {Tensor network states and algorithms in the presence of a global u (1)
  symmetry},}\ }\href@noop {} {\bibfield  {journal} {\bibinfo  {journal}
  {Physical Review B}\ }\textbf {\bibinfo {volume} {83}},\ \bibinfo {pages}
  {115125} (\bibinfo {year} {2011})}\BibitemShut {NoStop}%
\bibitem [{\citenamefont {Kleinberg}\ and\ \citenamefont
  {Tardos}(2006)}]{kleinberg2006algorithm}%
  \BibitemOpen
  \bibfield  {author} {\bibinfo {author} {\bibfnamefont {Jon}\ \bibnamefont
  {Kleinberg}}\ and\ \bibinfo {author} {\bibfnamefont {Eva}\ \bibnamefont
  {Tardos}},\ }\href@noop {} {\emph {\bibinfo {title} {Algorithm design}}}\
  (\bibinfo  {publisher} {Pearson Education India},\ \bibinfo {year}
  {2006})\BibitemShut {NoStop}%
\bibitem [{\citenamefont {Horowitz}\ and\ \citenamefont
  {Sahni}(1974)}]{horowitz1974computing}%
  \BibitemOpen
  \bibfield  {author} {\bibinfo {author} {\bibfnamefont {Ellis}\ \bibnamefont
  {Horowitz}}\ and\ \bibinfo {author} {\bibfnamefont {Sartaj}\ \bibnamefont
  {Sahni}},\ }\bibfield  {title} {\enquote {\bibinfo {title} {Computing
  partitions with applications to the knapsack problem},}\ }\href@noop {}
  {\bibfield  {journal} {\bibinfo  {journal} {Journal of the ACM (JACM)}\
  }\textbf {\bibinfo {volume} {21}},\ \bibinfo {pages} {277--292} (\bibinfo
  {year} {1974})}\BibitemShut {NoStop}%
\bibitem [{\citenamefont {Bellman}(1957)}]{bellman1957dynamic}%
  \BibitemOpen
  \bibfield  {author} {\bibinfo {author} {\bibfnamefont {RJNJ}\ \bibnamefont
  {Bellman}},\ }\bibfield  {title} {\enquote {\bibinfo {title} {Dynamic
  programming princeton university press princeton},}\ }\href@noop {}
  {\bibfield  {journal} {\bibinfo  {journal} {New Jersey Google Scholar}\ }
  (\bibinfo {year} {1957})}\BibitemShut {NoStop}%
\bibitem [{\citenamefont {Fishman}\ \emph {et~al.}(2020)\citenamefont
  {Fishman}, \citenamefont {White},\ and\ \citenamefont
  {Stoudenmire}}]{ITensor_julia}%
  \BibitemOpen
  \bibfield  {author} {\bibinfo {author} {\bibfnamefont {Matthew}\ \bibnamefont
  {Fishman}}, \bibinfo {author} {\bibfnamefont {Steven~R.}\ \bibnamefont
  {White}}, \ and\ \bibinfo {author} {\bibfnamefont {E.~Miles}\ \bibnamefont
  {Stoudenmire}},\ }\href {\doibase 10.48550/arxiv.2007.14822} {\enquote
  {\bibinfo {title} {The itensor software library for tensor network
  calculations},}\ } (\bibinfo {year} {2020})\BibitemShut {NoStop}%
\bibitem [{\citenamefont {Sachdev}(1999)}]{sachdev1999quantum}%
  \BibitemOpen
  \bibfield  {author} {\bibinfo {author} {\bibfnamefont {Subir}\ \bibnamefont
  {Sachdev}},\ }\bibfield  {title} {\enquote {\bibinfo {title} {Quantum phase
  transitions},}\ }\href@noop {} {\bibfield  {journal} {\bibinfo  {journal}
  {Physics world}\ }\textbf {\bibinfo {volume} {12}},\ \bibinfo {pages} {33}
  (\bibinfo {year} {1999})}\BibitemShut {NoStop}%
\bibitem [{\citenamefont {Gili}\ \emph {et~al.}(2022)\citenamefont {Gili},
  \citenamefont {Mauri},\ and\ \citenamefont
  {Perdomo-Ortiz}}]{gili2022evaluating}%
  \BibitemOpen
  \bibfield  {author} {\bibinfo {author} {\bibfnamefont {Kaitlin}\ \bibnamefont
  {Gili}}, \bibinfo {author} {\bibfnamefont {Marta}\ \bibnamefont {Mauri}}, \
  and\ \bibinfo {author} {\bibfnamefont {Alejandro}\ \bibnamefont
  {Perdomo-Ortiz}},\ }\bibfield  {title} {\enquote {\bibinfo {title}
  {Evaluating generalization in classical and quantum generative models},}\
  }\href {https://arxiv.org/abs/2201.08770} {\bibfield  {journal} {\bibinfo
  {journal} {arXiv:2201.08770}\ } (\bibinfo {year} {2022})}\BibitemShut
  {NoStop}%
\bibitem [{\citenamefont {Vidal}(2007)}]{vidal2007entanglement}%
  \BibitemOpen
  \bibfield  {author} {\bibinfo {author} {\bibfnamefont {Guifre}\ \bibnamefont
  {Vidal}},\ }\bibfield  {title} {\enquote {\bibinfo {title} {Entanglement
  renormalization},}\ }\href@noop {} {\bibfield  {journal} {\bibinfo  {journal}
  {Physical review letters}\ }\textbf {\bibinfo {volume} {99}},\ \bibinfo
  {pages} {220405} (\bibinfo {year} {2007})}\BibitemShut {NoStop}%
\bibitem [{\citenamefont {Hastie}\ \emph {et~al.}(2009)\citenamefont {Hastie},
  \citenamefont {Tibshirani}, \citenamefont {Friedman},\ and\ \citenamefont
  {Friedman}}]{hastie2009elements}%
  \BibitemOpen
  \bibfield  {author} {\bibinfo {author} {\bibfnamefont {Trevor}\ \bibnamefont
  {Hastie}}, \bibinfo {author} {\bibfnamefont {Robert}\ \bibnamefont
  {Tibshirani}}, \bibinfo {author} {\bibfnamefont {Jerome~H}\ \bibnamefont
  {Friedman}}, \ and\ \bibinfo {author} {\bibfnamefont {Jerome~H}\ \bibnamefont
  {Friedman}},\ }\href@noop {} {\emph {\bibinfo {title} {The elements of
  statistical learning: data mining, inference, and prediction}}},\
  Vol.~\bibinfo {volume} {2}\ (\bibinfo  {publisher} {Springer},\ \bibinfo
  {year} {2009})\BibitemShut {NoStop}%
\bibitem [{\citenamefont {Strashko}\ and\ \citenamefont
  {Stoudenmire}(2022)}]{strashko2022generalization}%
  \BibitemOpen
  \bibfield  {author} {\bibinfo {author} {\bibfnamefont {Artem}\ \bibnamefont
  {Strashko}}\ and\ \bibinfo {author} {\bibfnamefont {E~Miles}\ \bibnamefont
  {Stoudenmire}},\ }\bibfield  {title} {\enquote {\bibinfo {title}
  {Generalization and overfitting in matrix product state machine learning
  architectures},}\ }\href@noop {} {\bibfield  {journal} {\bibinfo  {journal}
  {arXiv preprint arXiv:2208.04372}\ } (\bibinfo {year} {2022})}\BibitemShut
  {NoStop}%
\bibitem [{\citenamefont {Garey}\ and\ \citenamefont
  {Johnson}(1979)}]{garey79}%
  \BibitemOpen
  \bibfield  {author} {\bibinfo {author} {\bibfnamefont {M.R.}\ \bibnamefont
  {Garey}}\ and\ \bibinfo {author} {\bibfnamefont {D.S.}\ \bibnamefont
  {Johnson}},\ }\href@noop {} {\emph {\bibinfo {title} {Computers and
  Intractability. A Guide to the Theory of NP-Completeness}}}\ (\bibinfo
  {publisher} {W.H. Freeman and Co., NY},\ \bibinfo {year} {1979})\BibitemShut
  {NoStop}%
\bibitem [{\citenamefont {Kellerer}\ \emph {et~al.}(2000)\citenamefont
  {Kellerer}, \citenamefont {Mansini},\ and\ \citenamefont
  {Speranza}}]{kellerer2000selecting}%
  \BibitemOpen
  \bibfield  {author} {\bibinfo {author} {\bibfnamefont {Hans}\ \bibnamefont
  {Kellerer}}, \bibinfo {author} {\bibfnamefont {Renata}\ \bibnamefont
  {Mansini}}, \ and\ \bibinfo {author} {\bibfnamefont {M~Grazia}\ \bibnamefont
  {Speranza}},\ }\bibfield  {title} {\enquote {\bibinfo {title} {Selecting
  portfolios with fixed costs and minimum transaction lots},}\ }\href@noop {}
  {\bibfield  {journal} {\bibinfo  {journal} {Annals of Operations Research}\
  }\textbf {\bibinfo {volume} {99}},\ \bibinfo {pages} {287--304} (\bibinfo
  {year} {2000})}\BibitemShut {NoStop}%
\bibitem [{\citenamefont {Deng}\ \emph {et~al.}(2017)\citenamefont {Deng},
  \citenamefont {Li},\ and\ \citenamefont {Sarma}}]{Deng_2017}%
  \BibitemOpen
  \bibfield  {author} {\bibinfo {author} {\bibfnamefont {Dong-Ling}\
  \bibnamefont {Deng}}, \bibinfo {author} {\bibfnamefont {Xiaopeng}\
  \bibnamefont {Li}}, \ and\ \bibinfo {author} {\bibfnamefont {S.~Das}\
  \bibnamefont {Sarma}},\ }\bibfield  {title} {\enquote {\bibinfo {title}
  {Machine learning topological states},}\ }\href {\doibase
  10.1103/physrevb.96.195145} {\bibfield  {journal} {\bibinfo  {journal}
  {Physical Review B}\ }\textbf {\bibinfo {volume} {96}} (\bibinfo {year}
  {2017}),\ 10.1103/physrevb.96.195145}\BibitemShut {NoStop}%
\bibitem [{\citenamefont {Huang}\ \emph {et~al.}(2021)\citenamefont {Huang},
  \citenamefont {Moore} \emph {et~al.}}]{huang2021neural}%
  \BibitemOpen
  \bibfield  {author} {\bibinfo {author} {\bibfnamefont {Yichen}\ \bibnamefont
  {Huang}}, \bibinfo {author} {\bibfnamefont {Joel~E}\ \bibnamefont {Moore}},
  \emph {et~al.},\ }\bibfield  {title} {\enquote {\bibinfo {title} {Neural
  network representation of tensor network and chiral states},}\ }\href@noop {}
  {\bibfield  {journal} {\bibinfo  {journal} {Physical Review Letters}\
  }\textbf {\bibinfo {volume} {127}},\ \bibinfo {pages} {170601} (\bibinfo
  {year} {2021})}\BibitemShut {NoStop}%
\bibitem [{\citenamefont {Choo}\ \emph {et~al.}(2018)\citenamefont {Choo},
  \citenamefont {Carleo}, \citenamefont {Regnault},\ and\ \citenamefont
  {Neupert}}]{choo2018symmetries}%
  \BibitemOpen
  \bibfield  {author} {\bibinfo {author} {\bibfnamefont {Kenny}\ \bibnamefont
  {Choo}}, \bibinfo {author} {\bibfnamefont {Giuseppe}\ \bibnamefont {Carleo}},
  \bibinfo {author} {\bibfnamefont {Nicolas}\ \bibnamefont {Regnault}}, \ and\
  \bibinfo {author} {\bibfnamefont {Titus}\ \bibnamefont {Neupert}},\
  }\bibfield  {title} {\enquote {\bibinfo {title} {Symmetries and many-body
  excitations with neural-network quantum states},}\ }\href@noop {} {\bibfield
  {journal} {\bibinfo  {journal} {Physical review letters}\ }\textbf {\bibinfo
  {volume} {121}},\ \bibinfo {pages} {167204} (\bibinfo {year}
  {2018})}\BibitemShut {NoStop}%
\bibitem [{\citenamefont {Dawid}\ \emph {et~al.}(2022)\citenamefont {Dawid},
  \citenamefont {Arnold}, \citenamefont {Requena}, \citenamefont {Gresch},
  \citenamefont {P{\l}odzie{\'n}}, \citenamefont {Donatella}, \citenamefont
  {Nicoli}, \citenamefont {Stornati}, \citenamefont {Koch}, \citenamefont
  {B{\"u}ttner} \emph {et~al.}}]{dawid2022modern}%
  \BibitemOpen
  \bibfield  {author} {\bibinfo {author} {\bibfnamefont {Anna}\ \bibnamefont
  {Dawid}}, \bibinfo {author} {\bibfnamefont {Julian}\ \bibnamefont {Arnold}},
  \bibinfo {author} {\bibfnamefont {Borja}\ \bibnamefont {Requena}}, \bibinfo
  {author} {\bibfnamefont {Alexander}\ \bibnamefont {Gresch}}, \bibinfo
  {author} {\bibfnamefont {Marcin}\ \bibnamefont {P{\l}odzie{\'n}}}, \bibinfo
  {author} {\bibfnamefont {Kaelan}\ \bibnamefont {Donatella}}, \bibinfo
  {author} {\bibfnamefont {Kim~A}\ \bibnamefont {Nicoli}}, \bibinfo {author}
  {\bibfnamefont {Paolo}\ \bibnamefont {Stornati}}, \bibinfo {author}
  {\bibfnamefont {Rouven}\ \bibnamefont {Koch}}, \bibinfo {author}
  {\bibfnamefont {Miriam}\ \bibnamefont {B{\"u}ttner}},  \emph {et~al.},\
  }\bibfield  {title} {\enquote {\bibinfo {title} {Modern applications of
  machine learning in quantum sciences},}\ }\href@noop {} {\bibfield  {journal}
  {\bibinfo  {journal} {arXiv preprint arXiv:2204.04198}\ } (\bibinfo {year}
  {2022})}\BibitemShut {NoStop}%
\bibitem [{\citenamefont {Shalev-Shwartz}\ \emph {et~al.}(2017)\citenamefont
  {Shalev-Shwartz}, \citenamefont {Shamir},\ and\ \citenamefont
  {Shammah}}]{shalev2017failures}%
  \BibitemOpen
  \bibfield  {author} {\bibinfo {author} {\bibfnamefont {Shai}\ \bibnamefont
  {Shalev-Shwartz}}, \bibinfo {author} {\bibfnamefont {Ohad}\ \bibnamefont
  {Shamir}}, \ and\ \bibinfo {author} {\bibfnamefont {Shaked}\ \bibnamefont
  {Shammah}},\ }\bibfield  {title} {\enquote {\bibinfo {title} {Failures of
  gradient-based deep learning},}\ }in\ \href@noop {} {\emph {\bibinfo
  {booktitle} {International Conference on Machine Learning}}}\ (\bibinfo
  {organization} {PMLR},\ \bibinfo {year} {2017})\ pp.\ \bibinfo {pages}
  {3067--3075}\BibitemShut {NoStop}%
\bibitem [{\citenamefont {Romero}\ \emph {et~al.}(2019)\citenamefont {Romero},
  \citenamefont {Mazzanti}, \citenamefont {Delgado},\ and\ \citenamefont
  {Buchaca}}]{romero2019weighted}%
  \BibitemOpen
  \bibfield  {author} {\bibinfo {author} {\bibfnamefont {Enrique}\ \bibnamefont
  {Romero}}, \bibinfo {author} {\bibfnamefont {Ferran}\ \bibnamefont
  {Mazzanti}}, \bibinfo {author} {\bibfnamefont {Jordi}\ \bibnamefont
  {Delgado}}, \ and\ \bibinfo {author} {\bibfnamefont {David}\ \bibnamefont
  {Buchaca}},\ }\bibfield  {title} {\enquote {\bibinfo {title} {Weighted
  contrastive divergence},}\ }\href@noop {} {\bibfield  {journal} {\bibinfo
  {journal} {Neural Networks}\ }\textbf {\bibinfo {volume} {114}},\ \bibinfo
  {pages} {147--156} (\bibinfo {year} {2019})}\BibitemShut {NoStop}%
\bibitem [{\citenamefont {Mont{\'u}far}\ \emph {et~al.}(2011)\citenamefont
  {Mont{\'u}far}, \citenamefont {Rauh},\ and\ \citenamefont
  {Ay}}]{montufar2011expressive}%
  \BibitemOpen
  \bibfield  {author} {\bibinfo {author} {\bibfnamefont {Guido~F}\ \bibnamefont
  {Mont{\'u}far}}, \bibinfo {author} {\bibfnamefont {Johannes}\ \bibnamefont
  {Rauh}}, \ and\ \bibinfo {author} {\bibfnamefont {Nihat}\ \bibnamefont
  {Ay}},\ }\bibfield  {title} {\enquote {\bibinfo {title} {Expressive power and
  approximation errors of restricted boltzmann machines},}\ }\href@noop {}
  {\bibfield  {journal} {\bibinfo  {journal} {Advances in neural information
  processing systems}\ }\textbf {\bibinfo {volume} {24}} (\bibinfo {year}
  {2011})}\BibitemShut {NoStop}%
\bibitem [{\citenamefont {Stokes}\ and\ \citenamefont
  {Terilla}(2019)}]{stokes2019probabilistic}%
  \BibitemOpen
  \bibfield  {author} {\bibinfo {author} {\bibfnamefont {James}\ \bibnamefont
  {Stokes}}\ and\ \bibinfo {author} {\bibfnamefont {John}\ \bibnamefont
  {Terilla}},\ }\bibfield  {title} {\enquote {\bibinfo {title} {Probabilistic
  modeling with matrix product states},}\ }\href
  {https://www.ncbi.nlm.nih.gov/pmc/articles/PMC7514580/} {\bibfield  {journal}
  {\bibinfo  {journal} {Entropy}\ }\textbf {\bibinfo {volume} {21}},\ \bibinfo
  {pages} {1236} (\bibinfo {year} {2019})}\BibitemShut {NoStop}%
\bibitem [{\citenamefont {Bradley}\ \emph {et~al.}(2020)\citenamefont
  {Bradley}, \citenamefont {Stoudenmire},\ and\ \citenamefont
  {Terilla}}]{bradley2020modeling}%
  \BibitemOpen
  \bibfield  {author} {\bibinfo {author} {\bibfnamefont {Tai-Danae}\
  \bibnamefont {Bradley}}, \bibinfo {author} {\bibfnamefont {E~Miles}\
  \bibnamefont {Stoudenmire}}, \ and\ \bibinfo {author} {\bibfnamefont {John}\
  \bibnamefont {Terilla}},\ }\bibfield  {title} {\enquote {\bibinfo {title}
  {Modeling sequences with quantum states: a look under the hood},}\ }\href
  {https://iopscience.iop.org/article/10.1088/2632-2153/ab8731} {\bibfield
  {journal} {\bibinfo  {journal} {Machine Learning: Science and Technology}\
  }\textbf {\bibinfo {volume} {1}},\ \bibinfo {pages} {035008} (\bibinfo {year}
  {2020})}\BibitemShut {NoStop}%
\bibitem [{\citenamefont {McCulloch}(2007)}]{mcculloch2007density}%
  \BibitemOpen
  \bibfield  {author} {\bibinfo {author} {\bibfnamefont {Ian~P}\ \bibnamefont
  {McCulloch}},\ }\bibfield  {title} {\enquote {\bibinfo {title} {From
  density-matrix renormalization group to matrix product states},}\ }\href@noop
  {} {\bibfield  {journal} {\bibinfo  {journal} {Journal of Statistical
  Mechanics: Theory and Experiment}\ }\textbf {\bibinfo {volume} {2007}},\
  \bibinfo {pages} {P10014} (\bibinfo {year} {2007})}\BibitemShut {NoStop}%
\bibitem [{\citenamefont {Singh}\ and\ \citenamefont
  {Vidal}(2012)}]{singh2012tensor}%
  \BibitemOpen
  \bibfield  {author} {\bibinfo {author} {\bibfnamefont {Sukhwinder}\
  \bibnamefont {Singh}}\ and\ \bibinfo {author} {\bibfnamefont {Guifre}\
  \bibnamefont {Vidal}},\ }\bibfield  {title} {\enquote {\bibinfo {title}
  {Tensor network states and algorithms in the presence of a global su (2)
  symmetry},}\ }\href@noop {} {\bibfield  {journal} {\bibinfo  {journal}
  {Physical Review B}\ }\textbf {\bibinfo {volume} {86}},\ \bibinfo {pages}
  {195114} (\bibinfo {year} {2012})}\BibitemShut {NoStop}%
\bibitem [{\citenamefont {Singh}\ \emph
  {et~al.}(2010{\natexlab{b}})\citenamefont {Singh}, \citenamefont {Zhou},\
  and\ \citenamefont {Vidal}}]{singh2010simulation}%
  \BibitemOpen
  \bibfield  {author} {\bibinfo {author} {\bibfnamefont {Sukhwinder}\
  \bibnamefont {Singh}}, \bibinfo {author} {\bibfnamefont {Huan-Qiang}\
  \bibnamefont {Zhou}}, \ and\ \bibinfo {author} {\bibfnamefont {Guifre}\
  \bibnamefont {Vidal}},\ }\bibfield  {title} {\enquote {\bibinfo {title}
  {Simulation of one-dimensional quantum systems with a global su (2)
  symmetry},}\ }\href@noop {} {\bibfield  {journal} {\bibinfo  {journal} {New
  Journal of Physics}\ }\textbf {\bibinfo {volume} {12}},\ \bibinfo {pages}
  {033029} (\bibinfo {year} {2010}{\natexlab{b}})}\BibitemShut {NoStop}%
\bibitem [{\citenamefont {Gasse}\ \emph {et~al.}(2019)\citenamefont {Gasse},
  \citenamefont {Ch{\'e}telat}, \citenamefont {Ferroni}, \citenamefont
  {Charlin},\ and\ \citenamefont {Lodi}}]{gasse2019exact}%
  \BibitemOpen
  \bibfield  {author} {\bibinfo {author} {\bibfnamefont {Maxime}\ \bibnamefont
  {Gasse}}, \bibinfo {author} {\bibfnamefont {Didier}\ \bibnamefont
  {Ch{\'e}telat}}, \bibinfo {author} {\bibfnamefont {Nicola}\ \bibnamefont
  {Ferroni}}, \bibinfo {author} {\bibfnamefont {Laurent}\ \bibnamefont
  {Charlin}}, \ and\ \bibinfo {author} {\bibfnamefont {Andrea}\ \bibnamefont
  {Lodi}},\ }\bibfield  {title} {\enquote {\bibinfo {title} {Exact
  combinatorial optimization with graph convolutional neural networks},}\
  }\href@noop {} {\bibfield  {journal} {\bibinfo  {journal} {Advances in Neural
  Information Processing Systems}\ }\textbf {\bibinfo {volume} {32}} (\bibinfo
  {year} {2019})}\BibitemShut {NoStop}%
\bibitem [{\citenamefont {Wang}\ \emph {et~al.}(2018)\citenamefont {Wang},
  \citenamefont {Hadfield}, \citenamefont {Jiang},\ and\ \citenamefont
  {Rieffel}}]{wang2018quantum}%
  \BibitemOpen
  \bibfield  {author} {\bibinfo {author} {\bibfnamefont {Zhihui}\ \bibnamefont
  {Wang}}, \bibinfo {author} {\bibfnamefont {Stuart}\ \bibnamefont {Hadfield}},
  \bibinfo {author} {\bibfnamefont {Zhang}\ \bibnamefont {Jiang}}, \ and\
  \bibinfo {author} {\bibfnamefont {Eleanor~G}\ \bibnamefont {Rieffel}},\
  }\bibfield  {title} {\enquote {\bibinfo {title} {Quantum approximate
  optimization algorithm for maxcut: A fermionic view},}\ }\href@noop {}
  {\bibfield  {journal} {\bibinfo  {journal} {Physical Review A}\ }\textbf
  {\bibinfo {volume} {97}},\ \bibinfo {pages} {022304} (\bibinfo {year}
  {2018})}\BibitemShut {NoStop}%
\bibitem [{\citenamefont {Niroula}\ \emph {et~al.}(2022)\citenamefont
  {Niroula}, \citenamefont {Shaydulin}, \citenamefont {Yalovetzky},
  \citenamefont {Minssen}, \citenamefont {Herman}, \citenamefont {Hu},\ and\
  \citenamefont {Pistoia}}]{niroula2022constrained}%
  \BibitemOpen
  \bibfield  {author} {\bibinfo {author} {\bibfnamefont {Pradeep}\ \bibnamefont
  {Niroula}}, \bibinfo {author} {\bibfnamefont {Ruslan}\ \bibnamefont
  {Shaydulin}}, \bibinfo {author} {\bibfnamefont {Romina}\ \bibnamefont
  {Yalovetzky}}, \bibinfo {author} {\bibfnamefont {Pierre}\ \bibnamefont
  {Minssen}}, \bibinfo {author} {\bibfnamefont {Dylan}\ \bibnamefont {Herman}},
  \bibinfo {author} {\bibfnamefont {Shaohan}\ \bibnamefont {Hu}}, \ and\
  \bibinfo {author} {\bibfnamefont {Marco}\ \bibnamefont {Pistoia}},\
  }\bibfield  {title} {\enquote {\bibinfo {title} {Constrained quantum
  optimization for extractive summarization on a trapped-ion quantum
  computer},}\ }\href@noop {} {\bibfield  {journal} {\bibinfo  {journal}
  {Scientific Reports}\ }\textbf {\bibinfo {volume} {12}},\ \bibinfo {pages}
  {1--14} (\bibinfo {year} {2022})}\BibitemShut {NoStop}%
\bibitem [{\citenamefont {Kondor}\ and\ \citenamefont
  {Trivedi}(2018)}]{kondor2018generalization}%
  \BibitemOpen
  \bibfield  {author} {\bibinfo {author} {\bibfnamefont {Risi}\ \bibnamefont
  {Kondor}}\ and\ \bibinfo {author} {\bibfnamefont {Shubhendu}\ \bibnamefont
  {Trivedi}},\ }\bibfield  {title} {\enquote {\bibinfo {title} {On the
  generalization of equivariance and convolution in neural networks to the
  action of compact groups},}\ }in\ \href@noop {} {\emph {\bibinfo {booktitle}
  {International Conference on Machine Learning}}}\ (\bibinfo {organization}
  {PMLR},\ \bibinfo {year} {2018})\ pp.\ \bibinfo {pages}
  {2747--2755}\BibitemShut {NoStop}%
\bibitem [{\citenamefont {Bronstein}\ \emph {et~al.}(2021)\citenamefont
  {Bronstein}, \citenamefont {Bruna}, \citenamefont {Cohen},\ and\
  \citenamefont {Veli{\v{c}}kovi{\'c}}}]{bronstein2021geometric}%
  \BibitemOpen
  \bibfield  {author} {\bibinfo {author} {\bibfnamefont {Michael~M}\
  \bibnamefont {Bronstein}}, \bibinfo {author} {\bibfnamefont {Joan}\
  \bibnamefont {Bruna}}, \bibinfo {author} {\bibfnamefont {Taco}\ \bibnamefont
  {Cohen}}, \ and\ \bibinfo {author} {\bibfnamefont {Petar}\ \bibnamefont
  {Veli{\v{c}}kovi{\'c}}},\ }\bibfield  {title} {\enquote {\bibinfo {title}
  {Geometric deep learning: Grids, groups, graphs, geodesics, and gauges},}\
  }\href@noop {} {\bibfield  {journal} {\bibinfo  {journal} {arXiv preprint
  arXiv:2104.13478}\ } (\bibinfo {year} {2021})}\BibitemShut {NoStop}%
\bibitem [{\citenamefont {Meyer}\ \emph {et~al.}(2022)\citenamefont {Meyer},
  \citenamefont {Mularski}, \citenamefont {Gil-Fuster}, \citenamefont {Mele},
  \citenamefont {Arzani}, \citenamefont {Wilms},\ and\ \citenamefont
  {Eisert}}]{meyer2022exploiting}%
  \BibitemOpen
  \bibfield  {author} {\bibinfo {author} {\bibfnamefont {Johannes~Jakob}\
  \bibnamefont {Meyer}}, \bibinfo {author} {\bibfnamefont {Marian}\
  \bibnamefont {Mularski}}, \bibinfo {author} {\bibfnamefont {Elies}\
  \bibnamefont {Gil-Fuster}}, \bibinfo {author} {\bibfnamefont {Antonio~Anna}\
  \bibnamefont {Mele}}, \bibinfo {author} {\bibfnamefont {Francesco}\
  \bibnamefont {Arzani}}, \bibinfo {author} {\bibfnamefont {Alissa}\
  \bibnamefont {Wilms}}, \ and\ \bibinfo {author} {\bibfnamefont {Jens}\
  \bibnamefont {Eisert}},\ }\bibfield  {title} {\enquote {\bibinfo {title}
  {Exploiting symmetry in variational quantum machine learning},}\ }\href@noop
  {} {\bibfield  {journal} {\bibinfo  {journal} {arXiv preprint
  arXiv:2205.06217}\ } (\bibinfo {year} {2022})}\BibitemShut {NoStop}%
\bibitem [{\citenamefont {Larocca}\ \emph {et~al.}(2022)\citenamefont
  {Larocca}, \citenamefont {Sauvage}, \citenamefont {Sbahi}, \citenamefont
  {Verdon}, \citenamefont {Coles},\ and\ \citenamefont
  {Cerezo}}]{Larocca_2022}%
  \BibitemOpen
  \bibfield  {author} {\bibinfo {author} {\bibfnamefont {Mart{\'i}n}\
  \bibnamefont {Larocca}}, \bibinfo {author} {\bibfnamefont
  {Fr{\'{e}}d{\'{e}}ric}\ \bibnamefont {Sauvage}}, \bibinfo {author}
  {\bibfnamefont {Faris~M.}\ \bibnamefont {Sbahi}}, \bibinfo {author}
  {\bibfnamefont {Guillaume}\ \bibnamefont {Verdon}}, \bibinfo {author}
  {\bibfnamefont {Patrick~J.}\ \bibnamefont {Coles}}, \ and\ \bibinfo {author}
  {\bibfnamefont {M.}~\bibnamefont {Cerezo}},\ }\bibfield  {title} {\enquote
  {\bibinfo {title} {Group-invariant quantum machine learning},}\ }\href
  {\doibase 10.1103/prxquantum.3.030341} {\bibfield  {journal} {\bibinfo
  {journal} {{PRX} Quantum}\ }\textbf {\bibinfo {volume} {3}} (\bibinfo {year}
  {2022}),\ 10.1103/prxquantum.3.030341}\BibitemShut {NoStop}%
\bibitem [{\citenamefont {Nguyen}\ \emph {et~al.}(2022)\citenamefont {Nguyen},
  \citenamefont {Schatzki}, \citenamefont {Braccia}, \citenamefont {Ragone},
  \citenamefont {Coles}, \citenamefont {Sauvage}, \citenamefont {Larocca},\
  and\ \citenamefont {Cerezo}}]{nguyen2022theory}%
  \BibitemOpen
  \bibfield  {author} {\bibinfo {author} {\bibfnamefont {Quynh~T}\ \bibnamefont
  {Nguyen}}, \bibinfo {author} {\bibfnamefont {Louis}\ \bibnamefont
  {Schatzki}}, \bibinfo {author} {\bibfnamefont {Paolo}\ \bibnamefont
  {Braccia}}, \bibinfo {author} {\bibfnamefont {Michael}\ \bibnamefont
  {Ragone}}, \bibinfo {author} {\bibfnamefont {Patrick~J}\ \bibnamefont
  {Coles}}, \bibinfo {author} {\bibfnamefont {Frederic}\ \bibnamefont
  {Sauvage}}, \bibinfo {author} {\bibfnamefont {Martin}\ \bibnamefont
  {Larocca}}, \ and\ \bibinfo {author} {\bibfnamefont {M}~\bibnamefont
  {Cerezo}},\ }\bibfield  {title} {\enquote {\bibinfo {title} {Theory for
  equivariant quantum neural networks},}\ }\href@noop {} {\bibfield  {journal}
  {\bibinfo  {journal} {arXiv preprint arXiv:2210.08566}\ } (\bibinfo {year}
  {2022})}\BibitemShut {NoStop}%
\bibitem [{\citenamefont {Ragone}\ \emph {et~al.}(2022)\citenamefont {Ragone},
  \citenamefont {Braccia}, \citenamefont {Nguyen}, \citenamefont {Schatzki},
  \citenamefont {Coles}, \citenamefont {Sauvage}, \citenamefont {Larocca},\
  and\ \citenamefont {Cerezo}}]{ragone2022representation}%
  \BibitemOpen
  \bibfield  {author} {\bibinfo {author} {\bibfnamefont {Michael}\ \bibnamefont
  {Ragone}}, \bibinfo {author} {\bibfnamefont {Paolo}\ \bibnamefont {Braccia}},
  \bibinfo {author} {\bibfnamefont {Quynh~T}\ \bibnamefont {Nguyen}}, \bibinfo
  {author} {\bibfnamefont {Louis}\ \bibnamefont {Schatzki}}, \bibinfo {author}
  {\bibfnamefont {Patrick~J}\ \bibnamefont {Coles}}, \bibinfo {author}
  {\bibfnamefont {Frederic}\ \bibnamefont {Sauvage}}, \bibinfo {author}
  {\bibfnamefont {Martin}\ \bibnamefont {Larocca}}, \ and\ \bibinfo {author}
  {\bibfnamefont {M}~\bibnamefont {Cerezo}},\ }\bibfield  {title} {\enquote
  {\bibinfo {title} {Representation theory for geometric quantum machine
  learning},}\ }\href@noop {} {\bibfield  {journal} {\bibinfo  {journal} {arXiv
  preprint arXiv:2210.07980}\ } (\bibinfo {year} {2022})}\BibitemShut {NoStop}%
\bibitem [{\citenamefont {Luo}\ \emph {et~al.}(2021)\citenamefont {Luo},
  \citenamefont {Chen}, \citenamefont {Hu}, \citenamefont {Zhao}, \citenamefont
  {Hur},\ and\ \citenamefont {Clark}}]{luo2021gauge}%
  \BibitemOpen
  \bibfield  {author} {\bibinfo {author} {\bibfnamefont {Di}~\bibnamefont
  {Luo}}, \bibinfo {author} {\bibfnamefont {Zhuo}\ \bibnamefont {Chen}},
  \bibinfo {author} {\bibfnamefont {Kaiwen}\ \bibnamefont {Hu}}, \bibinfo
  {author} {\bibfnamefont {Zhizhen}\ \bibnamefont {Zhao}}, \bibinfo {author}
  {\bibfnamefont {Vera~Mikyoung}\ \bibnamefont {Hur}}, \ and\ \bibinfo {author}
  {\bibfnamefont {Bryan~K}\ \bibnamefont {Clark}},\ }\bibfield  {title}
  {\enquote {\bibinfo {title} {Gauge invariant autoregressive neural networks
  for quantum lattice models},}\ }\href@noop {} {\bibfield  {journal} {\bibinfo
   {journal} {arXiv preprint arXiv:2101.07243}\ } (\bibinfo {year}
  {2021})}\BibitemShut {NoStop}%
\bibitem [{\citenamefont {Luo}\ \emph {et~al.}(2022)\citenamefont {Luo},
  \citenamefont {Yuan}, \citenamefont {Stokes},\ and\ \citenamefont
  {Clark}}]{luo2022gauge}%
  \BibitemOpen
  \bibfield  {author} {\bibinfo {author} {\bibfnamefont {Di}~\bibnamefont
  {Luo}}, \bibinfo {author} {\bibfnamefont {Shunyue}\ \bibnamefont {Yuan}},
  \bibinfo {author} {\bibfnamefont {James}\ \bibnamefont {Stokes}}, \ and\
  \bibinfo {author} {\bibfnamefont {Bryan~K}\ \bibnamefont {Clark}},\
  }\bibfield  {title} {\enquote {\bibinfo {title} {Gauge equivariant neural
  networks for 2+ 1d u (1) gauge theory simulations in hamiltonian
  formulation},}\ }\href@noop {} {\bibfield  {journal} {\bibinfo  {journal}
  {arXiv preprint arXiv:2211.03198}\ } (\bibinfo {year} {2022})}\BibitemShut
  {NoStop}%
\bibitem [{\citenamefont {Hibat-Allah}\ \emph {et~al.}(2020)\citenamefont
  {Hibat-Allah}, \citenamefont {Ganahl}, \citenamefont {Hayward}, \citenamefont
  {Melko},\ and\ \citenamefont {Carrasquilla}}]{Hibat2020}%
  \BibitemOpen
  \bibfield  {author} {\bibinfo {author} {\bibfnamefont {Mohamed}\ \bibnamefont
  {Hibat-Allah}}, \bibinfo {author} {\bibfnamefont {Martin}\ \bibnamefont
  {Ganahl}}, \bibinfo {author} {\bibfnamefont {Lauren~E.}\ \bibnamefont
  {Hayward}}, \bibinfo {author} {\bibfnamefont {Roger~G.}\ \bibnamefont
  {Melko}}, \ and\ \bibinfo {author} {\bibfnamefont {Juan}\ \bibnamefont
  {Carrasquilla}},\ }\bibfield  {title} {\enquote {\bibinfo {title} {Recurrent
  neural network wave functions},}\ }\href {\doibase
  10.1103/PhysRevResearch.2.023358} {\bibfield  {journal} {\bibinfo  {journal}
  {Phys. Rev. Res.}\ }\textbf {\bibinfo {volume} {2}},\ \bibinfo {pages}
  {023358} (\bibinfo {year} {2020})}\BibitemShut {NoStop}%
\bibitem [{\citenamefont {Morawetz}\ \emph {et~al.}(2021)\citenamefont
  {Morawetz}, \citenamefont {De~Vlugt}, \citenamefont {Carrasquilla},\ and\
  \citenamefont {Melko}}]{morawetz2021u}%
  \BibitemOpen
  \bibfield  {author} {\bibinfo {author} {\bibfnamefont {Stewart}\ \bibnamefont
  {Morawetz}}, \bibinfo {author} {\bibfnamefont {Isaac~JS}\ \bibnamefont
  {De~Vlugt}}, \bibinfo {author} {\bibfnamefont {Juan}\ \bibnamefont
  {Carrasquilla}}, \ and\ \bibinfo {author} {\bibfnamefont {Roger~G}\
  \bibnamefont {Melko}},\ }\bibfield  {title} {\enquote {\bibinfo {title} {U
  (1)-symmetric recurrent neural networks for quantum state reconstruction},}\
  }\href {https://journals.aps.org/pra/abstract/10.1103/PhysRevA.104.012401}
  {\bibfield  {journal} {\bibinfo  {journal} {Physical Review A}\ }\textbf
  {\bibinfo {volume} {104}},\ \bibinfo {pages} {012401} (\bibinfo {year}
  {2021})}\BibitemShut {NoStop}%
\bibitem [{\citenamefont {Rudolph}\ \emph
  {et~al.}(2022{\natexlab{a}})\citenamefont {Rudolph}, \citenamefont {Chen},
  \citenamefont {Miller}, \citenamefont {Acharya},\ and\ \citenamefont
  {Perdomo-Ortiz}}]{rudolph2022MPSdecomposition}%
  \BibitemOpen
  \bibfield  {author} {\bibinfo {author} {\bibfnamefont {Manuel~S}\
  \bibnamefont {Rudolph}}, \bibinfo {author} {\bibfnamefont {Jing}\
  \bibnamefont {Chen}}, \bibinfo {author} {\bibfnamefont {Jacob}\ \bibnamefont
  {Miller}}, \bibinfo {author} {\bibfnamefont {Atithi}\ \bibnamefont
  {Acharya}}, \ and\ \bibinfo {author} {\bibfnamefont {Alejandro}\ \bibnamefont
  {Perdomo-Ortiz}},\ }\bibfield  {title} {\enquote {\bibinfo {title}
  {Decomposition of matrix product states into shallow quantum circuits},}\
  }\href@noop {} {\bibfield  {journal} {\bibinfo  {journal} {arXiv preprint
  arXiv:2209.00595}\ } (\bibinfo {year} {2022}{\natexlab{a}})}\BibitemShut
  {NoStop}%
\bibitem [{\citenamefont {Rudolph}\ \emph
  {et~al.}(2022{\natexlab{b}})\citenamefont {Rudolph}, \citenamefont {Miller},
  \citenamefont {Chen}, \citenamefont {Acharya},\ and\ \citenamefont
  {Perdomo-Ortiz}}]{rudolph2022synergistic}%
  \BibitemOpen
  \bibfield  {author} {\bibinfo {author} {\bibfnamefont {Manuel~S.}\
  \bibnamefont {Rudolph}}, \bibinfo {author} {\bibfnamefont {Jacob}\
  \bibnamefont {Miller}}, \bibinfo {author} {\bibfnamefont {Jing}\ \bibnamefont
  {Chen}}, \bibinfo {author} {\bibfnamefont {Atithi}\ \bibnamefont {Acharya}},
  \ and\ \bibinfo {author} {\bibfnamefont {Alejandro}\ \bibnamefont
  {Perdomo-Ortiz}},\ }\bibfield  {title} {\enquote {\bibinfo {title} {Synergy
  between quantum circuits and tensor networks: Short-cutting the race to
  practical quantum advantage},}\ }\href {https://arxiv.org/abs/2208.13673}
  {\bibfield  {journal} {\bibinfo  {journal} {arXiv:2208.13673}\ } (\bibinfo
  {year} {2022}{\natexlab{b}})}\BibitemShut {NoStop}%
\bibitem [{\citenamefont {Benedetti}\ \emph {et~al.}(2019)\citenamefont
  {Benedetti}, \citenamefont {Garcia-Pintos}, \citenamefont {Perdomo},
  \citenamefont {Leyton-Ortega}, \citenamefont {Nam},\ and\ \citenamefont
  {Perdomo-Ortiz}}]{Benedetti2019}%
  \BibitemOpen
  \bibfield  {author} {\bibinfo {author} {\bibfnamefont {Marcello}\
  \bibnamefont {Benedetti}}, \bibinfo {author} {\bibfnamefont {Delfina}\
  \bibnamefont {Garcia-Pintos}}, \bibinfo {author} {\bibfnamefont {Oscar}\
  \bibnamefont {Perdomo}}, \bibinfo {author} {\bibfnamefont {Vicente}\
  \bibnamefont {Leyton-Ortega}}, \bibinfo {author} {\bibfnamefont {Yunseong}\
  \bibnamefont {Nam}}, \ and\ \bibinfo {author} {\bibfnamefont {Alejandro}\
  \bibnamefont {Perdomo-Ortiz}},\ }\bibfield  {title} {\enquote {\bibinfo
  {title} {A generative modeling approach for benchmarking and training shallow
  quantum circuits},}\ }\href
  {https://www.nature.com/articles/s41534-019-0157-8} {\bibfield  {journal}
  {\bibinfo  {journal} {npj Quantum Information}\ }\textbf {\bibinfo {volume}
  {5}},\ \bibinfo {pages} {45} (\bibinfo {year} {2019})}\BibitemShut {NoStop}%
\bibitem [{\citenamefont {Hemmecke}\ \emph {et~al.}(2010)\citenamefont
  {Hemmecke}, \citenamefont {K{\"o}ppe}, \citenamefont {Lee},\ and\
  \citenamefont {Weismantel}}]{hemmecke2010nonlinear}%
  \BibitemOpen
  \bibfield  {author} {\bibinfo {author} {\bibfnamefont {Raymond}\ \bibnamefont
  {Hemmecke}}, \bibinfo {author} {\bibfnamefont {Matthias}\ \bibnamefont
  {K{\"o}ppe}}, \bibinfo {author} {\bibfnamefont {Jon}\ \bibnamefont {Lee}}, \
  and\ \bibinfo {author} {\bibfnamefont {Robert}\ \bibnamefont {Weismantel}},\
  }\href@noop {} {\emph {\bibinfo {title} {Nonlinear integer programming}}}\
  (\bibinfo  {publisher} {Springer},\ \bibinfo {year} {2010})\BibitemShut
  {NoStop}%
\bibitem [{\citenamefont {Perez-Garcia}\ \emph {et~al.}(2007)\citenamefont
  {Perez-Garcia}, \citenamefont {Verstraete}, \citenamefont {Wolf},\ and\
  \citenamefont {Cirac}}]{garcia_mps}%
  \BibitemOpen
  \bibfield  {author} {\bibinfo {author} {\bibfnamefont {D}~\bibnamefont
  {Perez-Garcia}}, \bibinfo {author} {\bibfnamefont {F}~\bibnamefont
  {Verstraete}}, \bibinfo {author} {\bibfnamefont {MM}~\bibnamefont {Wolf}}, \
  and\ \bibinfo {author} {\bibfnamefont {JI}~\bibnamefont {Cirac}},\ }\bibfield
   {title} {\enquote {\bibinfo {title} {Matrix product state
  representations},}\ }\href@noop {} {\bibfield  {journal} {\bibinfo  {journal}
  {Quantum Information \& Computation}\ }\textbf {\bibinfo {volume} {7}},\
  \bibinfo {pages} {401--430} (\bibinfo {year} {2007})}\BibitemShut {NoStop}%
\end{thebibliography}%

\clearpage
\newpage


\appendix

\section{Introduction to \textsc{(0,1)-Integer Linear Programming}}\label{s:0-1}

\textsc{(0,1)-Integer Linear Programming} problems are optimization problems that involve binary decision variables, taking values 0 or 1. The objective is to find the combination of variables that maximize or minimize a given linear function subject to a set of linear constraints.

\subsection{Formulation of the Problem}

The general form of a \textsc{(0,1)-Integer Linear Programming} problem can be stated as follows \cite{schrijver1998theory}: \\

\textsc{Maximize or minimize } $\vec{c}\cdot \vec{x}$\\

\textsc{subject to } $A\vec{x}\leq \vec{b}$,\\

where $c_i$ and $A_{ij}$ are integer coefficients of the objective function and the constraints, respectively. The variables $x_i$, take on binary values of 0 or 1, and $b_i$ are integers. This class of problems belong to one of Karps' list of 21 \textsc{NP-complete} problems. Hence, solving instances of this problem takes in general exponential computational resources in the problem size.

There exist extensions of the \textsc{(0,1)-Integer Linear Programming} to the case when some of the variables are let to be arbitrary reals and the rest arbitrary integers, in which case this class of problems goes by the name of \textsc{Mixed Integer Linear Programming} (MILP) problems. This class of problems are well-known to be \textsc{NP-Hard} as well in general. Other extensions include the case when the objective function and/or constraints are not linear functions (and some variables are still integer valued). The complexity class corresponding to \textit{nonlinear} integer programming is also \textsc{NP-Hard} and expected to be in general much harder to solve than their linear counterparts.

We will briefly discuss some standard methods for solving \textsc{(0,1)-Integer Linear Programming} problems (we refer the reader to Ref. \cite{hemmecke2010nonlinear} discussing some of their extensions to the presence of nonlinear cost functions and linear constraints, which is the class of problems our TN algorithm is able to handle).

\subsection{Solution Methods}

The solution to a \textsc{(0,1)-Integer Linear Programming} problem can be found using a variety of methods. We distinguish two classes of methods: \textit{exact} methods, including branch and bound, cutting plane, and branch and cut algorithms; as well as \textit{heuristic} methods, including simulated annealing, parallel tempering, and genetic algorithms. The difference between these two classes is that the former provides guarantees of optimality of the solutions found, but do not scale well with problem size in general. The second class of methods, being heuristic do not have guarantees of optimality but do not suffer from the same scalability issues. Here we review the former class of methods since they are the ones most frequently used in most MILP solvers \cite{gleixner2021miplib}. This class of methods involve iteratively solving a series of linear programming problems, where the decision variables are not restricted to binary values. All of these methods are discussed in detail in standard references such as Ref. \cite{schrijver1998theory}.

\textit{Branch and bound -- .} The branch and bound algorithm works by dividing the search space into smaller subspaces (or \textit{branches}), and using bounds on the objective function to eliminate subspaces that are guaranteed to contain suboptimal solutions. The algorithm then selects the subspace with the best solution and divides it again until the optimal solution is found.

\textit{Cutting plane --.} The cutting plane algorithm involves adding additional linear constraints to the problem to exclude certain solutions. These additional constraints are based on the solution to the current linear programming problem and are designed to tighten the feasible region of the search space.

\textit{Branch and cut --.} The branch and cut algorithm combines the techniques of branch and bound and cutting plane. The algorithm begins by using branch and bound to divide the search space, and then uses cutting plane to tighten the feasible region of each subspace.

\section{Symmetric Tensor Networks}\label{s:symmetric_TNs_app}
In this Appendix we give a brief introduction to Matrix Product States and their $U(1)$ symmetric versions, as done in \cite{singh2010tensor, singh2011tensor}.
\subsection{Vanilla Tensor Networks and Matrix Product States}~\label{s:intro_MPS_app}
Consider a system of $N$ sites where each local site is characterized by a local complex vector space $\mathbb{V}$. A wavefunction for this system is given as
\begin{equation} \label{eq_wf_expansion}
|\Psi\rangle = \sum_{i_1,i_2,\cdots,i_N} (\Psi)_{i_1,i_2,\cdots,i_N}|i_1,i_2,\cdots,i_N\rangle,
\end{equation}
with $|i_v\rangle \in \mathbb{V}$. The coefficient $(\Psi)_{i_1,i_2,\cdots,i_N}$ is in fact a \textit{tensor}, of rank $N$. The main idea behind tensor networks is to decompose this big tensor into tensors of smaller rank contracted with each other such that \cite{singh2011tensor}
\begin{equation}
(\Psi)_{i_1,i_2,\cdots,i_N} = \text{t Tr}\left(\bigotimes_{v=1}^N T^{[v]i_v}\right),
\end{equation}
where $\text{t Tr}$ stands for tensor trace which contracts all repeated tensor indices of tensors $T^{[v]}$. A tensor network of special relevance is the \textit{Matrix Product State} (MPS) which is given as
\begin{equation}
(\Psi)_{i_1,i_2,\cdots,i_N} = \text{Tr}(T^{[1]i_1}T^{[2]i_2}\cdots T^{[N]i_N}),
\end{equation}
where the trace now is a conventional matrix trace, where each tensor $T^{[v]i_v}$ is a $\chi_{v-1}\times \chi_{v}$ matrix with components $T^{[v]i_v}_{l_{v-1},l_v}$. For our purposes, we shall be considering the MPS with \textit{open boundary conditions} (OBC) \cite{garcia_mps} so that the left- and right-most tensors are actually vectors  so that $\chi_{0}=\chi_{N}=1$ (so that the trace is not needed). A perspective that is useful when dealing with symmetry group transformations on tensors, is to view tensors as linear maps from an \textit{input} vector space $\mathbb{V}^{\rm in}$ to \textit{output} vector space $\mathbb{V}^{\rm out}$. A significant simplification in many tensor network algorithms involving MPS is to use the canonical form \cite{schollwock2011density}, which is used throughout in this work. When put in canonical form, tensors to the left of the canonical center correspond to left isometries, satisfying $|\beta\rangle=\sum_{a,\alpha}T_{\alpha,\beta}^a|a,\alpha\rangle$ with $\sum_{a}(T^a)^\dagger T^a=\mathbb{1}$ and where $|a, \alpha\rangle \in \mathbb{V}^{\rm out}$ and $|\beta\rangle \in \mathbb{V}^{\rm in}$, while tensors to the right of the canonical center correspond to right isometries, satisfying $|\alpha\rangle=\sum_{a,\beta}T_{\alpha,\beta}^a|a,\beta\rangle$ with $\sum_{a}T^a (T^a)^{\dagger}=\mathbb{1}$ and where $|\alpha\rangle \in \mathbb{V}^{\rm in}$ and $|a,\beta\rangle \in \mathbb{V}^{\rm out}$. Each \textit{link} or \textit{virtual} state can be represented as $|l_v\rangle \in \{|1\rangle, |2\rangle, \cdots, |\chi_v\rangle\}$. The size of each link vector space is sometimes known as bond dimension, and its magnitude controls the amount of entanglement shared by connected tensors. The states associated to the upper index $i_v$ will be referred to as \textit{physical} states, since they are the ones that appear in the original wavefunction expansion (\ref{eq_wf_expansion}), at variance with the virtual states just discussed.
\subsection{$U(1)$ Symmetric Matrix Product States}
Since tensors are linear transformations, the tensor network ansatz makes it very convenient to exploit the toolbox of representation theory. In essence, by requiring that our state $|\Psi\rangle$ be \textit{invariant} under a \textit{global} group transformation $\mathcal{G}$ represented by a unitary matrix $U_g$, with $g\in \mathcal{G}$ s.t. \cite{singh2010tensor}
\begin{equation} \label{eq_Ug_psi}
    (U_g)^{\otimes N}|\Psi\rangle = |\Psi\rangle,
\end{equation}
we may constrain the form of each tensor in the MPS as dictated by the transformation
\begin{equation} \label{eq_constraint_tensors}
    \sum_{a',b',c'} (U_g)_{c,c'}(V_g)_{b,b'}T_{a',c'}^{b'}(W_g)^{\dagger}_{a',a}=T^{b}_{a,c},
\end{equation}
where $V_g, W_g$ are unitary representations of the same group $\mathcal{G}$. The upshot of this is that each tensor will be constrained to have nonzero entries only at those entries fulfilling Eq. (\ref{eq_constraint_tensors}). Our focus here will be on tensors that are \textit{symmetric} under $\mathcal{G}=U(1)$, where representations are one dimensional and parameterized as $U_g=e^{-i\hat{n}\phi}$, with $\hat{n}$ the so-called \textit{charge} operator and $\phi \in [0,2\pi)$. This means that our state must be invariant under this group transformation up to a phase:
\begin{equation} \label{eq_symmetric_U1}
    e^{-i \hat{n}\phi}|\Psi\rangle = e^{-in\phi}|\Psi\rangle,
\end{equation}
where the state $|\Psi\rangle$ is an eigenstate of the charge operator $\hat{n}$ with eigenvalue $n$,
\begin{equation}
    \hat{n}|\Psi\rangle = n |\Psi\rangle.
\end{equation}
Note that, the case $n=0$ corresponds to  the state being invariant (as in (\ref{eq_Ug_psi})), while for $n\neq 0$ one obtains a \textit{covariant} state. Such phase factor only arises for states that are covariant under Abelian group transformations like $U(1)$ (in the case of non-Abelian group transformations, covariant states transform instead under a unitary matrix of dimension 2 or higher). The different charges $n$ label different irreps of $U(1)$ so that for each vector space $\mathbb{V}$ we may decompose this as
\begin{equation}
    \mathbb{V}\simeq \bigoplus_n \mathbb{V}_n,
\end{equation}
where $\mathbb{V}_n$ are irreps of fixed charge $n$, each of dimension (or \textit{degeneracy}) $d_n$. Thus, any arbitrary vector $|v\rangle \in \mathbb{V}$ can be expanded as a linear combination of $|n,t_n\rangle \in \mathbb{V}_n$ vectors of fixed charge and degeneracy label $t_n$, where $t_n=1,\cdots, d_n$, with $d_n$ the degeneracy of charge $n$. By going to this basis of well-defined charge, the invariance condition on the tensors, (\ref{eq_constraint_tensors}), translates into
\begin{equation} \label{eq_invariant_tensors}
    T^{b}_{a,c}=(T^{n_b}_{n_a,n_c})^{t_{n_b}}_{t_{n_a},t_{n_c}}\delta_{N_{\rm in} , N_{\rm out}},
\end{equation}
where
\begin{equation}
N_{\rm in}=\sum_{i\in \mathcal{I}} n_i, \hspace{0.1in} N_{\rm out}=\sum_{i \in \mathcal{O}} n_i,
\end{equation}
where $\mathcal{I}$ ($\mathcal{O}$) denotes the set of incoming (outgoing) indices. From (\ref{eq_invariant_tensors}) we see that each tensor in the MPS factorizes into the product of two tensors. The first one, $(T^{n_b}_{n_a,n_c})^{t_b}_{t_a,t_c}$, gives the components of tensor $T$ w.r.t. the degeneracy vectors $\{|t_n\rangle\}$, reason why this tensor goes by the name of \textit{degeneracy tensor}. The second tensor, $\delta_{N_{\rm in},N_{\rm out}}$ depends solely on the conservation of $U(1)$ charge, and it is referred to as \textit{structural tensor} (for an arbitrary symmetry group $\mathcal{G}$ the structural tensor can take a rather nontrivial form). For fixed physical charge (fixed $n_b$ appearing in (\ref{eq_invariant_tensors})), the resulting matrix $T^b_{a,c}$ will be block diagonal, with blocks of size $d_{n_a}\times d_{n_c}$. These blocks are sometimes referred to as \textit{quantum number} (QN) blocks. A symmetric MPS will be in general not only composed of invariant tensors fulfilling (\ref{eq_invariant_tensors}) but also covariant tensors of well defined charge $n\neq 0$ and fulfilling
\begin{equation}
    \sum_{a',b',c'} (U_g)_{c,c'}(V_g)_{b,b'}T_{a',c'}^{b'}(W_g)^{\dagger}_{a',a}=e^{-in\phi}T^{b}_{a,c}.
\end{equation}
Covariant tensors can be most compactly written in the basis of well-defined charge as
\begin{equation} \label{eq_covariant_tensors}
    T^{b}_{a,c}=(T^{n_b}_{n_a,n_c})^{t_{n_b}}_{t_{n_a},t_{n_c}}\delta_{N_{\rm in} +n, N_{\rm out}}.
\end{equation}
Alternatively, we may transform a covariant tensor to an invariant tensor upon introducing an extra index carrying charge $n$ and of dimension 1. The charge $n$ in a covariant tensor is also sometimes known as the \textit{flux}. For a given MPS, there is at least one covariant tensor, and this is usually taken to be located at the canonical center of the MPS. Alternatively, we may decide to spread the total charge across different sites. It is however preferred to concentrate all flux in one single site as it is easier for bookkeeping purposes (we do not have to keep track of which sites have flux and which do not) and the resulting complexity (as measured by the size of the tensors) is not affected by this choice.

Lastly, we remark that for the purposes of our work where we have multiple arbitrary equality constraints, the above formalism still carries through. The major difference is that now we require the state to be invariant under multiple global group transformations of the form $U_g=e^{-i\sum_j \alpha_j \hat{n}_j \phi}$. Since each of the operators $\hat{n}_j$ acts locally on each tensor, it is straightforward to generalize the above formulas to the case of arbitrary number of charges (i.e. translation invariance is not a requirement).
\section{Exact MPS construction for assignment type problems} \label{sec:assignment}
\begin{figure}
    \centering
    \includegraphics[width=\linewidth, scale=0.5]{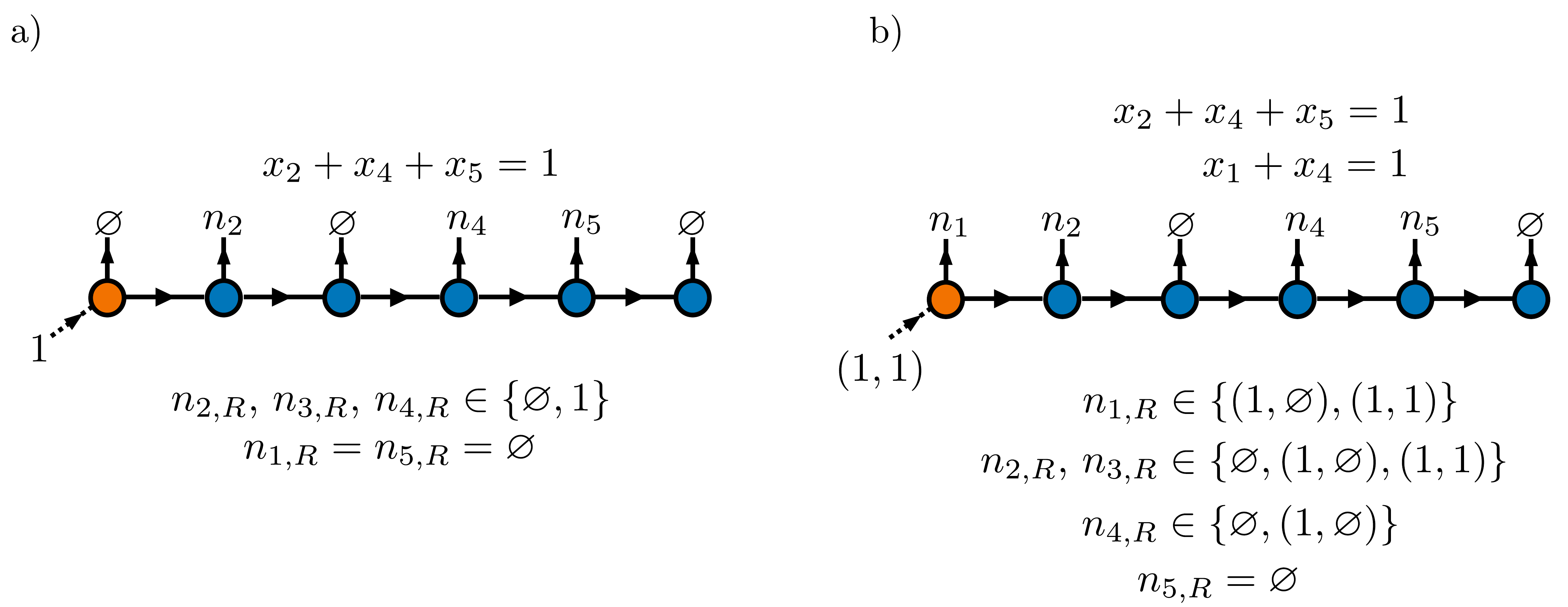}
    \caption{\textbf{Charges in an MPS in the presence of assignment type equalities.} a) One equation. b) Two equations. }
    \label{fig:assignment}
\end{figure}
In Sec. \ref{subs_constr_u1} we gave an exact construction of the MPS in the presence of a cardinality constraint. Here we give the derivation for another widely present equality constraint in combinatorial optimization problems, namely, an \textit{assignment} type constraint. This is of the form
\begin{equation}
    \sum_{i \in \mathcal{A}} x_i = 1,
\end{equation}
with $\mathcal{A}$ some set of site indices. The valid space for this constraint can be encoded in an MPS of $\chi=2$, with link charges $n_{i,R}\in\{\varnothing, 1\}$. See Fig. \ref{fig:assignment} for an illustration for a single equality. For $m$ number of assignment type equalities, the link charges take values in $n_{i,R}\in\{\varnothing, 1\}^m$, see Fig. \ref{fig:assignment} for an example of two equalities. Thus at worst the bond dimension grows exponentially with the number of assignments.

\section{Degeneracy counting for negative separation cost function}\label{s:degeneracy}
The negative separation cost function together with the cardinality constraint as introduced in Sec. \ref{sec_quality_II} has $\kappa-1$ bitstrings with minimum cost function, which is $\mathcal{C}_{\rm min}=-N+\kappa-1$. This corresponds to having a single connected block of zeros of size $N-\kappa$, with ones at each edge. For the remainder we shall refer to an arbitrary-sized block of zeros with ones at each side as a \textit{domain wall} (to connect with the statistical physics literature). Here we wish to characterize the degeneracy (that is, number of bitstrings) for arbitrary value of the cost function and for any system size. This will be useful as a performance benchmark of the trained s-TNBM against our baseline which is a uniform sampler on the space of bitstrings with the right cardinality $\kappa$. Since for large system sizes $N\gtrsim 30$ it becomes numerically challenging to extract the degeneracy factors, we aim at extracting these analytically.

The problem of counting such configurations is a straightforward problem of combinatorics. The approach we take here is recursive. We present results for the case when there is a single dominant domain wall, meaning one domain wall is bigger than the rest in a given bitstring (this is analogous to a low-energy approximation above the absolute minimum when viewing the cost function as an energy function). Our results are as follows.

Let $a \in \mathbb{Z}$ with $2a < N-\kappa$, where $\kappa$ is the Hamming weight (cardinality). The degeneracy in the number of bitstrings with cost $\mathcal{C}=-N+\kappa+a-1$ is given as
\begin{widetext}
\begin{equation}
  |\mathcal{S}|(a,\kappa)=
    \begin{cases}
    2\sum_{i=a}^{\kappa+a-2}\binom{i}{a}+2\sum_{j=1}^{\frac{a-1}{2}}\sum_{i=a-j}^{\kappa+a-2-j}\binom{i}{a-j}\binom{\kappa+a-2-i}{j} ,\hspace{0.1in} &\text{if } a \in 2\mathbb{Z}+1, \\
    2\sum_{i=a}^{\kappa+a-2}\binom{i}{a}+2\sum_{j=1}^{\frac{a}{2}-1}\sum_{i=a-j}^{\kappa+a-2-j}\binom{i}{a-j}\binom{\kappa+a-2-i}{j}+\sum_{i=\frac{a}{2}}^{\kappa-2+\frac{a}{2}}\binom{i}{\frac{a}{2}}\binom{\kappa+a-2-i}{\frac{a}{2}} ,\hspace{0.1in} &\text{if } a \in 2\mathbb{Z}.
    \end{cases}
\end{equation}
\end{widetext}
The way to derive this result consists in a simple counting of the number of configurations of zeros at each side of the main domain wall. In particular, for the experiment of Fig. \ref{fig:utility_evol}, we have $|\mathcal{S}|(6,25)/|\mathcal{S}| = \mathcal{O}(10^{-7})$, where $|\mathcal{S}|=\binom{50}{25}$ the number of solutions of fixed cardinality.

\end{document}